\documentclass[preprint]{aastex63}

\usepackage{amsbsy}
\usepackage{amsmath}
\usepackage{textcomp}
\usepackage{amsmath}
\usepackage{color}

\usepackage{graphicx}
\usepackage{epstopdf}

\DeclareMathAlphabet{\mathsfsl}{OT1}{cmss}{m}{sl}

\begin{document}
	
	\title{Bayesian Cross-Matching of High Proper Motion Stars in \textit{Gaia} DR2 and Photometric Metallicities for $\sim$1.7 million K and M Dwarfs}
	
	\author{Ilija Medan}
	\affiliation{Department of Physics and Astronomy, Georgia State University, Atlanta, GA 30302, USA}
	
	\author{S\'{e}bastien L\'{e}pine}
	\affiliation{Department of Physics and Astronomy, Georgia State University, Atlanta, GA 30302, USA}

	\author{Zachary Hartman}
	\affiliation{Department of Physics and Astronomy, Georgia State University, Atlanta, GA 30302, USA}	
	\affiliation{Lowell Observatory, 1400 W Mars Hill Road, Flagstaff, AZ 86001}
	
\begin{abstract}
We present a Bayesian method to cross-match 5,827,988 high proper motion \textit{Gaia} sources ($\mu>40 \ mas \ yr^{-1}$) to various photometric surveys: 2MASS, AllWISE, GALEX, RAVE, SDSS and Pan-STARRS. To efficiently associate these objects across catalogs, we develop a technique that compares the multidimensional distribution of all sources in the vicinity of each \textit{Gaia} star to a reference distribution of random field stars obtained by extracting all sources in a region on the sky displaced 2$^\prime$. This offset preserves the local field stellar density and magnitude distribution allowing us to characterize the frequency of chance alignments. The resulting catalog with Bayesian probabilities $>$95\% has a marginally higher match rate than current internal \textit{Gaia} DR2 matches for most catalogs. However, a significant improvement is found with Pan-STARRS, where $\sim$99.8\% of the sample within the Pan-STARRS footprint is recovered, as compared to a low $\sim$20.8\% in \textit{Gaia} DR2. Using these results, we train a Gaussian Process Regressor to calibrate two photometric metallicity relationships. For dwarfs of $3500<T_{eff}<5280$ K, we use metallicity values of 4,378 stars from APOGEE and \citet{hejazi2020} to calibrate the relationship, producing results with a $1\sigma$ precision of 0.12 dex and few systematic errors. We then indirectly infer the metallicity of 4,018 stars with $2850<T_{eff}<3500$ K, that are wide companions of primaries whose metallicities are estimated with our first regressor, to produce a relationship with a $1\sigma$ precision of 0.21 dex and significant systematic errors. Additional work is needed to better remove unresolved binaries from this sample to reduce these systematic errors.
\end{abstract}

\keywords{catalog, astrometry, stars: abundances}

\received{Sep. 22, 2020}

\accepted{Feb. 17, 2021}

\submitjournal{The Astronomical Journal}

\section{Introduction}

Over the past couple of decades, the stellar astronomical community has created multiple sky surveys that chart the positions and measure the brightness of stars in the local Milky Way in different wavelength regimes, down to different brightness limits and for varying regions of the sky. By combining these various catalogs, astronomers are able to derive a wealth of knowledge about the stellar properties and kinematics for hundreds of millions of stars. The recent \textit{Gaia} Data Release 2 \citep{gaiadr2} is arguably the most important catalog to be released in recent years, as it not only provides accurate photometry for an unprecedented numbers of stars, but it also provides the most accurate position and motion measurements to date.

With the number of very large stellar catalogs on the rise, developing methods to accurately match stars between catalogs becomes very important. Distant, bright stars, which make up most of the stars in traditional catalogs, are typically straightforward to match as their brightness make them prominent compared with fainter field stars in their vicinity, and because their position does not vary much over time. Of particular interest, however, are the nearby, low-mass stars, which are typically faint on the sky, and have large parallaxes and proper motions, now readily available from \textit{Gaia}. Due to their faintness, they are more easily confused with background stars in the field, and because of their significant change in position over time, these stars become much more difficult to match correctly between catalogs.

In recent years this problem has been addressed through multiple methods, some statistical in nature. \citet{tamas2008} created a Bayesian framework that mostly relies on the angular separation between objects in order to get a probability of a match. This method can be extended to include other parameters (such as photometry) by assuming these distributions to be independent from the probability distributions for the angular separation between catalog objects. This framework however does not set up procedures for dealing with catalogs that have significantly different epochs of observation, when large proper motions are involved. Additionally, the assumption of independence between distributions in angular separation and brightness may not hold, as we will show in this work.

Additionally, many large catalogs are attempting to address this issue by including their own cross-matches against popular external catalogs. \textit{Gaia} includes such attempted cross-matches using a custom, multi-filter algorithm created to find the best matches between \textit{Gaia} and several external catalogs based on the \textit{Gaia} astrometry \citep{gaia_cross_match}. While this does expand on the \citet{tamas2008} approach by considering astrometric motions, the method does not explicitly consider the photometry of the sources in either of the catalogs during the match. 

This study aims to address both of the components missing in the above two methods by combining astrometry and photometry to create a Bayesian framework to match high proper motion ($\mu>40 \ mas \ yr^{-1}$) stars in \textit{Gaia} with existing catalogs that span the wavelength regime from the UV to the Infrared.

Complete and accurate photometry for stars across a larger portion of the EM spectrum is very important for those interested in studying the physical properties of stars. A catalog of high proper motion stars, in particular, focuses on low-mass field stars in the solar vicinity, whose magnitudes and colors are less affected by interstellar reddening and potentially can be used to estimate their masses and chemical compositions. This is important as low mass stars are the most abundant stars in the Galaxy \citep{bonchanski2010} and their long main sequence lifetimes mean they have the potential to trace the entirety of the Galactic star-formation history, and provide clues to understand the structure and evolution of the Milky Way. 

Additionally, low mass stars are of great interest for the exoplanet community, as these sources are the best targets for detecting terrestrial planets in habitable zones, due to such planets having shorter orbital periods, and creating deeper transits and producing larger Doppler shifts in their host stars \citep{carmenes}. With missions like the Transiting Exoplanet Survey Satellite \citep[TESS;][]{tess}, knowledge of the chemistry of low mass stars allows for constraints to be made on the physical properties of the orbiting planets and their formation history.

To get an approximation of physical properties with minimal work for large number of stars, photometry has been used in place of spectroscopy by using calibrated relationships between various photometric bands and physical stellar properties. This concept has been used to successfully calibrate relationships for hotter dwarfs of spectral types F,G and K to estimate effective temperature \citep[e.g.][]{gonzalez2009,casagrande2010} and metallicity \citep[e.g.][]{ramirez2005}. Calibrations of K and M dwarfs have proved more difficult as spectra needed for such a calibration required high signal-to-noise to accurately measure equivalent widths of atomic lines. The use of large telescopes have however allowed for such measurements for limited subsets of very nearby low-mass stars \citep[e.g.][]{woolf2005,woolf2006}, which have provided tentative photometric calibrations for stellar parameters of low mass dwarfs \citep[e.g.][]{bonfils2005,casagrande2008,dittmann2016}. 

These relationships however rely on photometric band measurements (e.g. Johnson–Cousins band-passes) that are not common in the most extensive modern astronomical surveys, which means they are not applicable to the largest datasets now available. To update these calibrations, \citet{Schmidt2016} combined SDSS and AllWISE photometry with the derived stellar parameters of 3,834 stars from APOGEE spectra to calibrate new relationships for K and early M dwarfs metallicities down to temperatures of 3500 K with a precision of $\sim$0.18 dex. This calibration is however limited by the fact that: $(1)$ APOGEE spectra do not allow for metallicity determinations for lower mass objects, and $(2)$ that it relies on SDSS colors, which limits the sky coverage over which this relationship can be used. More recently, \citet{davenport2019} used \textit{Gaia}, 2MASS and AllWISE photometry to calibrate a color-metallicity relationship with a precision of $\sim$0.11 dex. This relationship however only applies to more massive FGK dwarfs, and also appears to suffer from systematic errors. In their relationship, metal-poor stars have systematically overestimated photometric metallicities, while metal-rich stars have underestimated photometric metallicities.

Despite the large amount of work done to calibrate photometric relationships for dwarf stars down to temperatures of $\sim$3500 K, these relationships still are not applicable to cooler M dwarfs. Reliable metallicity measurements from the spectra of M dwarfs have long been a challenge due to the complex molecular bands in their spectra that depress the continuum and hinder precise metallicity determinations \citep{allard1997}. Recent studies have made progress in improving metallicity precision for M dwarfs by using moderate resolution spectra of common proper motion pairs with an M dwarf component to develop empirical relationships \citep[e.g.][]{rojas2010,terrien2012,newton2014}, while others have been successful with deriving parameters for comparison with improved synthetic spectra \citep[e.g.][]{lindgren2016,hejazi2020}. Despite these advances, high resolution spectra of M dwarfs are still not widely available, making direct calibration of photometric metallicity relationships for M dwarfs difficult.

 In this study, we use the photometric results from our detailed, multi-catalog cross-match and derive an improved color-metallicity calibration for K and early M dwarfs. We also develop a new relationship to predict metallicity for M dwarfs of $T_{eff}<3500$ K by using M dwarfs in wide binary systems as calibrators. Our relationship uses filter bandpasses that are available over large regions of the sky, which enable us to estimate metallicities for 1,708,194 nearby low-mass stars, with a typical precision of $\pm0.12$ dex for stars of $T_{eff}>3500$ K and $\pm0.28$ dex for stars of $T_{eff}<3500$ K.

\section{Data sets}

\subsection{High Proper Motion \textit{Gaia} Subset}

\textit{Gaia} sources that have larger proper motions can be the most challenging to match accurately due to their large apparent motions. We extract from \textit{Gaia} DR2 a subset of 5,827,988 stars with recorded proper motions $\mu>40 \ mas \ yr^{-1}$; this will be our primary catalog for the subsequent cross match. We emphasize that no ``cleaning" has been done to this subset, such that all \textit{Gaia} sources with proper motions listed with $\mu>40\ mas \ yr^{-1}$ are included regardless of source magnitude or uncertainty in the astrometry and/or photometry. All coordinate entries in \textit{Gaia} are recorded at an epoch of $J=$2015.5 using the measured proper motions.

\subsection{External Catalogs}

\subsubsection{2MASS}

The Two Micron All Sky Survey (2MASS) was conducted between June 1997 and February 2001 and collected photometry of stars for the entirety of the sky in three infrared bands: $J$ (1.25 $\mu$m), $H$ (1.65 $\mu$m), and $K_s$ (2.16 $\mu$m) \citep{2mass}. The final point source catalog includes 470,992,970 individual sources where, with a 10 $\sigma$ point-source detection level, objects are detected down to a limit of 15.8, 15.1, and 14.3 mag for the $J$, $H$, and $K_s$ bands, respectively.

\subsubsection{AllWISE}

The  Wide-field  Infrared Survey  Explorer  (WISE) operated between January 2010 and July 2010 and collected photometry for the entirety of the sky in four infrared bands: $W1$ (3.4 $\mu$m), $W2$ (4.6 $\mu$m), $W3$ (12 $\mu$m), and $W4$ (22 $\mu$m) \citep{wise}. For this study, we have opted to use the updated version of the catalog, AllWISE, which has enhanced sensitivity and accuracy compared with earlier WISE data releases \citep{allwise}. This release consists of 747,634,026 individual sources. Readers should be aware that the AllWISE release results in saturation issues for photometric measurements of W1 $<8$ and W2 $<7$\footnote{\url{https://wise2.ipac.caltech.edu/docs/release/allwise/expsup/sec2_1.html}}. For the high proper motion sample used in this study, such stars make up a small portion of the overall sample ($<1.5\%$).

\subsubsection{GALEX DR5}

The GALEX  All-Sky  Imaging  Survey in its fifth data release has collected photometry for sources in 21,435 square degrees of the sky between May 2003 and February 2012 in two bands: $FUV$ (1344-1786 \AA) and $NUV$ (1771-2831 \AA) \citep{galex}.  This data release consists of 65.3 million objects where objects are detected down to $FUV=19.9$ and $NUV=20.8$ at a 5 $\sigma$ detection level.

\subsubsection{RAVE DR5}

The Radial Velocity Experiment (RAVE) is a magnitude limited ($9<I<12$) radial velocity survey of bright stars randomly selected in the southern hemisphere \citep{ravedr5} that operated between 2003 and 2013. Certain extinction limits were imposed during the target selection in order to bias the survey towards giants. The fifth data release includes radial velocities for 457,588 unique stars. In addition, this data release includes stellar parameters for these unique objects.

\subsubsection{SDSS DR12}

The twelfth release of the Sloan Digital Sky Survey \citep[SDSS DR12;][]{sdss_dr12} consists of photometric and spectroscopic data from SDSS-III \citep{sdss_III}, collected from August 2008 to July 2014, along with all previous SDSS campaigns beginning in 2000. The total unique area covered by this phase of the survey was 14,555 square degrees, with observations from 469,053,874 primary sources (i.e. sources that have been deemed as non-duplicate observations). For this study, we are mainly interested in the photometric part of the survey, which consists of measurements in five optical bands: $u$ (3543 \AA), $g$ (4770 \AA), $r$ (6231 \AA), $i$ (7625 \AA) and $z$ (9134 \AA).

\subsubsection{Pan-STARRS}

The Panoramic Survey Telescope and Rapid Response System \citep[Pan-STARRS;][]{ps1} is imaging 30,000 square degrees of the sky in five optical bands: $g_{P1}$ (4776 \AA), $r_{P1}$ (6130 \AA), $i_{P1}$ (7485 \AA), $z_{P1}$ (8658 \AA) and $y_{P1}$ (9603 \AA). The first data release from the survey consists of 1,919,106,885 sources down to a 5 $\sigma$ detection, which corresponds to limiting magnitudes of 23.3, 23.2, 23.1, 22.3, and 21.4 mag for the $g_{P1}$, $r_{P1}$, $i_{P1}$, $z_{P1}$, and $y_{P1}$ bands, respectively.

\section{Catalog Cross-Match}

\subsection{Initial Catalog Query}\label{inital_query}

To reduce the number of sources to be included in the detailed search, the full external catalogs are initially queried at the coordinates around the sources in the \textit{Gaia} subset. As these sources of interest are (allegedly) high proper motion stars, their motion must be taken into account during the initial query due to the significant epoch difference between catalogs. With a \textit{Gaia} epoch of 2015.5, the location of a \textit{Gaia} source, $G$ at the mean epoch of an external catalog, $J_{mean}$, is given by:
\begin{equation}\label{mean_RA}
    \alpha_{G,mean}=\alpha_{G,2015.5}+2.7778\times 10^{-7}\left(J_{mean}-2015.5\right)\mu_{G,\alpha}/cos(\delta)
\end{equation}
\begin{equation}\label{mean_DE}
    \delta_{G,mean}=\delta_{G,2015.5}+2.7778\times 10^{-7}\left(J_{mean}-2015.5\right)\mu_{G,\delta}
\end{equation}
where $\alpha$ and $\delta$ are measured in degrees, and $\mu$ in $mas \ yr^{-1}$. The mean epochs for 2MASS, AllWISE, GALEX DR5, RAVE DR5, SDSS DR12 and Pan-STARRS were assumed to be 2000.16, 2010.80, 2004.87, 1999.57, 2004.54 and 2014.23, respectively. All sources within 30 arcseconds of the mean location of a given \textit{Gaia} source at the epoch of each survey are retrieved and comprise the initial external catalogs for our cross-match.

\subsection{Initial Cross-Match}\label{intial_corss_match}

Starting with the initial external catalogs outlined in the section above, a more detailed cross-match is conducted. For each \textit{Gaia} source, all sources from the external catalogs with $|x_G-x_c|<0.01$, $|y_G-y_c|<0.01$ and $|z_G-z_c|<0.01$ are first selected, where:
\begin{equation}
    x=cos(\alpha)cos(\delta)
\end{equation}
\begin{equation}
    y=sin(\alpha)cos(\delta)
\end{equation}
\begin{equation}
    z=sin(\delta)
\end{equation}
and where $(x_c,y_c,z_c)$ are based on the coordinates listed in the external catalog and $(x_G,y_G,z_G)$ are based on the location of the \textit{Gaia} sources at the mean epoch of the external catalog (as described by eqs. \ref{mean_RA} and \ref{mean_DE}). This coordinate system is used to provide more accurate separations near the poles, where other appropriations can have large errors.

Following this initial cut, we extrapolate the positions of all sources in the external catalogs assuming they have the same proper motion as the \textit{Gaia} sources of interest:
\begin{equation}
    \alpha_{c,2015.5}=\alpha_{c,J}+2.7778\times 10^{-7}\left(2015.5-J\right)\mu_{G,\alpha}/cos(\delta)
\end{equation}
\begin{equation}
    \delta_{c,2015.5}=\delta_{c,J}+2.7778\times 10^{-7}\left(2015.5-J\right)\mu_{G,\delta}
\end{equation}
All sources with extrapolated locations less than 15 arcseonds from the \textit{Gaia} source location are then included in the final cut of the cross match. GALEX DR5 does not specifically list observed epochs for their sources, so the mean epochs for this survey (see above) are assigned to all sources. Also, while AllWISE does provide epochs for individual bands, the mean epochs for this survey (see above) are also assigned to these sources. This is done for two reasons. First, these epochs are for individual bands, and do not describe the epoch for the coordinates listed in the catalog. Second, even if there are differences between the mean epoch for this survey and the epochs listed for individual bands, due to the very short baseline of the AllWISE measurements ($\sim$6 months), there would be very little difference in the epoch corrected positions compared to our current assumptions, especially compared with the errors on the AllWISE positions. Using these assumptions, all sources that pass this final cut are considered potential matches to the \textit{Gaia} source for the remainder of the analysis.

In the next step, we account for the possibility that the epochs of the external catalogs might be inaccurate or unreliable, and for each source we extrapolate a set of positions using epochs ranging from 1950 and 2050 in increments of 1/12 year. The epoch that minimizes the angular separation between the external source and the \textit{Gaia} source is then recorded as the optimal epoch for that cross-match.

\begin{figure*}
	\epsscale{1.1}
	\plotone{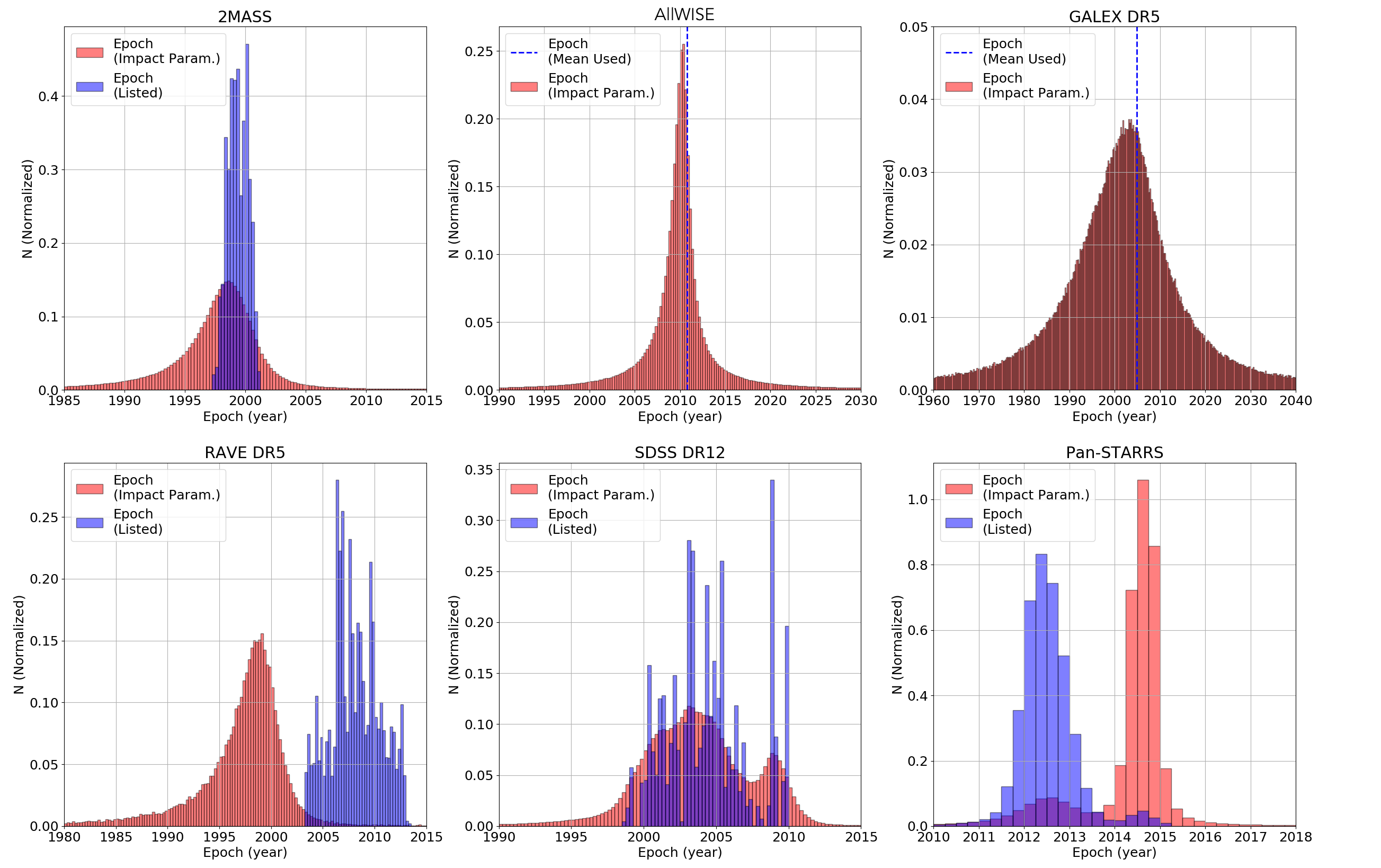}
	\caption{Distributions for optimal epochs (red histograms) as compared to the listed epochs (blue histograms) for matches with minimzed angular separations between the external source and the \textit{Gaia} source. For external catalogs that do not list individual epochs for each source (AllWISE and GALEX DR5), the mean epoch assumed for the survey is shown as a blue dotted line.}
	\label{mean_epochs}
\end{figure*}

As a check, we examine the distribution of optimal epochs for all the cross-matches from each external catalog; these are shown in Figure \ref{mean_epochs}. We use these distributions to verify the effective mean epoch of each of the external catalogs. Sources with best epochs equal to 1950 or 2050 (the edges of the epoch grid) are excluded as they are unlikely to be correct matches. The widths of the optimal epoch distributions are generally consistent with the astrometric precision of each external catalog, where catalogs like GALEX DR5 demonstrate a much lower angular resolution than a catalog like Pan-STARRS. Additionally, differences between the means of the optimal and listed epoch distributions are indicative of systematic errors in sky coordinates. Whereas catalogs like 2MASS and SDSS DR12 demonstrate relativity small offsets, catalogs like RAVE DR5 and Pan-STARRS show large systematics. Differences in the RAVE DR5 epochs are simply due to the positions of the sources being listed at their 2000.0 epoch and not at the epoch of the spectroscopic observations, which is the value listed in the RAVE catalog. The Pan-STARRS difference, on the other hand, may indicate larger errors in the catalog itself.

Finally, the angular separations of the stars that pass the final cut in the initial search at these mean epochs are calculated and will be used in the subsequent Bayesian analysis. 

\subsection{Displaced Sample}

All sources from an external catalog passing the final angular separation cut are considered possible matches to a specific \textit{Gaia} source. However, at least some of these matches are bound to be random field stars instead of true counterparts. In order to to evaluate the likelihood that a source is a random field star, we create a ``displaced" sample of sources from each external catalog, which is meant to represent a sample of purely random field stars. 

This displaced sample is created by repeating the procedure in Sections \ref{inital_query} and \ref{intial_corss_match}, but by first displacing all the \textit{Gaia} declinations by $\pm 2$ arcminutes (depending if the source is in the northern or southern hemisphere). This creates an initial catalog of cross-matched objects consisting of solely random field stars. Due to the small shift in declination, the field density of the displaced sample is in most cases statistically comparable to the field density of the actual sample used in the cross-match \citep{lepine2007}. When correcting all sources in the displaced sample to a mean epoch, the mean epoch from the true distribution is used to determine the angular separations used in the subsequent Bayesian analysis.

\subsection{Bayesian Analysis}\label{sec:bayes}

Generally, for some vector of observations, $\vec{x}$, the probability of some hypothesis $H_1$ given that  $\vec{x}$ is true is given by:
\begin{equation}
    P(H_1|\vec{x})=\frac{P(\vec{x}|H_1)P(H_1)}{P(\vec{x})}
\end{equation}
Given an alternative hypothesis to $H_1$, $H_0$, where $P(H_1)+P(H_0)=1$, the above expression can be rewritten as:
\begin{equation}\label{eq:bayes}
    P(H_1|\vec{x})=\frac{P(\vec{x}|H_1)P(H_1)}{P(\vec{x}|H_1)P(H_1)+P(\vec{x}|H_0)P(H_0)}
\end{equation}
For a cross-match, the initial hypothesis, $H_1$, is that the external catalog entry is a match to the \textit{Gaia} source being considered, or associated with the \textit{Gaia} source, while the alternate hypothesis, $H_0$, is that the external catalog entry is a random field star unrelated to the \textit{Gaia} source. We include the caveat of ``associated with the \textit{Gaia} source" in the statement for the hypothesis $H_1$ to account for any image or pipeline artifact associated with the \textit{Gaia} source or a possible binary star companion to the \textit{Gaia} source, as these would not be present in the displaced sample. We do expect that these artifacts/companions will be lower probability, secondary matches in the final result. All Bayesian probabilities in general will be a function of the mean angular separation of the external source, but also of the magnitude of the external source, the $G$ magnitude of the \textit{Gaia} source, the Galactic latitude of the \textit{Gaia} source being matched, and possibly other parameters.

In order to increase the dimensionality of the problem, while keeping the probability distributions reasonably simple (e.g. two-dimensions), we will build statistics for different subsets defined by various cuts in e.g. the $G$ magnitude and Galactic latitude of the \textit{Gaia} source, and reduce the dimensions of the probability distributions to two factors: angular separation, and difference between the external catalog magnitude and the \textit{Gaia} source $G$ magnitude, or ``quasi-color". The quasi-color parameter will act a spectral type check when paired with examining probability distributions in finite bins of $G$ magnitude, as we would expect the true matches of similar spectral type to center around a determined value in this quasi-color space. For this study, the following bins in \textit{Gaia} magnitude $G$ and Galactic latitude \textit{b} are used: $G<10$, $10\leq G<12.5$, $12.5\leq G<15$, $15\leq G<17.5$, $17.5\leq G<20$ and $G\geq 20$, and $|b|<19.5$, $19.5\leq |b|<41.8$ and $|b|\geq 41.8$. The latitude cuts are chosen such that that they cover regions of approximately equal solid angle on the sky. Due to low number statistics, however, these cuts were not applied to the RAVE DR5 sample.

Based on these inputs, the probability of the possible match being a random field star, $P(\vec{x}|H_0)$, can be directly inferred from the results of the cross match with the displaced sample. This is done by creating a two-dimensional frequency distribution with the displaced sample results for each of the surveys using the same cuts in $G$ and \textit{b} above. 

Frequency distributions are also determined for stars in the true sample, but, as mentioned, these distributions are contaminated by the presence of random field stars. In order to account for the components of the distribution due to these random field stars, the frequency distributions of the displaced samples are used. Under ideal circumstances, the number of random field stars should be the same in both samples, and simple subtraction of the two distributions should suffice in inferring the component of the distribution for the true cross-match that is equivalent to $P(\vec{x}|H_1)$.

\begin{figure}
    \plotone{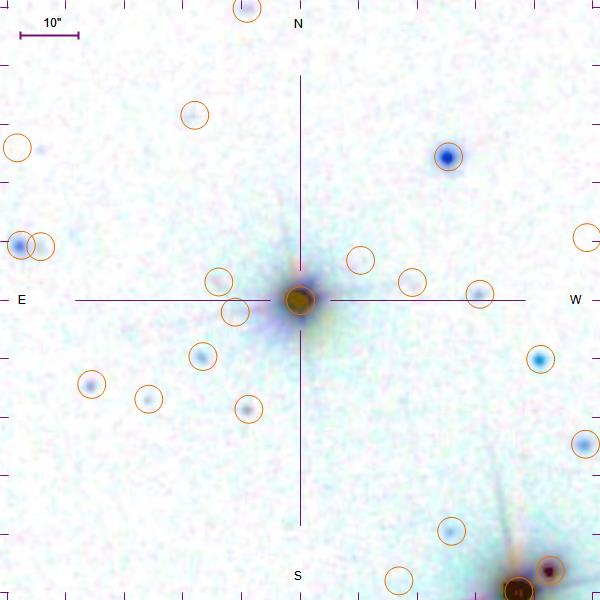}
    \caption{SDSS image of the source SDSS J000000.21+091332.1 (centered onthe crosshairs), which is matched with \textit{Gaia} source 2747168838957855104. The orange circles in the image indicate other photometric objects in the SDSS DR12 catalog.}
    \label{fig:bright_source_ex}
\end{figure}

\begin{figure*}
    \plotone{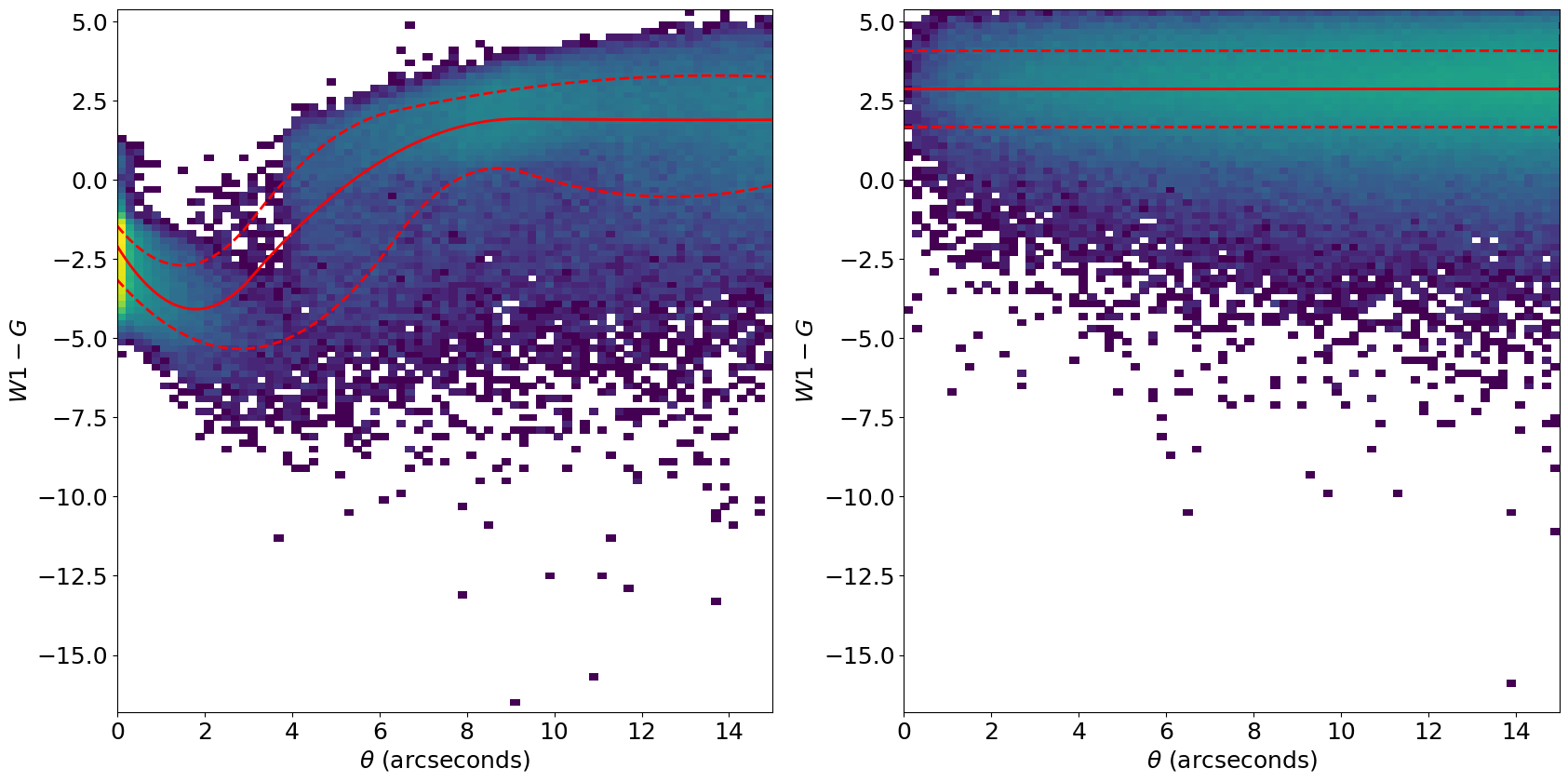}
    \caption{Figure showing the 2D distributions of $W1-G$ color versus angular separation for the AllWISE initial cross match (left panel) and for the displaced sample (right panel), in both cases for \textit{Gaia} sources with $12.5<G\leq 15$ and $|b|\geq 41.8$. The colormap of the distribution shows the number density of sources in a bin, on a logarithmic scale; both panels have the same colormap range. The median and 68th percentile of $W1-G$ as a function of $\theta$ are shown as the red solid and red dashed lines, respectively, where quantiles are found using COBS \citep{cobs}. At small angular separations, the distribution in the left panel is dominated by the counterparts and artifacts/companions in the external catalog, while at larger angular separations the distribution is due to random field stars.}
    \label{fig:zone_ex}
\end{figure*}

This is not necessarily the case for our distributions though. When performing the search for the displaced sample, all pointings are randomized and are most likely not falling in the immediate vicinity of a bright source. This is however not the case for the true cross-match, since many of the $Gaia$ sources are matched with bright sources in these external catalogs. In these cases, the PSFs of these bright sources outshines some of these random field stars and leaves them missing from the catalog and from our inferred statistical distribution. This ``blind spot" effect can be seen in Figure \ref{fig:bright_source_ex}, which shows an SDSS image made using the SDSS DR12 Image Tool\footnote{https://skyserver.sdss.org/dr12/en/tools/chart/listinfo.aspx}. In this image, the PSF effectively extends nearly 10 arcseconds from the center of the object. This effect is present in all external catalogs and has an obvious effect on the frequency distributions for the true cross-match. An example of the resulting effect is shown in Figure \ref{fig:zone_ex}, where, when comparing the two distributions, a large gap in dim stars at short angular separations is observed. In Figure \ref{fig:zone_ex}, the median and 68th percentile of $W1-G$ as a function of $\theta$ is shown, where quantiles are found using COBS \citep{cobs}. These quantiles demonstrate that only at very large angular separations does the distribution of the true sample begin to match that of the displaced sample due to this ``outshining" effect.

\begin{figure*}
	\plotone{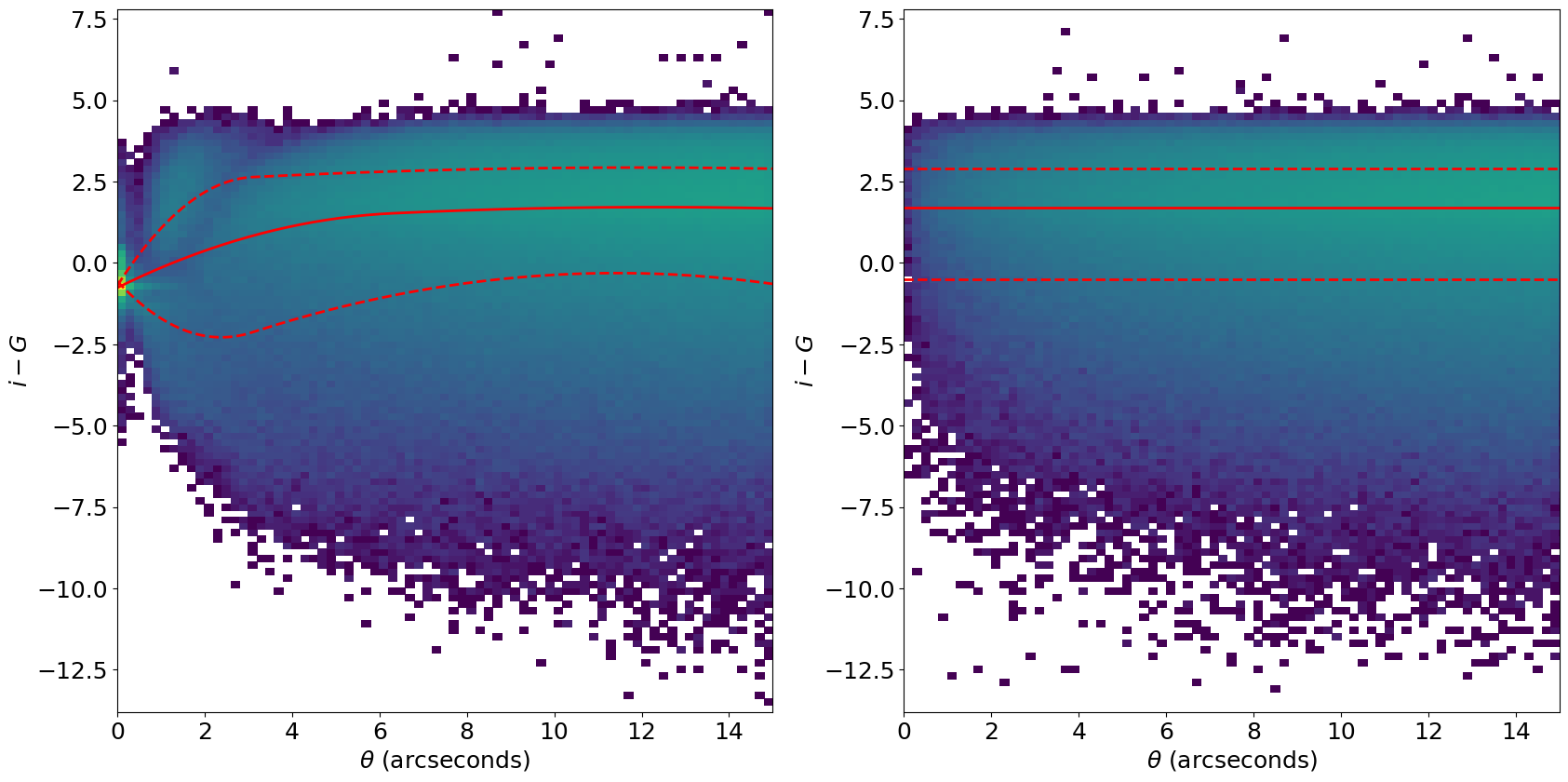}
	\caption{Figure showing the 2D distributions of $i-G$ color versus angular separation for the Pan-STARRS initial cross match (left panel) and for the displaced sample (right panel), in both cases for \textit{Gaia} sources with $17.5<G\leq 20$ and $19.5\leq |b|< 41.8$. The colormap of the distribution shows the number density of sources in a bin, on a logarithmic scale; both panels have the same colormap range. The median and 68th percentile of $i-G$ as a function of $\theta$ are shown as the red solid and red dashed lines, respectively, where quantiles are found using COBS \citep{cobs}. At small angular separations, the distribution in the left panel is dominated by the counterparts and artifacts/companions in the external catalog, while at larger angular separations the distribution is due to random field stars.}
	\label{fig:large_sep_ex}
\end{figure*}

For distributions in bins of fainter $G$ magnitude however, the component due to random field stars in the true distribution becomes more apparent at smaller angular separations as compared to distributions for bright sources (Figure \ref{fig:large_sep_ex}). This is seen clearly when comparing the quantiles of the two distributions in Figure \ref{fig:large_sep_ex}. It is at these larger angular separations that the displaced distribution is used to model the random field star component. For the brighter bins though, the ``outshining" effect discussed above and random fluctuations in the number of random field stars, this distribution at larger angular separations does not always have the same amplitude.

\begin{figure}
    \plotone{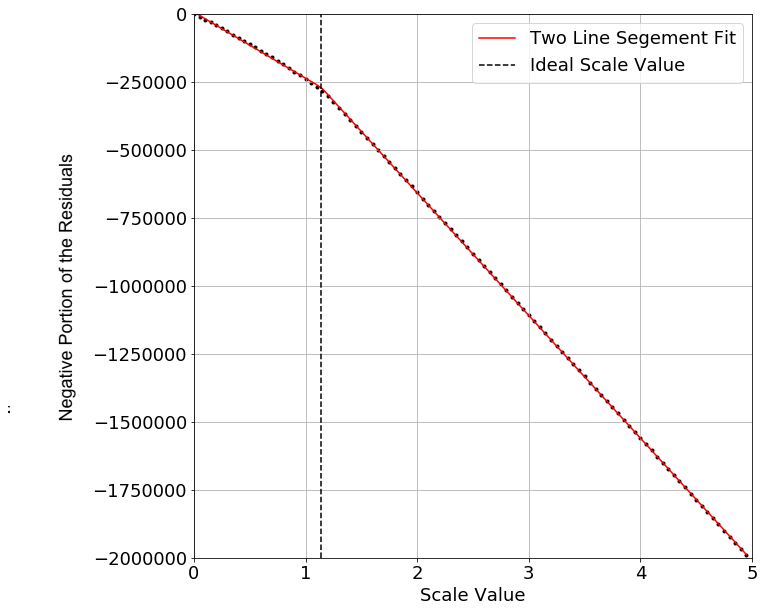}
    \caption{Plot of the integral of the negative part of the residuals between the true and displaced distribution functions (see eq. \ref{scale_eq}), represented as black dots. The example above is for a cross-match of our \textit{Gaia} subset with the Pan-STARRS catalog, with $12.5\leq G<15$ and $19.5\leq |b|<41.8$. The black dashed line shows the intersect of the two model line segments (red lines), which is determined to be the ideal scaling factor for the displaced sample in this region.}
    \label{fig:elbow_ex}
\end{figure}

To determine the correct scaling factor and properly subtract the distribution of field stars represented by the displaced sample, we examine the negative part of the residuals. Figure \ref{fig:elbow_ex} shows the integral of the negative part of the residuals of the inferred frequency distribution for the hypothesis $H_1$ after subtracting the frequency distribution of the displaced sample multiplied by a scaling factor. The curve always shows two segments with differing slopes. This change in slope occurs at the point where the scaling factor over-estimates the number of stars in the displaced sample that may have been masked by the PSF of the source in the true sample, causing an over-subtraction to occur. The appropriate scale factor for each distribution, then, is found at the scaling factor where this slope change occurs. We find that this value is usually around $1.0$, where deviations from $1.0$ are due to two reasons. First, deviations from $S=1.0$ can be attributed to random fluctuations in the number of random field stars in both the true and displaced samples. Second, deviations are observed for brighter sources with extended PSFs where the scaling factor is increased to account for the field stars hidden by the PSF.

\begin{figure*}
	\plotone{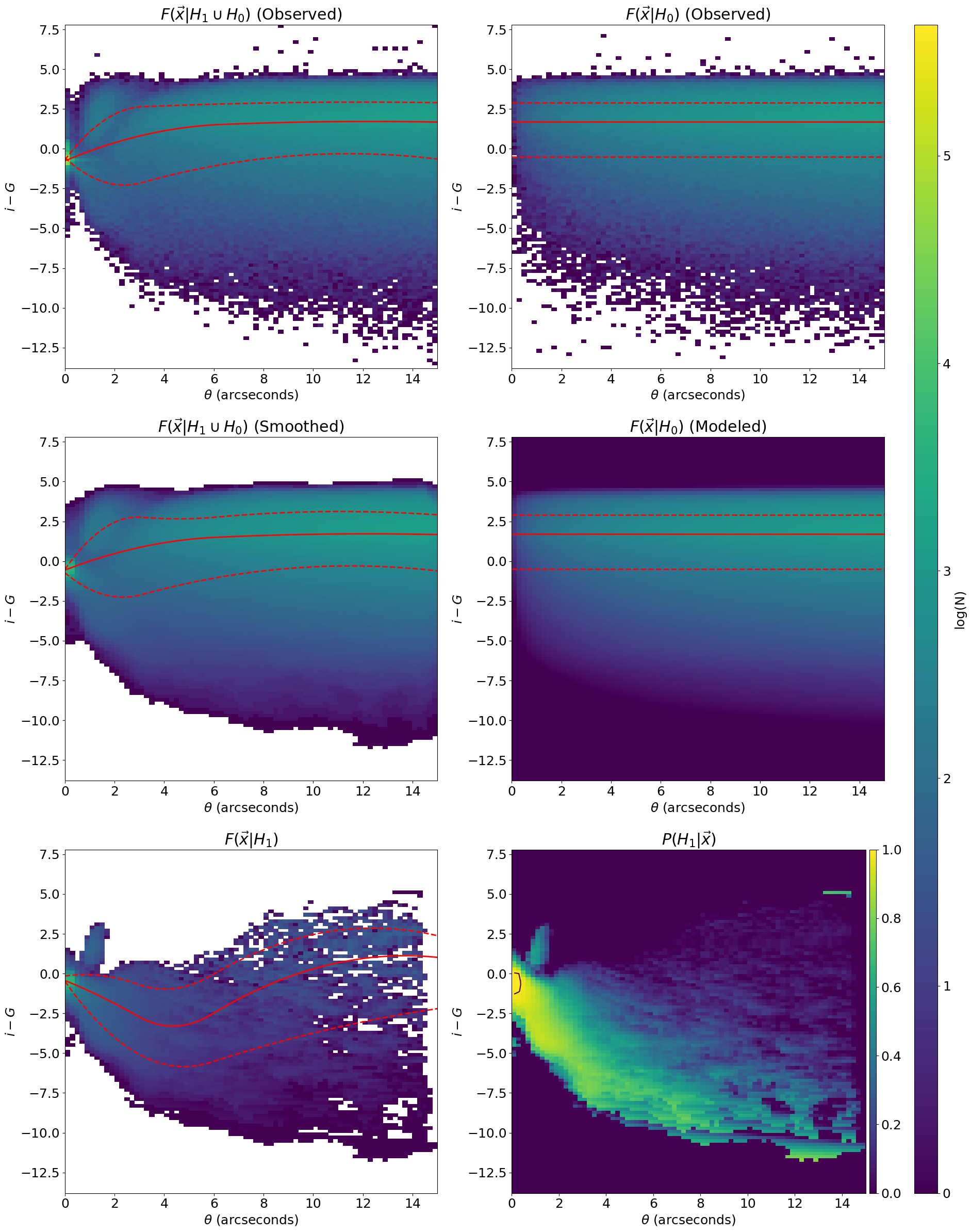}
	\caption{Cross-match frequency distributions discussed in Section \ref{sec:bayes} for Pan-STARRS counterparts of \textit{Gaia} with $17.5<G\leq 20$ and $19.5\leq |b|<41.8$. The description of individual panels is as follows: upper left is the frequency distribution for the ``true" sample, upper right is the frequency distribution for the ``displaced sample", middle left is the smoothed frequency distribution for the ``true" sample, middle right is the modeled frequency distribution for the ``displaced" sample, lower left is the difference between the smoothed ``true" frequency distribution and the modeled ``displaced" frequency distribution using the correct scaling factor, and the lower right is the resulting Bayesian probability distribution. All plots, except for the distribution for $P(H_1|\vec{x})$, show the number of sources per bin on a logarithmic scale and they all share the same colormap range (indicated by the colorbar on the far right). The median and 68th percentile of $i-G$ as a function of $\theta$ are shown on these plots as the red solid and red dashed lines, respectively, where quantiles are found using COBS \citep{cobs}. The distribution for $P(H_1|\vec{x})$ is shown for a range of 0 to 1 and the black solid line overlaid shows the 99\% line.}
	\label{fig:result_ex}
\end{figure*}

In summary, the frequency distribution for the hypothesis $H_1$ for a distribution based on some magnitude, $m$, some cut in Gaia $G$ and some cut in Galactic Latitude, \textit{b}, is given by:
\begin{equation}\label{scale_eq}
    F(\vec{x}|H_1)_{m,G,b}=\begin{cases}
    f = F(\vec{x}|H_1 \cup H_0)_{m,G,b}-S_{m,G,b}\times F(\vec{x}|H_0)_{m,G,b}, & \text{if $f>0$}\\
    0, & \text{if $f\leq0$}
    \end{cases}
\end{equation}
where $S$ is the scale factor determined from above. The conditional probabilities for a distribution of $m-G$ vs. $\theta$ based on some magnitude, $m$, some cut in Gaia $G$ and some cut in Galactic Latitude, \textit{b}, are then given by the likelihoods:
\begin{equation}\label{eq:condition}
    P(\vec{x}|H_1)_{m,G,b}=\frac{F(\vec{x}|H_1)_{m,G,b}}{\sum_{\vec{x}=(\theta,m-G)} F(\vec{x}|H_1)_{m,G,b}}
\end{equation}
\begin{equation}
    P(\vec{x}|H_0)_{m,G,b}=\frac{F(\vec{x}|H_0)_{m,G,b}}{\sum_{\vec{x}=(\theta,m-G)} F(\vec{x}|H_0)_{m,G,b}}
\end{equation}
Finally, the prior probabilities are based on the expected number of matches and random field stars:
\begin{equation}
    P(H_1)_{m,G,b}=\frac{\sum_{\vec{x}=(\theta,m-G)} F(\vec{x}|H_1)_{m,G,b}}{\sum_{\vec{x}=(\theta,m-G)} F(\vec{x}|H_1 \cup H_0)_{m,G,b}}
\end{equation}
\begin{equation}\label{eq:prior0}
    P(H_0)_{m,G,b}=1-P(H_1)_{m,G,b}
\end{equation}
Using the definitions laid out in this section, we can now determine a discrete 2D Bayesian probability distribution based on eq. \ref{eq:bayes} for all combinations of our $G$ magnitude and Galactic latitude bins, and for every external catalog that we match. An example of the various cross-match distributions discussed above are shown in Figure \ref{fig:result_ex}, where the labels above each panel match the functional definitions outlined in eq. \ref{eq:bayes} and eqs. \ref{eq:condition}-\ref{eq:prior0}. In the bottom right panel of Figure \ref{fig:result_ex}, some artifacts are present, such as the line around $(\theta, i-G) \approx (14,5)$. Such artifacts are caused by the smoothing of the true distribution, which can cause an overestimation around the edges of the distribution. When you pair this overestimation along the edge with the lower number of field stars in the displaced model, sometimes these artifacts will arise in the probability distribution. These artificial probabilities are much lower than the best match threshold we use for this study though, so it will not affect the final results.

\subsection{Distribution Smoothing and Modeling}

One of the issues with the above procedure is the statistical noise present in these distributions. To remedy this, while keeping relatively small bin sizes for the frequency distributions, a percentile smoothing filter was applied and the priors of the discrete and smoothed distributions are then found to be comparable.

Another issue with the procedure is that in low occurrence regimes, the under-sampling of the displaced sample distribution can cause artificially high Bayesian probabilities. Fortunately, while the true cross-match distribution is too complex to model in most cases, the displaced sample distribution is fairly easily modeled. This is because the placement of fields stars is inherently random, and the angular separation and magnitudes differences can be considered independent variables, and their distributions modeled separately.

\begin{figure*}
    \plotone{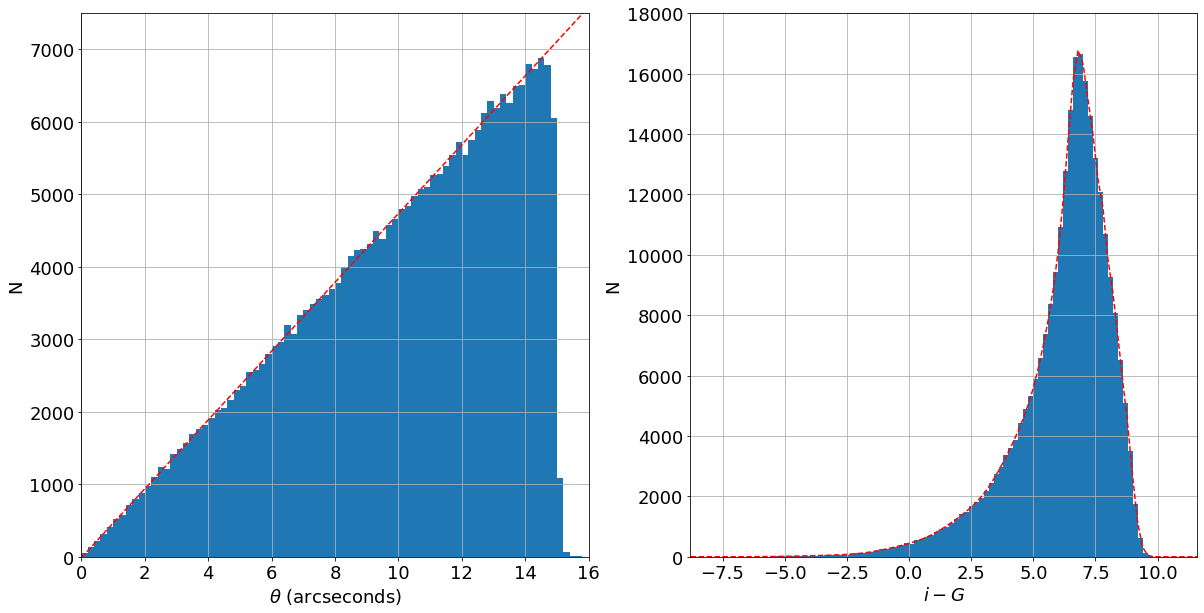}
    \caption{1D distributions (blue bins) of angular separation (left panel) and $i-G$ (right panel) for cross-matched pairs in the Pan-STARRS displaced sample with $12.5\leq G<15$ and $19.5\leq |b|<41.8$. The best models based on eqs. \ref{eq:1d_ang_sep} and \ref{eq:1d_mag} are shown as the red dashed lines.}
    \label{fig:dis_1d_dist_ex}
\end{figure*}

Representative models of these one-dimensional, independent distributions are shown in Figure \ref{fig:dis_1d_dist_ex}. As can be seen from the left panel, the angular separations form a distribution that increases linearly at higher separations:
\begin{equation}\label{eq:1d_ang_sep}
    F(x=\theta)=mx+b
\end{equation}
The quasi-color distribution (right panel), on the other hand, can be modeled as a sum of normal distributions with varying weights:
\begin{equation}\label{eq:1d_mag}
    F(y=m-G)=\sum_i^n W_i e^{-\frac{(y-\mu_i)^2}{2\sigma_i^2}}
\end{equation}
The number of components is varied from $n=2$ to $n=10$, and the ideal number of components is based on the function that minimizes the Bayesian information criterion (BIC):
\begin{equation}
    \text{BIC}=n \ ln\left[\frac{\sum (F(x)_{measured}-F(x)_{model})^2}{n} \right] + k \ ln(n)
\end{equation}
Where $n$ is the number of data points used for the model and $k$ is the number of components.

\begin{figure*}
    \plotone{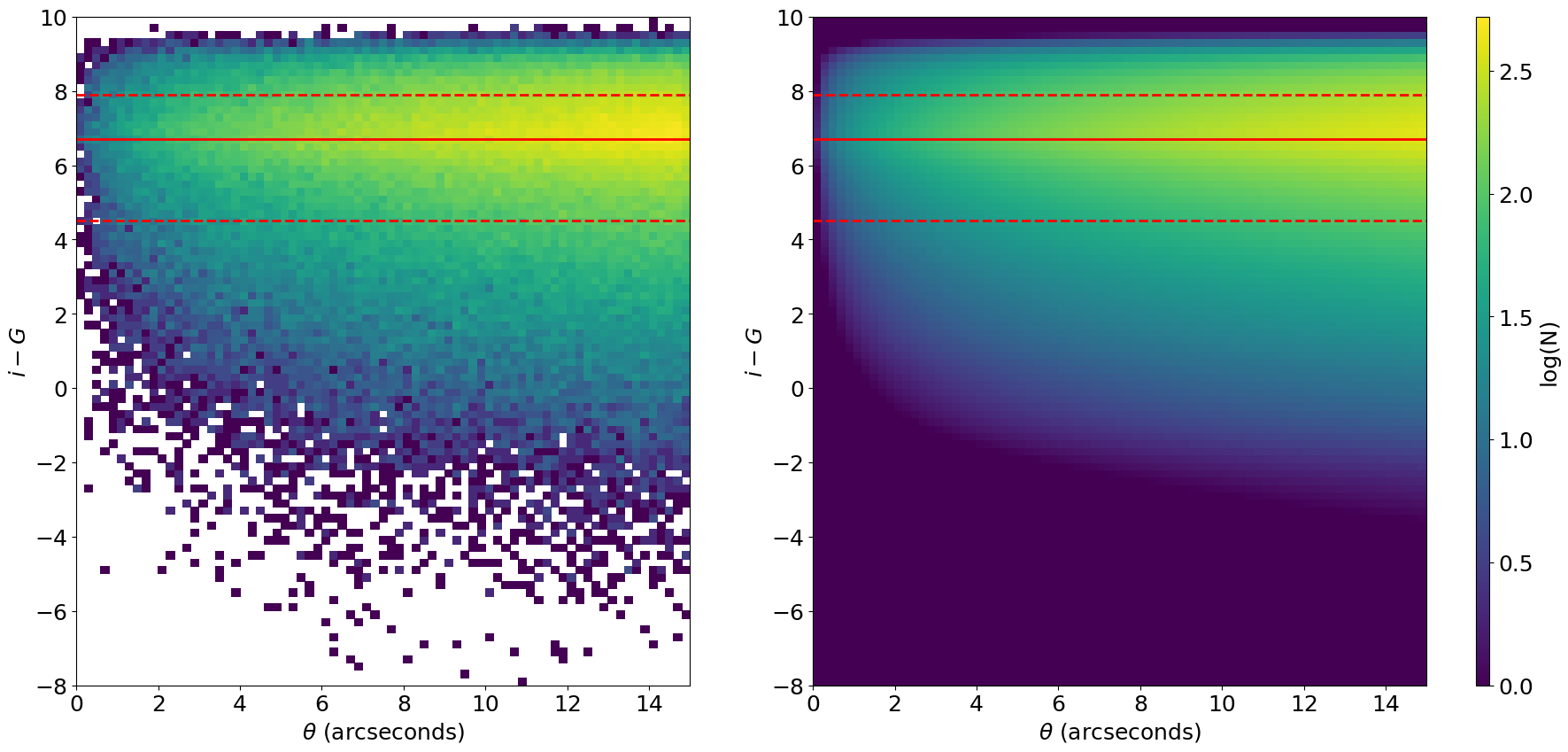}
    \caption{2D distributions of $i-G$ versus angular separation for the Pan-STARRS displaced sample with $12.5\leq G<15$ and $19.5\leq |b|<41.8$. The colormap of the distribution shows the number of sources in a bin on a logarithmic scale and both panels have the same colormap range. The left panel shows the observed displaced distribution while the right panel shows the modeled distribution. Additionally, the median and 68th percentile of $i-G$ as a function of $\theta$ are shown as the red solid and red dashed lines, respectively, where quantiles are found using COBS \citep{cobs}.}
    \label{fig:dis_2d_mod_ex}
\end{figure*}

After fitting the above equations to the 1D distributions for each of the displaced samples, the 2D distribution of field stars from the displaced sample can generally be modeled by:
\begin{equation}\label{eq:back_model}
    F(\vec{x}|H_0)=F(x=\theta,y=m-G)=C (mx+b) \sum_i^n W_i e^{-\frac{(y-\mu_i)^2}{2\sigma_i^2}}
\end{equation}
where the scaling constant $C$ is added to ensure the scale of the modeled distribution matches that of the observed distribution. Best-fitting parameters are evaluated separately for each distribution; one example is shown on the right panel in Figure \ref{fig:dis_2d_mod_ex}, where it is compared to its associated observed distribution on the left. As a note, this modelling procedure is completed for 18 different distributions per photometric band in each survey, to account for all combinations of \textit{Gaia} $G$ and Galactic latitude bins described in Section \ref{sec:bayes}, so the parameters in eq. \ref{eq:back_model} will be different for each of these distributions.

\subsection{High Probability Matches}

In order to determine the best matches from the external catalogs to the \textit{Gaia} sources, we calculate the total probability defined as the product of the probabilities for each magnitude in a given catalog, when there is a valid magnitude measurement for the source. For SDSS DR12 sources not part of the primary catalog (i.e. are flagged as duplicates in the SDSS DR12) we force their probability to be 0\%. It should be noted that this does imply that for catalogs with a larger number of photometric measurements, the individual probabilities must be higher to reach this $95\%$ threshold. For example, for a catalog like 2MASS that has three photometric measurements, each individual magnitude must have a probability $>98.30\%$ on average to reach the $95\%$ threshold, while a catalog like Pan-STARRS that has five photometric measurements would require each individual magnitude to have a probability $>98.98\%$ on average. This small difference should not cause a large difference in the final catalog of high probability matches as a large number of the stars with high probability have probabilities $>99\%$ (see black line in lower right panel of Figure \ref{fig:result_ex}). Best matches are then defined as those possible sources with a total probability $>95\%$, and if multiple entries in a catalog have a total probability $>95\%$ for a particular \textit{Gaia} source the catalog entry with the largest total probability is designated as the best match. Table \ref{tab:best_match_results} shows the percentage of matches to \textit{Gaia} sources that meet this criterion. Low match percentages for catalogs like SDSS DR12 and Pan-STARRS can be attributed to the difference in sky coverage, while low match probabilities in the other catalogs can be attributed to the limiting magnitudes of the external catalogs.

\begin{deluxetable}{lc}
	\tablecaption{Number of sources matched with a total probability $>95\%$\label{tab:best_match_results}}
	\tablehead{\colhead{External Catalog} & \colhead{\% Matched}}
	\startdata
	2MASS & 75.40\% \\
	AllWISE & 75.17\% \\
	GALEX DR5 & 8.36\% \\
	RAVE DR5 & 0.63\% \\
	SDSS DR12 & 30.69\% \\
	Pan-STARRS & 68.63\% \\
	\enddata
\end{deluxetable}

The catalog of ``best matches", i.e. where the \textit{Gaia} sources have a counterpart with total probability $>95\%$, is presented in Table \ref{tab:best_match_results_descrip}. As a supplement, Tables \ref{tab:2MASS_all_matches}-\ref{tab:RAVE_all_matches} give all possible matches for each of the cross matches such that the user can vary the tolerances and obtain alternate and/or additional matches to the \textit{Gaia} sources. The code needed to reproduce these results, or perform additional cross-matches, can also be found on our GitHub repository\footnote{\url{https://github.com/imedan/bayes_match}}.

\begin{longrotatetable}
	\begin{deluxetable}{lllp{15cm}}
		\tabletypesize{\scriptsize}
		\tablecaption{Best matches found in this study with a Bayesian probability greater than 95\% for all cross matches in this study.\label{tab:best_match_results_descrip}}
		\tablehead{\colhead{Column Number} & \colhead{Label} & \colhead{Units} & \colhead{Column Description}}
		\startdata
		1   & GaiaDR2\_ID           & \nodata & Gaia DR2 Identifier                                                                          \\
		2   & Gmag\_GaiaDR2         & mag                    & Gaia DR2 G-band magnitude\tablenotemark{1}                                                                \\
		3   & BPmag\_GaiaDR2        & mag                    & Gaia DR2 BP-band magnitude\tablenotemark{2}                                                                   \\
		4   & RPmag\_GaiaDR2        & mag                    & Gaia DR2 RP-band magnitude\tablenotemark{3}                                                                   \\
		5   & RA\_ICRS\_GaiaDR2     & deg                    & Right ascension at Epoch=2015.5 from Gaia DR2                                                \\
		6   & DE\_ICRS\_GaiaDR2     & deg                    & Declination at Epoch=2015.5 from Gaia DR2                                                    \\
		7   & PLX\_GaiaDR2          & mas                    & Parallax measurment from Gaia DR2                                                            \\
		8   & PLXerr\_GaiaDR2       & mas                    & Parallax measurment error from Gaia DR2                                                      \\
		9   & PMRA\_GaiaDR2         & mas/yr                 & Proper motion in right ascension direction (PMRA*cosDE) from Gaia DR2\tablenotemark{4}                        \\
		10  & PMDE\_GaiaDR2         & mas/yr                 & Proper motion in declination direction from Gaia DR2\tablenotemark{5}                                         \\
		11  & SDSSDR12\_ID          & \nodata & SDSS DR12 Identifier                                                                         \\
		12  & umag\_SDSSDR12        & mag                    & u-band magntiude on AB scale from SDSS DR12                                                  \\
		13  & umag\_err\_SDSSDR12   & mag                    & u-band magntiude error on AB scale from from SDSS DR12                                       \\
		14  & gmag\_SDSSDR12        & mag                    & g-band magntiude on AB scale from from SDSS DR12                                             \\
		15  & gmag\_err\_SDSSDR12   & mag                    & g-band magntiude error on AB scale from from SDSS DR12                                       \\
		16  & rmag\_SDSSDR12        & mag                    & r-band magntiude on AB scale from from SDSS DR12                                             \\
		17  & rmag\_err\_SDSSDR12   & mag                    & r-band magntiude error on AB scale from from SDSS DR12                                       \\
		18  & imag\_SDSSDR12        & mag                    & i-band magntiude on AB scale from from SDSS DR12                                             \\
		19  & imag\_err\_SDSSDR12   & mag                    & i-band magntiude error on AB scale from from SDSS DR12                                       \\
		20  & zmag\_SDSSDR12        & mag                    & z-band magntiude on AB scale from from SDSS DR12                                             \\
		21  & zmag\_err\_SDSSDR12   & mag                    & z-band magntiude error on AB scale from from SDSS DR12                                       \\
		22  & RA\_ICRS\_SDSSDR12    & deg                    & Right ascension from SDSS DR12 (ICRS)                                                        \\
		23  & DE\_ICRS\_SDSSDR12    & deg                    & Declination from SDSS DR12 (ICRS)                                                            \\
		24  & Bayes\_Prob\_SDSSDR12 & \nodata & Total bayesian probability of SDSS DR12 object match to Gaia DR2 Object                      \\
		25  & Ang\_Sep\_SDSSDR12    & arcsec                 & Angular separation between SDSS DR12 object and Gaia DR2 object at mean epoch of SDSS DR12   \\
		26  & 2MASS\_ID             & \nodata & 2MASS Identifier\tablenotemark{6}                                                                             \\
		27  & Jmag\_2MASS           & mag                    & J-band magntiude from 2MASS\tablenotemark{7}                                                                  \\
		28  & Jmag\_err\_2MASS      & mag                    & J-band magntiude error from 2MASS\tablenotemark{8}                                                            \\
		29  & Hmag\_2MASS           & mag                    & H-band magntiude from 2MASS\tablenotemark{7}                                                                  \\
		30  & Hmag\_err\_2MASS      & mag                    & H-band magntiude error from 2MASS\tablenotemark{8}                                                            \\
		31  & Ksmag\_2MASS          & mag                    & Ks-band magntiude from 2MASS\tablenotemark{7}                                                                 \\
		32  & Ksmag\_err\_2MASS     & mag                    & Ks-band magntiude error from 2MASS\tablenotemark{8}                                                           \\
		33  & RA\_J2000\_2MASS      & deg                    & Right ascension from 2MASS (J2000)                                                           \\
		34  & DE\_J2000\_2MASS      & deg                    & Declination from 2MASS (J2000)                                                               \\
		35  & Bayes\_Prob\_2MASS    & \nodata & Total bayesian probability of 2MASS object match to Gaia DR2 Object                          \\
		36  & Ang\_Sep\_2MASS       & arcsec                 & Angular separation between 2MASS object and Gaia DR2 object at mean epoch of 2MASS           \\
		37  & AllWISE\_ID           & \nodata & AllWISE Identifier                                                                          \\
		38  & W1mag\_AllWISE        & mag                    & W1-band magnitude from AllWISE                                                              \\
		39  & W1mag\_err\_AllWISE   & mag                    & W1-band magntiude error from AllWISE                                                        \\
		40  & W2mag\_AllWISE        & mag                    & W2-band magnitude from AllWISE                                                              \\
		41  & W2mag\_err\_AllWISE   & mag                    & W2-band magntiude error from AllWISE                                                        \\
		42  & W3mag\_AllWISE        & mag                    & W3-band magnitude from AllWISE                                                              \\
		43  & W3mag\_err\_AllWISE   & mag                    & W3-band magntiude error from AllWISE                                                        \\
		44  & W4mag\_AllWISE        & mag                    & W4-band magnitude from AllWISE                                                              \\
		45  & W4mag\_err\_AllWISE   & mag                    & W4-band magntiude error from AllWISE                                                        \\
		46  & RA\_J2000\_AllWISE    & deg                    & Right ascension from AllWISE (J2000)                                                        \\
		47  & DE\_J2000\_AllWISE    & deg                    & Declination from AllWISE (J2000)                                                            \\
		48  & Bayes\_Prob\_AllWISE  & \nodata & Total bayesian probability of AllWISE object match to Gaia DR2 Object                       \\
		49  & Ang\_Sep\_AllWISE     & arcsec                 & Angular separation between AllWISE object and Gaia DR2 object at mean epoch of AllWISE     \\
		50  & GALEXDR5\_ID          & \nodata & GALEX DR5 Identifier                                                                      \\
		51  & FUVmag\_GALEXDR5      & mag                    & FUV-band magnitufe from GALEX DR5                                                            \\
		52  & FUVmag\_err\_GALEXDR5 & mag                    & FUV-band magntiude error from GALEX DR5                                                      \\
		53  & NUVmag\_GALEXDR5      & mag                    & NUV-band magnitufe from GALEX DR5                                                            \\
		54  & NUVmag\_err\_GALEXDR5 & mag                    & NUV-band magntiude error from GALEX DR5                                                      \\
		55  & RA\_J2000\_GALEXDR5   & deg                    & Right ascension from GALEX DR5 (J2000)\tablenotemark{9}                                                      \\
		56  & DE\_J2000\_GALEXDR5   & deg                    & Declination from GALEX DR5 (J2000)\tablenotemark{9}                                                         \\
		57  & Bayes\_Prob\_GALEXDR5 & \nodata & Total bayesian probability of GALEX DR5 object match to Gaia DR2 Object                      \\
		58  & Ang\_Sep\_GALEXDR5    & arcsec                 & Angular separation between GALEX DR5 object and Gaia DR2 object at mean epoch of GALEX DR5   \\
		59  & PS1\_ID               & \nodata & Pan-STARRS Identifier                                                                        \\
		60  & gmag\_PS1             & mag                    & g-band magntiude on AB scale from Pan-STARRS                                                 \\
		61  & gmag\_err\_PS1        & mag                    & g-band magntiude error on AB scale from Pan-STARRS                                           \\
		62  & rmag\_PS1             & mag                    & r-band magntiude on AB scale from Pan-STARRS                                                 \\
		63  & rmag\_err\_PS1        & mag                    & r-band magntiude error on AB scale from Pan-STARRS                                           \\
		64  & imag\_PS1             & mag                    & i-band magntiude on AB scale from Pan-STARRS                                                 \\
		65  & imag\_err\_PS1        & mag                    & i-band magntiude error on AB scale from Pan-STARRS                                           \\
		66  & zmag\_PS1             & mag                    & z-band magntiude on AB scale from Pan-STARRS                                                 \\
		67  & zmag\_err\_PS1        & mag                    & z-band magntiude error on AB scale from Pan-STARRS                                           \\
		68  & ymag\_PS1             & mag                    & y-band magntiude on AB scale from Pan-STARRS                                                 \\
		69  & ymag\_err\_PS1        & mag                    & y-band magntiude error on AB scale from Pan-STARRS                                           \\
		70  & RA\_J2000\_PS1        & deg                    & Right Ascension from Pan-STARRS (J2000)\tablenotemark{10}                                                     \\
		71  & DE\_J2000\_PS1        & deg                    & Declination from Pan-STARRS (J2000)\tablenotemark{10}                                                         \\
		72  & Bayes\_Prob\_PS1      & \nodata & Total bayesian probability of Pan-STARRS object match to Gaia DR2 Object                     \\
		73  & Ang\_Sep\_PS1         & arcsec                 & Angular separation between Pan-STARRS object and Gaia DR2 object at mean epoch of Pan-STARRS \\
		74  & RAVEDR5\_ID           & \nodata & RAVE DR5 Identifier                                                                          \\
		75  & BTmag\_RAVEDR5        & mag                    & Tycho-2 BT magnitude listed in RAVE DR5                                                      \\
		76  & BTmag\_err\_RAVEDR5   & mag                    & Tycho-2 BT magnitude error listed in RAVE DR5                                                \\
		77  & VTmag\_RAVEDR5        & mag                    & Tycho-2 VT magnitude listed in RAVE DR5                                                      \\
		78  & VTmag\_err\_RAVEDR5   & mag                    & Tycho-2 VT magnitude error listed in RAVE DR5                                                \\
		79  & RA\_J2000\_RAVEDR5    & deg                    & Right ascension fr0m RAVE DR5 (J2000)                                                        \\
		80  & DE\_J2000\_RAVEDR5    & deg                    & Declination from RAVE DR5 (J2000)                                                            \\
		81  & Bayes\_Prob\_RAVEDR5  & \nodata & Total bayesian probability of RAVE DR5 object match to Gaia DR2 Object                       \\
		82  & Ang\_Sep\_RAVEDR5     & arcsec                 & Angular separation between RAVE DR5 object and Gaia DR2 object at mean epoch of RAVE DR5    
		\\
		\enddata
		\tablecomments{This table is published in its entirety in the machine-readable format in the electric version of this manuscript. The description of its columns are shown here for guidance regarding its form and content.}
		\tablenotetext{1}{This is computed from the G-band mean flux applying the magnitude zero-point in the Vega scale. No error is provided for this quantity as the 
			error distribution is only symmetric in flux space. This converts to an asymmetric error distribution in magnitude space which cannot be represented by
			a single error value \citep{gaiadr2}.}
		\tablenotetext{2}{Mean magnitude in the integrated BP band. This is computed from the BP-band mean flux applying the magnitude zero-point in the Vega scale. No
			error is provided for this quantity as the error distribution is only symmetric in flux space. This converts to an asymmetric error distribution in
			magnitude space which cannot be represented by a single error value \citep{gaiadr2}.}
		\tablenotetext{3}{Mean magnitude in the integrated RP band. This is computed from the RP-band mean flux applying the magnitude zero-point in the Vega scale. No
			error is provided for this quantity as the error distribution is only symmetric in flux space. This converts to an asymmetric error distribution in
			magnitude space which cannot be represented by a single error value \citep{gaiadr2}.}
		\tablenotetext{4}{Proper motion in right ascension of the source in ICRS at the reference epoch. This is the tangent plane projection of the proper motion vector in
			the direction of increasing right ascension \citep{gaiadr2}.}
		\tablenotetext{5}{Proper motion in declination of the source at the reference epoch. This is the tangent plane projection of the proper motion vector in the
			direction of increasing declination \citep{gaiadr2}.}
		\tablenotetext{6}{Sexagesimal, equatorial position-based source name in the form: hhmmssss+ddmmsss[ABC...] \cite{2mass}.}
		\tablenotetext{7}{This is the selected "default" magnitude for each band, [JHK]. If the source is not detected in the band, this is the 95\% confidence upper limit
			derived from a 4" radius aperture measurement taken at the position of the source on the Atlas Image. The origin of the default magnitude is given by
			the first character of the Rflg value (Rflg) in the original catalog. This column is null if the source is nominally detected in the band, but no useful
			brightness estimate could be made (Rflg="9") \cite{2mass}.}
		\tablenotetext{8}{Combined, or total photometric uncertainty [JHK]msigcom for the default magnitude in that band. The combined uncertainty is derived from the
			following relation: e\_[JHK]mag = sqrt([JHK]cmsig2 + [JHK]zperr2 + fferr2 + [r1normrms2]), where cmsig = Corrected band photometric uncertainty, zperr = Nightly photometric zero point uncertainty = 0.011 mag, fferr = Flat-fielding residual error = 0.005 mags, r1normrms = R1 normalization uncertainty = 0.012 mags (applied only for sources with Rflg="1"). This column is null if the default magnitude is a 95\% confidence upper limit (i.e. the source is not detected, or inconsistently deblended in the band)) \cite{2mass}.}
		\tablenotetext{9}{Position of the NUV source, or FUV if seen in FUV only \citep{galex}.}
		\tablenotetext{10}{Coordinates from single epoch detections (weighted mean) in equinox J2000 at the mean epoch given by mean epoch from object \citep{ps1}.}
	\end{deluxetable}
\end{longrotatetable}

\begin{longrotatetable}
	\begin{deluxetable}{lllllllllllllll}
		\tabletypesize{\tiny}
		\tablecaption{All of the possible matches for the cross match between \textit{Gaia} and 2MASS. For more detailed descriptions of the columns in this table, look at the corresponding columns in Table \ref{tab:best_match_results_descrip}.\label{tab:2MASS_all_matches}}
		\tablehead{\colhead{\textit{Gaia} ID}            & \colhead{G}       & \colhead{$\alpha_{Gaia}$}               & \colhead{$\delta_{Gaia}$}            & \colhead{2MASS ID}        & \colhead{$J$}           & \colhead{$\sigma_J$}          & \colhead{$H$}           & \colhead{$\sigma_H$}          & \colhead{$K$}           & \colhead{$\sigma_K$}          & \colhead{$\alpha_{2MASS}$}             & \colhead{$\delta_{2MASS}$}           & \colhead{Bayes Prob.}       & \colhead{$\Delta\theta$} \\
			\colhead{}            & \colhead{[mag]}       & \colhead{[deg]}               & \colhead{[deg]}            & \colhead{}        & \colhead{[mag]}           & \colhead{[mag]}          & \colhead{[mag]}           & \colhead{[mag]}          & \colhead{[mag]}           & \colhead{[mag]}          & \colhead{[deg]}             & \colhead{[deg]}           & \colhead{}       & \colhead{[arcsec]}}
		\startdata
		429379051804727296  & 16.9667 & 0.00021864741 & 60.89771179398     & 00000009+6053531 & 13.876   & 0.024       & 13.308   & 0.032       & 13.055   & 0.026       & 0.00039   & 60.898087  & 0.9993495    & 0.1131188     \\
		429379051804727296  & 16.9667 & 0.00021864741 & 60.89771179398     & 00000116+6054047 & 16.228   & 0.1         & 15.771   & 0.156       & 15.436   & 0.195       & 0.004842  & 60.901306  & 0.0          & 13.893606     \\
		384318827606840448  & 17.9638 & 0.00047931984 & 42.19465632646     & 00000004+4211408 & 14.897   & 0.038       & 14.253   & 0.045       & 13.992   & 0.051       & 0.000202  & 42.194679  & 0.9998343    & 0.0210179     \\
		2772104109112208256 & 18.4284 & 0.00051294158 & 15.565175017629997 & 00000015+1533550 & 15.539   & 0.061       & 15.081   & 0.093       & 14.763   & 0.0970      & 0.00065   & 15.565296  & 0.999958     & 0.1038959     \\
		4922439138656081024 & 16.1428 & 0.00054775339 & -57.26154422458    & 00000021-5715408 & 13.517   & 0.025       & 12.957   & 0.028       & 12.713   & 0.0290      & 0.000914  & -57.261356 & 0.9999721    & 0.2605187     \\
		2738617902668633344 & 16.053  & 0.00059265319 & 1.41569388809      & 00000009+0124562 & 14.005   & 0.028       & 13.378   & 0.029       & 13.19    & 0.034       & 0.00041   & 1.415625   & 0.9999895    & 0.0681481     \\
		422597225421620992  & 19.7116 & 0.00062329596 & 57.67086338256     & 00000006+5740156 & 16.327   & 0.119       & 15.766   & 0.143       & 15.057   & \nodata        & 0.000289  & 57.671021  & 0.9888028    & 0.2357019     \\
		2420684907087327104 & 16.4742 & 0.000644395   & -14.44069806218    & 00000012-1426271 & 14.273   & 0.032       & 13.616   & 0.033       & 13.47    & 0.037       & 0.000518  & -14.440871 & 0.9999867    & 0.0757191     \\
		2305877746911854336 & 16.9816 & 0.00064631606 & -41.07994573153    & 00000010-4104482 & 14.338   & 0.043       & 13.627   & 0.038       & 13.532   & 0.045       & 0.000449  & -41.080074 & 0.9999833    & 0.1291896     \\
		2305877746911854336 & 16.9816 & 0.00064631606 & -41.07994573153    & 00000046-4104378 & 16.126   & 0.089       & 15.426   & 0.111       & 15.211   & 0.152       & 0.00193   & -41.07719  & 0.0744052    & 11.2620835   \\
		\enddata
		\tablecomments{This table is published in its entirety in the machine-readable format. A portion is shown here for guidance regarding its form and content.}
	\end{deluxetable}
\end{longrotatetable}

\begin{longrotatetable}
	\movetabledown=0.65in
	\begin{deluxetable}{lllllllllllllllll}
		\tabletypesize{\tiny}
		\tablecaption{All of the possible matches for the cross match between \textit{Gaia} and AllWISE. For more detailed descriptions of the columns in this table, look at the corresponding columns in Table \ref{tab:best_match_results_descrip}.\label{tab:ALLWISE_all_matches}}
		\tablehead{\colhead{\textit{Gaia} ID}            & \colhead{G}       & \colhead{$\alpha_{Gaia}$}               & \colhead{$\delta_{Gaia}$}            & \colhead{AllWISE ID}        & \colhead{$W1$}           & \colhead{$\sigma_{W1}$}          & \colhead{$W2$}           & \colhead{$\sigma_{W2}$}          & \colhead{$W3$}           & \colhead{$\sigma_{W3}$} & \colhead{$W4$}           & \colhead{$\sigma_{W4}$}          & \colhead{$\alpha_{AllWISE}$}             & \colhead{$\delta_{AllWISE}$}           & \colhead{Bayes Prob.}       & \colhead{$\Delta\theta$} \\
			\colhead{}            & \colhead{[mag]}       & \colhead{[deg]}               & \colhead{[deg]}            & \colhead{}        & \colhead{[mag]}           & \colhead{[mag]}          & \colhead{[mag]}           & \colhead{[mag]}          & \colhead{[mag]}           & \colhead{[mag]}         & \colhead{[mag]}  & \colhead{[mag]}   & \colhead{[deg]}             & \colhead{[deg]}           & \colhead{}       & \colhead{[arcsec]}}
		\startdata
		429379051804727296  & 16.9667 & 0.00021864741 & 60.89771179398  & J000000.06+605352.1 & 12.938      & 0.026          & 12.789      & 0.027          & 12.975      & \nodata           & 9.619       & \nodata           & 0.0002866   & 60.8978205  & 0.9990087      & 0.0652974       \\
		4991418100136496000 & 16.9943 & 0.00032430948 & -45.44935581167 & J000000.05-452657.5 & 12.97       & 0.024          & 12.826      & 0.026          & 11.821      & \nodata           & 8.402       & \nodata           & 0.0002164   & -45.4493089 & 0.0            & 0.0829052       \\
		2314289015157585664 & 20.407  & 0.00047449546 & -32.67650819595 & J000000.10-324035.5 & 17.812      & 0.216          & 17.295      & 0.493          & 12.047      & \nodata           & 8.683       & \nodata           & 0.0004466   & -32.6765412 & 0.995448       & 0.3449281       \\
		384318827606840448  & 17.9638 & 0.00047931984 & 42.19465632646  & J000000.09+421140.7 & 13.882      & 0.026          & 13.769      & 0.032          & 12.516      & \nodata           & 9.007       & \nodata           & 0.0003918   & 42.1946548  & 0.9994221      & 0.0253729       \\
		2772104109112208256 & 18.4284 & 0.00051294158 & 15.56517501763  & J000000.13+153354.8 & 14.6        & 0.031          & 14.366      & 0.055          & 12.449      & \nodata           & 8.658       & \nodata           & 0.0005666   & 15.5652269  & 0.9994498      & 0.0618302       \\
		4922439138656081024 & 16.1428 & 0.00054775339 & -57.26154422458 & J000000.16-571541.2 & 12.651      & 0.024          & 12.452      & 0.023          & 12.385      & \nodata           & 9.173       & \nodata           & 0.0006692   & -57.2614716 & 0.9998207      & 0.0245727       \\
		2738617902668633344 & 16.053  & 0.00059265319 & 1.41569388809   & J000000.13+012456.4 & 13.117      & 0.024          & 13.174      & 0.03           & 11.927      & \nodata           & 8.819       & \nodata           & 0.0005457   & 1.4156679   & 0.9997752      & 0.0552086       \\
		2420684907087327104 & 16.4742 & 0.000644395   & -14.44069806218 & J000000.14-142626.6 & 13.351      & 0.023          & 13.354      & 0.031          & 11.984      & \nodata           & 8.513       & \nodata           & 0.0006046   & -14.4407494 & 0.9994174      & 0.0174426       \\
		2305877746911854336 & 16.9816 & 0.00064631606 & -41.07994573153 & J000000.16-410448.0 & 13.25       & 0.024          & 13.155      & 0.027          & 12.562      & 0.477          & 9.173       & \nodata           & 0.0006781   & -41.0800067 & 0.9993026      & 0.2645646       \\
		2305877746911854336 & 16.9816 & 0.00064631606 & -41.07994573153 & J000000.50-410436.8 & 15.163      & 0.052          & 15.112      & 0.084          & 12.574      & \nodata           & 9.047       & \nodata           & 0.0021152   & -41.0768938 & 0.0            & 11.915629      
		\\
		\enddata
		\tablecomments{This table is published in its entirety in the machine-readable format. A portion is shown here for guidance regarding its form and content.}
	\end{deluxetable}
\end{longrotatetable}

\begin{longrotatetable}
	\begin{deluxetable}{lllllllllllll}
		\tabletypesize{\tiny}
		\tablecaption{All of the possible matches for the cross match between \textit{Gaia} and GALEX DR5. For more detailed descriptions of the columns in this table, look at the corresponding columns in Table \ref{tab:best_match_results_descrip}.\label{tab:GALEX_all_matches}}
		\tablehead{\colhead{\textit{Gaia} ID}            & \colhead{G}       & \colhead{$\alpha_{Gaia}$}               & \colhead{$\delta_{Gaia}$}            & \colhead{GALEX DR5 ID}        & \colhead{$FUV$}           & \colhead{$\sigma_{FUV}$}          & \colhead{$NUV$}           & \colhead{$\sigma_{NUV}$}      &     \colhead{$\alpha_{GALEX}$}             & \colhead{$\delta_{GALEX}$}           & \colhead{Bayes Prob.}       & \colhead{$\Delta\theta$} \\
			\colhead{}            & \colhead{[mag]}       & \colhead{[deg]}               & \colhead{[deg]}            & \colhead{}        & \colhead{[mag]}           & \colhead{[mag]}                  & \colhead{[mag]}           & \colhead{[mag]}          & \colhead{[deg]}             & \colhead{[deg]}           & \colhead{}       & \colhead{[arcsec]}}
		\startdata
		4991418100136496000 & 16.9943 & 0.00032430948 & -45.44935581167 & 6384813650104815341 & \nodata       & \nodata          & 22.495     & 0.464         & 0.005075  & -45.448452 & 0.2391745    & 13.0376701    \\
		2747168838957855104 & 12.9883 & 0.00100436367 & 9.22566353979   & 6376052767224496578 & \nodata       & \nodata          & 18.421     & 0.058         & 0.00088   & 9.225952   & 0.9999961    & 1.0644309     \\
		2739983083793578368 & 12.7572 & 0.00113295198 & 3.02061233516   & 6381013735771736390 & \nodata       & \nodata          & 18.946     & 0.077         & 0.001213  & 3.020992   & 0.9999957    & 0.5896474     \\
		4976553527564650496 & 16.0853 & 0.00113599787 & -50.7069141397  & 6384813684464554035 & \nodata       & \nodata          & 21.502     & 0.22          & 0.000586  & -50.706711 & 0.9997666    & 0.7214425     \\
		2422910181182822784 & 13.6938 & 0.00181091004 & -9.91118104462  & 6380415555275327592 & \nodata       & \nodata          & 22.407     & 0.49          & 0.001584  & -9.910674  & 0.9993253    & 1.510795      \\
		2341416475275983872 & 14.3727 & 0.00212583048 & -21.25877753606 & 6380556326049682051 & \nodata       & \nodata          & 23.007     & 0.399         & 0.000507  & -21.260026 & 0.0          & 6.5963465     \\
		4702977442386044032 & 16.3353 & 0.00371708969 & -70.18979943279 & 6384884009185313203 & \nodata       & \nodata          & 21.812     & 0.28          & 0.003073  & -70.189192 & 0.9945253    & 1.9620924     \\
		2875011834762675200 & 20.2727 & 0.00390831131 & 33.17729257174  & 6376756383799313130 & \nodata       & \nodata          & 22.337     & 0.463         & 0.005806  & 33.180352  & 0.0          & 11.9387993    \\
		2341335283213230464 & 10.9516 & 0.00419603514 & -21.68609418402 & 6380556326051776169 & 20.541     & 0.116         & 14.589     & 0.004         & 0.004004  & -21.686302 & 1.0000001    & 0.3253531     \\
		2335082979062194304 & 16.9159 & 0.00480314620 & -25.16797866075 & 6380556367925610339 & \nodata       & \nodata          & 21.476     & 0.347         & 0.004921  & -25.167562 & 0.9998403    & 0.8272201     \\
		\enddata
		\tablecomments{This table is published in its entirety in the machine-readable format. A portion is shown here for guidance regarding its form and content.}
	\end{deluxetable}
\end{longrotatetable}

\begin{longrotatetable}
	\movetabledown=0.7in
	\begin{deluxetable}{lllllllllllllllllll}
		\tabletypesize{\tiny}
		\tablecaption{All of the possible matches for the cross match between \textit{Gaia} and Pan-STARRS. For more detailed descriptions of the columns in this table, look at the corresponding columns in Table \ref{tab:best_match_results_descrip}.\label{tab:PS1_all_matches}}
		\tablehead{\colhead{\textit{Gaia} ID}            & \colhead{G}       & \colhead{$\alpha_{Gaia}$}               & \colhead{$\delta_{Gaia}$}            & \colhead{Pan-STARRS ID}        & \colhead{$g$}           & \colhead{$\sigma_{g}$}          & \colhead{$r$}           & \colhead{$\sigma_{r}$}          & \colhead{$i$}           & \colhead{$\sigma_{i}$} & \colhead{$z$}           & \colhead{$\sigma_{z}$}          &\colhead{$y$}           & \colhead{$\sigma_{y}$}          & \colhead{$\alpha_{PS1}$}             & \colhead{$\delta_{PS1}$}           & \colhead{Bayes Prob.}       & \colhead{$\Delta\theta$} \\
			\colhead{}            & \colhead{[mag]}       & \colhead{[deg]}               & \colhead{[deg]}            & \colhead{}        & \colhead{[mag]}           & \colhead{[mag]}          & \colhead{[mag]}           & \colhead{[mag]}          & \colhead{[mag]}           & \colhead{[mag]}         & \colhead{[mag]}  & \colhead{[mag]}         & \colhead{[mag]}  & \colhead{[mag]}   & \colhead{[deg]}             & \colhead{[deg]}           & \colhead{}       & \colhead{[arcsec]}}
		\startdata
		429379051804727296  & 16.9667 & 0.00021864741 & 60.89771179398  & 181070000002178144 & 18.9908 & 0.0032    & 17.7993 & 0.0042    & 16.2296 & 0.0028    & 15.5105 & 0.0014    & 15.1478 & 0.0045    & 0.00022122 & 60.89773213  & 0.9960894          & 0.0443499   \\
		429379051804727296  & 16.9667 & 0.00021864741 & 60.89771179398  & 181070000017386176 & \nodata    & \nodata      & \nodata    & \nodata      & 21.503  & 0.1446    & 20.7487 & 0.1538    & \nodata    & \nodata      & 0.00180374 & 60.89620386  & 0.0188694          & 6.1896178   \\
		429379051804727296  & 16.9667 & 0.00021864741 & 60.89771179398  & 181070000029977216 & \nodata    & \nodata      & \nodata    & \nodata      & 21.0219 & 0.0667    & 20.7423 & 0.0955    & \nodata    & \nodata      & 0.00298512 & 60.89703229  & 0.0159218          & 5.460592    \\
		429379051804727296  & 16.9667 & 0.00021864741 & 60.89771179398  & 181070000048379136 & 21.7554 & 0.2015    & 20.4099 & 0.0365    & 19.7856 & 0.0142    & 19.4149 & 0.0327    & 19.1697 & 0.035     & 0.00481773 & 60.89865394  & 0.0                & 8.6744328   \\
		429379051804727296  & 16.9667 & 0.00021864741 & 60.89771179398  & 181080000049152160 & 18.8686 & 0.0183    & 18.0388 & 0.0056    & 17.6127 & 0.008     & 17.3777 & 0.0068    & 17.1979 & 0.0084    & 0.00491942 & 60.90115514  & 0.0                & 14.7724567  \\
		2314289015157585664 & 20.407  & 0.00047449546 & -32.67650819595 & 68780000004548592  & \nodata    & \nodata      & \nodata    & \nodata      & 19.8489 & 0.0183    & \nodata    & \nodata      & \nodata    & \nodata      & 0.00047162 & -32.67650276 & 0.9996824000000001 & 0.0394867   \\
		384318827606840448  & 17.9638 & 0.00047931984 & 42.19465632646  & 158630000003914240 & 19.9025 & 0.0054    & 18.6328 & 0.0068    & 17.2093 & 0.0023    & 16.5534 & 0.0032    & 16.2281 & 0.0048    & 0.00046683 & 42.19465832  & 0.99864            & 0.0246682   \\
		2772104109112208256 & 18.4284 & 0.00051294158 & 15.56517501763  & 126670000005728752 & 20.2047 & 0.0216    & 19.0564 & 0.0123    & 17.6737 & 0.0047    & 17.0499 & 0.0045    & 16.7519 & 0.0051    & 0.00052103 & 15.56518018  & 0.9995842          & 0.0219664   \\
		2738617902668633344 & 16.053  & 0.00059265319 & 1.41569388809   & 109690000004829264 & 17.2422 & 0.0034    & 16.0862 & 0.0039    & 15.5176 & 0.0022    & 15.2649 & 0.0051    & 15.0979 & 0.0055    & 0.00057851 & 1.4156878    & 0.9999516          & 0.0064351   \\
		2738617902668633344 & 16.053  & 0.00059265319 & 1.41569388809   & 109690000006064288 & 21.7947 & 0.2067    & 21.7223 & 0.0567    & 21.5412 & 0.0293    & 20.7805 & 0.1455    & \nodata    & \nodata      & 0.00072035 & 1.41159149   & 0.0                & 14.7581938    
		\\
		\enddata
		\tablecomments{This table is published in its entirety in the machine-readable format. A portion is shown here for guidance regarding its form and content.}
	\end{deluxetable}
\end{longrotatetable}

\begin{longrotatetable}
	\movetabledown=0.7in
	\begin{deluxetable}{lllllllllllllllllll}
		\tabletypesize{\tiny}
		\tablecaption{All of the possible matches for the cross match between \textit{Gaia} and SDSS DR12. For more detailed descriptions of the columns in this table, look at the corresponding columns in Table \ref{tab:best_match_results_descrip}.\label{tab:SDSS_all_matches}}
		\tablehead{\colhead{\textit{Gaia} ID}            & \colhead{G}       & \colhead{$\alpha_{Gaia}$}               & \colhead{$\delta_{Gaia}$}            & \colhead{SDSS DR12 ID}        &\colhead{$u$}           & \colhead{$\sigma_{u}$}& \colhead{$g$}           & \colhead{$\sigma_{g}$}          & \colhead{$r$}           & \colhead{$\sigma_{r}$}          & \colhead{$i$}           & \colhead{$\sigma_{i}$} & \colhead{$z$}           & \colhead{$\sigma_{z}$}                    & \colhead{$\alpha_{SDSS}$}             & \colhead{$\delta_{SDSS}$}           & \colhead{Bayes Prob.}       & \colhead{$\Delta\theta$} \\
			\colhead{}            & \colhead{[mag]}       & \colhead{[deg]}               & \colhead{[deg]}            & \colhead{}        & \colhead{[mag]}           & \colhead{[mag]}          & \colhead{[mag]}           & \colhead{[mag]}          & \colhead{[mag]}           & \colhead{[mag]}         & \colhead{[mag]}  & \colhead{[mag]}         & \colhead{[mag]}  & \colhead{[mag]}   & \colhead{[deg]}             & \colhead{[deg]}           & \colhead{}       & \colhead{[arcsec]}}
		\startdata
		2772104109112208256 & 18.4284 & 0.00051294158 & 15.56517501763 & J000000.15+153354.9 & 23.126  & 0.421      & 20.441  & 0.023      & 19.025  & 0.011      & 17.637  & 0.007      & 16.912  & 0.011      & 0.000635 & 15.565261 & 0.9997612   & 0.0421463    \\
		2772104109112208256 & 18.4284 & 0.00051294158 & 15.56517501763 & J000000.15+153354.9 & 22.831  & 0.297      & 20.454  & 0.022      & 19.048  & 0.011      & 17.655  & 0.007      & 16.941  & 0.01       & 0.000641 & 15.565276 & 0.0         & 0.0975381    \\
		2772104109112208256 & 18.4284 & 0.00051294158 & 15.56517501763 & J000000.30+153408.3 & 24.246  & 0.701      & 23.369  & 0.217      & 22.907  & 0.201      & 23.235  & 0.346      & 22.838  & 0.428      & 0.001256 & 15.568985 & 0.0         & 13.6115789   \\
		2772104109112208256 & 18.4284 & 0.00051294158 & 15.56517501763 & J000000.30+153408.4 & 23.284  & 0.491      & 23.025  & 0.176      & 22.679  & 0.175      & 22.568  & 0.278      & 23.536  & 0.605      & 0.001261 & 15.569004 & 0.0         & 13.6818764   \\
		2738617902668633344 & 16.053  & 0.00059265319 & 1.41569388809  & J000000.12+012456.3 & 20.117  & 0.042      & 17.428  & 0.005      & 16.074  & 0.004      & 15.493  & 0.004      & 15.171  & 0.005      & 0.000531 & 1.41566   & 0.9999276   & 0.2760008    \\
		2738617902668633344 & 16.053  & 0.00059265319 & 1.41569388809  & J000000.17+012441.5 & 24.972  & 1.399      & 22.099  & 0.131      & 21.326  & 0.092      & 20.721  & 0.083      & 20.653  & 0.269      & 0.000739 & 1.411541  & 0.0         & 14.8164078   \\
		2873403867726399232 & 19.0963 & 0.00089948698 & 30.80569762321 & J000000.17+304820.6 & 22.965  & 0.295      & 20.334  & 0.018      & 19.103  & 0.011      & 18.554  & 0.01       & 18.258  & 0.024      & 0.000712 & 30.805743 & 0.9997258   & 0.0623462    \\
		2873403867726399232 & 19.0963 & 0.00089948698 & 30.80569762321 & J000000.19+304820.5 & 22.385  & 0.258      & 20.356  & 0.019      & 19.064  & 0.011      & 18.537  & 0.011      & 18.21   & 0.024      & 0.000833 & 30.805711 & 0.0         & 0.4533442    \\
		2747168838957855104 & 12.9883 & 0.00100436367 & 9.22566353979  & J000000.21+091332.1 & 15.214  & 0.005      & 14.593  & 0.005      & 13.118  & 0.002      & 12.915  & 0.002      & 13.1    & 0.005      & 0.000893 & 9.225608  & 0.9999519   & 0.1975248    \\
		2747168838957855104 & 12.9883 & 0.00100436367 & 9.22566353979  & J000000.96+091330.2 & 23.84   & 1.334      & 23.15   & 0.271      & 22.441  & 0.257      & 21.988  & 0.243      & 23.428  & 0.883      & 0.004033 & 9.225064  & 0.0         & 11.4517712    
		\\
		\enddata
		\tablecomments{This table is published in its entirety in the machine-readable format. A portion is shown here for guidance regarding its form and content.}
	\end{deluxetable}
\end{longrotatetable}

\begin{longrotatetable}
	\begin{deluxetable}{lllllllllllll}
		\tabletypesize{\tiny}
		\tablecaption{All of the possible matches for the cross match between \textit{Gaia} and RAVE DR5. For more detailed descriptions of the columns in this table, look at the corresponding columns in Table \ref{tab:best_match_results_descrip}.\label{tab:RAVE_all_matches}}
		\tablehead{\colhead{\textit{Gaia} ID}            & \colhead{G}       & \colhead{$\alpha_{Gaia}$}               & \colhead{$\delta_{Gaia}$}            & \colhead{RAVE DR5 ID}        & \colhead{$B_T$}           & \colhead{$\sigma_{B_T}$}          & \colhead{$V_T$}           & \colhead{$\sigma_{V_T}$}      &     \colhead{$\alpha_{RAVE}$}             & \colhead{$\delta_{RAVE}$}           & \colhead{Bayes Prob.}       & \colhead{$\Delta\theta$} \\
			\colhead{}            & \colhead{[mag]}       & \colhead{[deg]}               & \colhead{[deg]}            & \colhead{}        & \colhead{[mag]}           & \colhead{[mag]}                  & \colhead{[mag]}           & \colhead{[mag]}          & \colhead{[deg]}             & \colhead{[deg]}           & \colhead{}       & \colhead{[arcsec]}}
		\startdata
		4919074289477561472 & 11.6106 & 0.00780738967 & -58.71847417652 & J000001.9-584307 & 12.533   & 0.154       & 11.706   & 0.103       & 0.00808  & -58.71853 & 0.9934741   & 0.2496141    \\
		4919427610667433600 & 16.5728 & 0.03675557788 & -57.48719495688 & J000009.0-572913 & \nodata     & \nodata        & \nodata     & \nodata        & 0.03729  & -57.48683 & 0.0         & 2.1838843    \\
		4919427610667433728 & 13.0809 & 0.03783936489 & -57.4868416914  & J000009.0-572913 & \nodata     & \nodata        & \nodata     & \nodata        & 0.03729  & -57.48683 & 0.0         & 0.2650074    \\
		4973451770902209920 & 13.2093 & 0.05450144802 & -51.61164961089 & J000013.0-513642 & \nodata     & \nodata        & \nodata     & \nodata        & 0.05404  & -51.61158 & 0.0         & 0.2256993    \\
		2413944625931002624 & 12.148  & 0.06847151987 & -19.10762290314 & J000016.3-190627 & \nodata     & \nodata        & \nodata     & \nodata        & 0.068    & -19.10758 & 0.0         & 0.2232123    \\
		2448187747347233664 & 11.0992 & 0.06966855007 & -2.94569133915  & J000016.7-025652 & 10.927   & 0.048       & 10.002   & 0.032       & 0.06971  & -2.94781  & 0.6104921   & 7.8678634    \\
		2448187747347250048 & 9.7686  & 0.0698486806  & -2.94784184175  & J000016.7-025652 & 10.927   & 0.048       & 10.002   & 0.032       & 0.06971  & -2.94781  & 0.9946625   & 0.3243487    \\
		4634503362901736576 & 10.6905 & 0.07271168199 & -80.26747298433 & J000017.1-801603 & 11.828   & 0.079       & 11.019   & 0.069       & 0.07125  & -80.26739 & 0.9946625   & 0.1978637    \\
		2420804032300281728 & 12.7836 & 0.07347788125 & -13.57904040546 & J000017.7-133444 & \nodata     & \nodata        & \nodata     & \nodata        & 0.07354  & -13.57889 & 0.0         & 0.2641875    \\
		2414164459536733952 & 9.9923  & 0.0744390513  & -18.17238322596 & J000017.7-181021 & 11.006   & 0.046       & 10.204   & 0.033       & 0.07392  & -18.1725  & 0.9946625   & 0.045774    
		\\
		\enddata
		\tablecomments{This table is published in its entirety in the machine-readable format. A portion is shown here for guidance regarding its form and content.}
	\end{deluxetable}
\end{longrotatetable}

\section{Photometric Metallicity Determination}

With the cross-matched data discussed above, we can now assemble a catalog of stars that includes not only a large variety of photometric bands for all objects, but also spectra for a subset of $\sim$21,000 stars thanks to surveys like SDSS. Using the derived properties from these spectra we now attempt to uncover a relationship between stellar metallicity and broadband photometry that could be used to infer metallicity values for all stars. We focus on finding such a relationship specifically for K and M dwarfs, as these are by far the must abundant stellar subtype in a high proper motion sample such as this one. 

\subsection{K and Early M Dwarf Relationship}\label{sec:KM_metals}

Thanks to the SDSS cross match discussed in the previous section, our database comprises spectroscopic data for $\sim$21,000 stars. Of particular interest are the APOGEE spectra \citep{apogee} of the matched K and early M dwarfs in our sample, as APOGEE provides metallicity measurements. We choose to focus on M dwarfs with temperatures $>$3500 K as below this value the stellar parameters from APOGEE are derived from a coarser grid of synthetic spectra, leading to more inaccuracies in the derived metallicities. To determine which stars in our high proper motion sample are K/M dwarfs with $T_{eff}>3500$ K, we use the relationships between \textit{Gaia} photometry and spectral type from \citet{mamajek2013}. To complement the APOGEE spectra at both the lower mass end and the lower metallicity end, we also use the catalog from \citet{hejazi2020}, who derived stellar parameters from low-resolution spectra of 1544 high proper motion M dwarfs and subdwarfs, all of which are included in our database. Additionally, we focus on stars that have grizy photometry from Pan-STARRS, as it has been shown that low-mass stars show large variations in their optical colors due to molecular band opacities, and that some of these variations are tied to differences in chemical composition \citep{lepine2007metal}. We also require that the stars have counterparts in 2MASS and AllWISE, as it has also been shown that optical colors become degenerate in relation to metallicity at the very low mass end \citep[e.g.][]{Schmidt2016}. Due to saturation issues with the Pan-STARRS photometric bands, we only include sources with $g>13.5$ in our sample. Finally, we correct all photometry for extinction using the 3D dust map from \citet{baystar19}. We exclude all stars from the calibration sample with $E(B-V)>0.05$. Values of $A_\lambda$ are calculated for Pan-STARRS, assuming $R_V=3.1$, using the results from \citet{schlafly2011}, and $A_\lambda$ values for 2MASS and AllWISE using the results from \cite{davenport2014}. These values of $A_\lambda$ are used to correct all photometric measurements for extinction. This yields a training sample of 6197 stars with APOGEE spectra, plus 173 stars from \citet{hejazi2020} that meet all criteria and have listed magnitudes in all bands being considered for the calibration (i.e. g, r ,i, z, y, J, H, K$_s$, W1 and W2).

To determine a relationship between photometry and metallicity we use a Gaussian Process Regressor implemented in \textit{scikit-learn} \citep{sklearn} with an RBF kernel and a white-noise kernel proportional to the average error of the derived abundances in an iterative manner. In the first iteration, the sample includes all of the K and early M dwarfs with APOGEE spectra that have $g>13.5$. To determine the optimal set of colors and absolute magnitudes for the metallicity prediction, we start out by performing the regression on a training subset consisting of 60\% of the overall sample, determined by splitting the sample into five metallicity bins and selecting 60\% from each bin to guarantee proportional representation in all color combinations and absolute magnitudes for the chosen photometry. The regression is then evaluated using the mean squared error of the regression on the testing subset consisting of the remaining 40\% of the overall sample. Next, to determine the color or absolute magnitude that least affects the regression, each color and absolute magnitude is individually removed in turn, the regression performed again, and the new mean squared error compared to the initial value obtained when all colors and absolute magnitudes are in use. The color or absolute magnitude that produces the smallest change in the mean squared error is then omitted for the remaining iterations, as it has the least effect on the overall regression. This process is repeated until one color or absolute magnitude remains. The optimal set of colors and absolute magnitudes that provide the best metallicity prediction are determined to be the set of $N$ colors and magnitudes that minimize the mean squared error for the entire process.

\begin{figure*}
	\plotone{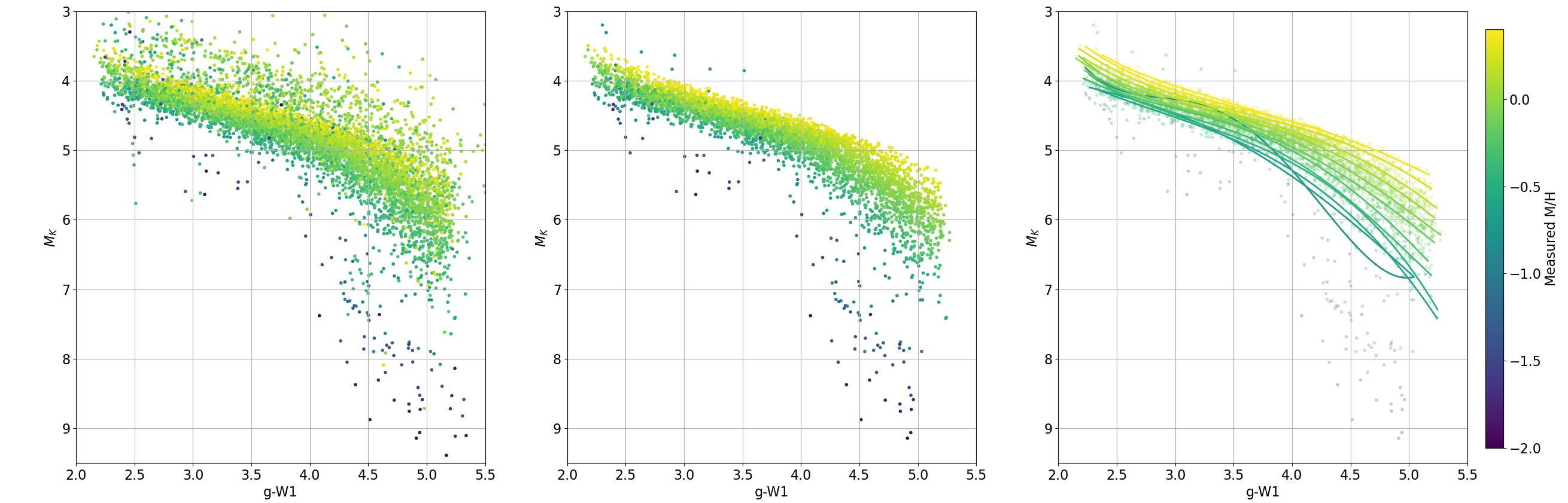}
	\caption{HR diagrams of $M_{K}$ vs. $g-W1$ for our learning sample of K and early M dwarfs with APOGEE spectra or abundances from \cite{hejazi2020} and $g>13.5$. The left panel is the original subset, the middle panel is the subset after unresolved binaries have been removed, using the method described in Section \ref{sec:KM_metals}, and the right panel shows the polynomial fits from Table \ref{tab:KM_poly} overlayed on the subset after the removal of unresolved binaries, where the color of the line corresponds tp the bin of the polynomial fit.}
	\label{fig:KM_single_star}
\end{figure*}

With this first iteration completed, we perform additional cleaning of our sample by attempting to remove unresolved binaries, as their derived metallicities in spectroscopic surveys will be inaccurate due to their properties being derived from synthetic spectra of single stars. As unresolved binaries appear over-luminous on an HR diagram compared with stars of similar metallicity, we use the best related color and absolute magnitude combination from the first iteration to clean the sample in the following manner: we plot subsets of stars from each of 13 metallicity bins of 0.1 dex in width, defined in the range $-0.8 < [M/H] < 0.5$; to each bin we fit a single-star main sequence using a fourth degree polynomial fit, where the fit is performed iteratively by removing stars with $d_{M,i}>\overline{d_M}+1.95 \sigma_{d_M}$ (i.e. stars significantly above the single-star main sequence on a HR diagram) and $d_{M,i}<\overline{d_M}-3 \sigma_{d_M}$ (i.e. stars significantly below the single-star main sequence on a HR diagram), where $d_{M,i}$ is the distance in absolute magnitude from the fitted relationship of the $i$th star in the sample, defined as $d_{M,i}=M_{fit}-M_i$, also where $\overline{d_M}$ is the mean distance in absolute magnitude from the fitted relationship for all stars in each step, and finally where $\sigma_{d_M}$ is the standard deviation in absolute magnitude around the fitted relationship for all stars in each step; iterations continue until no more stars are removed. Due to low number statistics, we are however unable to fit color-magnitude relationships for stars with $M/H\leq -0.8$, so we exclude objects with $d_M>\overline{d_M}$ using the color-magnitude relationship polynomial relationship for stars with $-0.8<M/H\leq-0.7$ bin. The polynomial fits to the single star main sequence are shown in Table \ref{tab:KM_poly} along with the mean and standard deviation distance from the relationship used to qualify stars as over- or under-luminous. This results in a cleaned sample of 4296 K and early M dwarfs from APOGEE and 82 dwarfs from \citet{hejazi2020}. This cleaned sample is shown in the middle panel of Figure \ref{fig:KM_single_star}, which demonstrates how this ``single star filter" cleans the sample significantly as compared to the original sample (left panel). Additionally, the polynomial fits from Table \ref{tab:KM_poly} are overlayed on the ``cleaned" sample in the right panel of Figure \ref{fig:KM_single_star}. All polynomials seem to have converged onto the single star sequence, with the exception of the polynomial for $-0.8<M/H\leq -0.7$ due to the small number of stars in this bin. This lack of convergence means some metal-poor over-luminous stars remain in our sample, but due to their small numbers they will not significantly change the results.

\begin{deluxetable*}{cccccccc}
	\tabletypesize{\small}
	\tablecaption{Polynomial fits to the single star main sequence for thirteen bins of metallicity for K and early M dwarfs. Coefficients of the polynomial fits describe the relationship $y=a_0+a_1x+a_2x^2+a_3x^3+a_4x^4$ where $y=M_{K}$ and $x=g-W1$. The last two columns of the table give the mean offset and standard deviation, in absolute magnitude, from the relationship; these are used to identify stars deemed over- or under-luminous. \label{tab:KM_poly}}
	\tablehead{\colhead{M/H Range} & \colhead{$a_0$} & \colhead{$a_1$} & \colhead{$a_2$} & \colhead{$a_3$} & \colhead{$a_4$} & \colhead{$\overline{d_M}$} & \colhead{$\sigma_{d_M}$}}
	\startdata
	$-0.8<M/H\leq -0.7$ & -43.19 & 57.48 & -25.74 & 5.004 & -0.3510 & $-1.170\times10^{-13}$ & 0.3731 \\
	$-0.7<M/H\leq -0.6$ & 1.824 & 2.537 & -1.192 & 0.2692 & -0.01857 & $9.992\times10^{-16}$ & 0.1659 \\
	$-0.6<M/H\leq -0.5$ & -0.3690 & 4.037 & -1.296 & 0.1782 & -0.004553 & $-3.303\times10^{-15}$ & 0.06606 \\
	$-0.5<M/H\leq -0.4$ & 2.118 & 0.4164 & 0.5289 & -0.2076 & 0.02430 & $-4.996\times10^{-16}$ & 0.08184 \\
	$-0.4<M/H\leq -0.3$ & -12.04 & 17.25 & -6.821 & 1.188 & -0.07299 & $-2.626\times10^{-14}$ & 0.07402 \\
	$-0.3<M/H\leq -0.2$ & -4.396 & 8.858 & -3.415 & 0.5781 & -0.03274 & $-1.782\times10^{-14}$ & 0.08891 \\
	$-0.2<M/H\leq -0.1$ & -6.768 & 11.16 & -4.278 & 0.7296 & -0.04357 & $3.678\times10^{-14}$ & 0.07739 \\
	$-0.1<M/H\leq 0.0$ & -11.87 & 16.33 & -6.188 & 1.034 & -0.06152 & $-3.089\times10^{-14}$ & 0.06604 \\
	$0.0<M/H\leq 0.1$ & -5.712 & 9.465 & -3.394 & 0.5390 & -0.02949 & $1.132\times10^{-14}$ & 0.06488 \\
	$0.1<M/H\leq 0.2$ & -4.380 & 7.557 & -2.479 & 0.3569 & -0.01679 & $7.633\times10^{-16}$ & 0.05740 \\
	$0.2<M/H\leq 0.3$ & -3.690 & 6.418 & -1.909 & 0.2425 & -0.009016 & $7.730\times10^{-15}$ & 0.05395 \\
	$0.3<M/H\leq 0.4$ &-5.631 & 8.614 & -2.874 & 0.4313 & -0.02284 & $1.929\times10^{-14}$ & 0.04636 \\
	$0.4<M/H\leq 0.5$ & -11.19 & 16.37 & -6.857 & 1.316 & -0.09463 & $3.941\times10^{-15}$ & 0.04425 \\
	\enddata
\end{deluxetable*}

\begin{figure*}
	\epsscale{0.95}
	\plotone{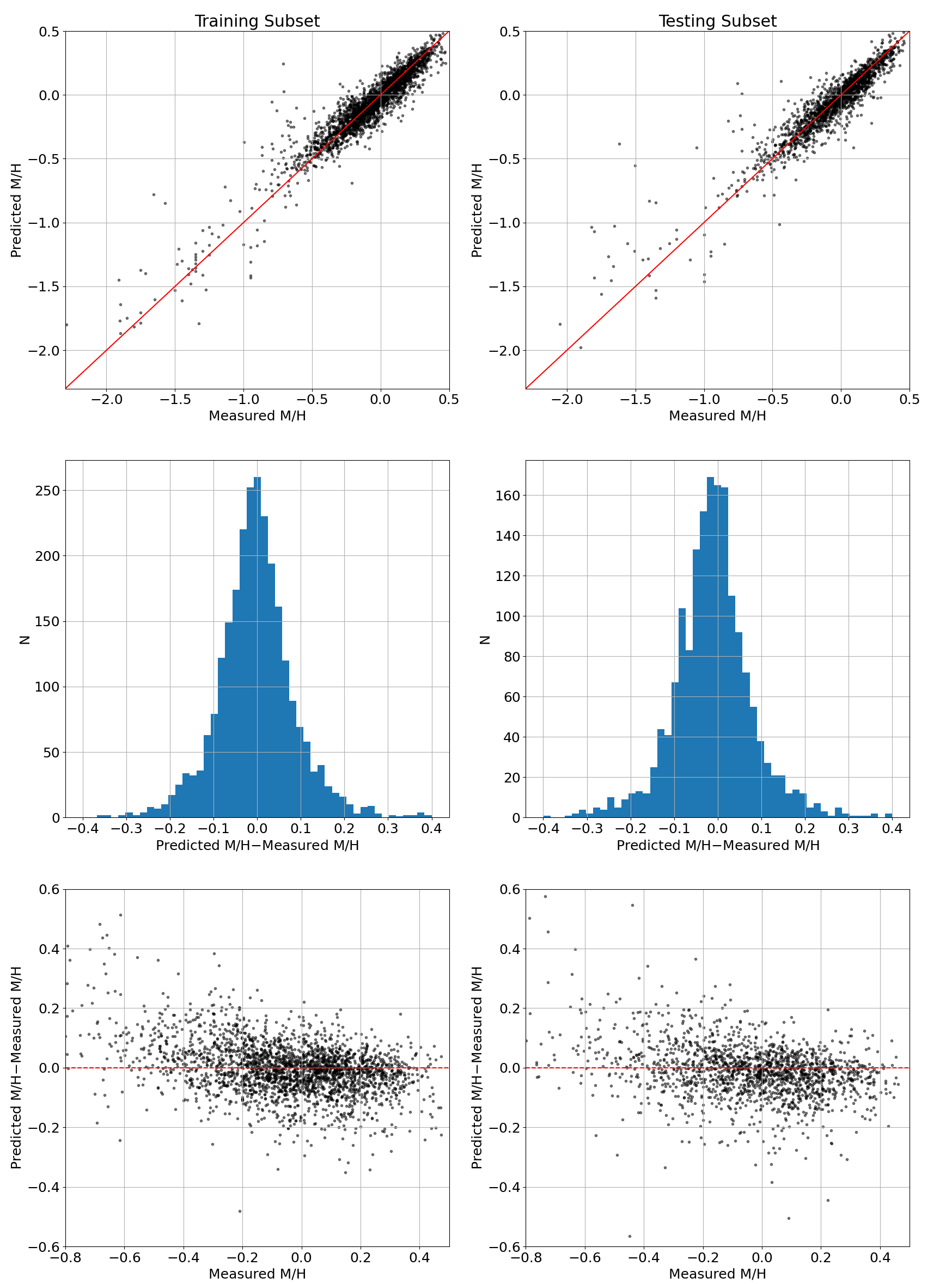}
	\caption{Results of the Gaussian Process Regressor on the single-star data set for K and early M dwarfs using the following combination of colors and absolute magnitudes: $M_g$, $g-y$, $y-W2$, $J-W2$, $W1-W2$. The first column shows the result for the training subset (the subset used to train the regressor) and the second column the testing subset (the subset used to evaluate the regressor). The top row compares the predicted metallicity from the regressor to the actual value from APOGEE or \cite{hejazi2020}, the middle row shows the distribution of the differences between the predicted and APOGEE metallicity and the bottom rows show the residual error in metallicity versus the measured metallicity.}
	\label{fig:KM_regress_results}
\end{figure*}

Finally, we use stars in this cleaned sample to perform our final regression. To determine the optimal colors and absolute magnitudes, we use the same iterative method, as discussed above. To reiterate, in this method inputs are iteratively removed, based on the color/absolute magnitude that produces the smallest change in the mean squared error when removed, until only two inputs remain. The optimal colors and absolute magnitudes are then the combination of $N$ inputs that provided the minimum mean squared error for the testing subset during the entire process. For the cleaned sample, the optimal set in this case consists of $M_g$, $g-y$, $y-W2$, $J-W2$, $W1-W2$. The results of the regressor for the training subset (the subset used to train the regressor) and the testing subset (the subset used to evaluate the regressor) are shown in Figure \ref{fig:KM_regress_results}. A comparison of predicted and observed metallicity (Figure \ref{fig:KM_regress_results}, second row) values suggests that the regressor is capable of predicting the metallicities from the input colors and absolute magnitudes to a precision of $\sim$0.12 dex based on the $1\sigma$ scatter. Additionally, the residual error in predicted vs. measured metallicity (Figure \ref{fig:KM_regress_results}, bottom row) demonstrates that this scatter is roughly constant over the metallicity range under consideration and that systematic offsets are only apparent at the very metal poor (and sparse) end of the distribution.

The actual regressor can be loaded via a Python script; the file with the trained \textit{scikit-learn} regressor, along with example code to load and use the regressor to predict metallicites, are hosted on our GitHub repository (\url{https://github.com/imedan/gpr_metallicity_relationship}).

\subsection{M Dwarf Relationship}\label{sec:M_metals}

Next we attempt to determine a photometry-metallicity relationship for M dwarfs cooler than 3500 K, in the same manner as for the K and early M dwarfs (see Section \ref{sec:KM_metals}). The main challenge is to identify a training set for these cooler stars. As discussed above, the stellar parameters for later-type M dwarfs determined from APOGEE are not as reliable as for the K dwarfs. However, since we now have a reliable photometric relationship for the K and early M dwarfs, we can exploit the wide binaries in our high proper motion sample and infer the metallicities of the cooler M dwarf secondaries by proxy of their K dwarf primary. To do this, we use the SUPERWIDE catalog of wide binaries from \citet{hartman2020}, which lists $\sim$100,000 high proper motion, wide binaries identified in \textit{Gaia} through a Bayesian method. Similar to above, the photometry of the stars in each binary is corrected for extinction prior to the analysis. From this catalog, we photometrically select all pairs with a K and early M dwarf primary that also have an M dwarf secondary cooler than 3500 K, again based on the relationship between \textit{Gaia} photometry and spectral type from \citet{mamajek2013}. Metallicities for the primaries are determined using the final regressor from the Section \ref{sec:KM_metals} above, and all unresolved binaries for the primary sources are removed using the polynomial relationships and cuts determined for the single star main sequence found in the previous section. This results in a set of 4,859 primary stars, used as proxy for the metallicity values of their cooler M dwarf secondaries. Additionally, we complement this sample with 364 cool, single-star M dwarfs ($<$3500 K) with metallicity estimates from \cite{hejazi2020}.

\begin{figure*}
	\plotone{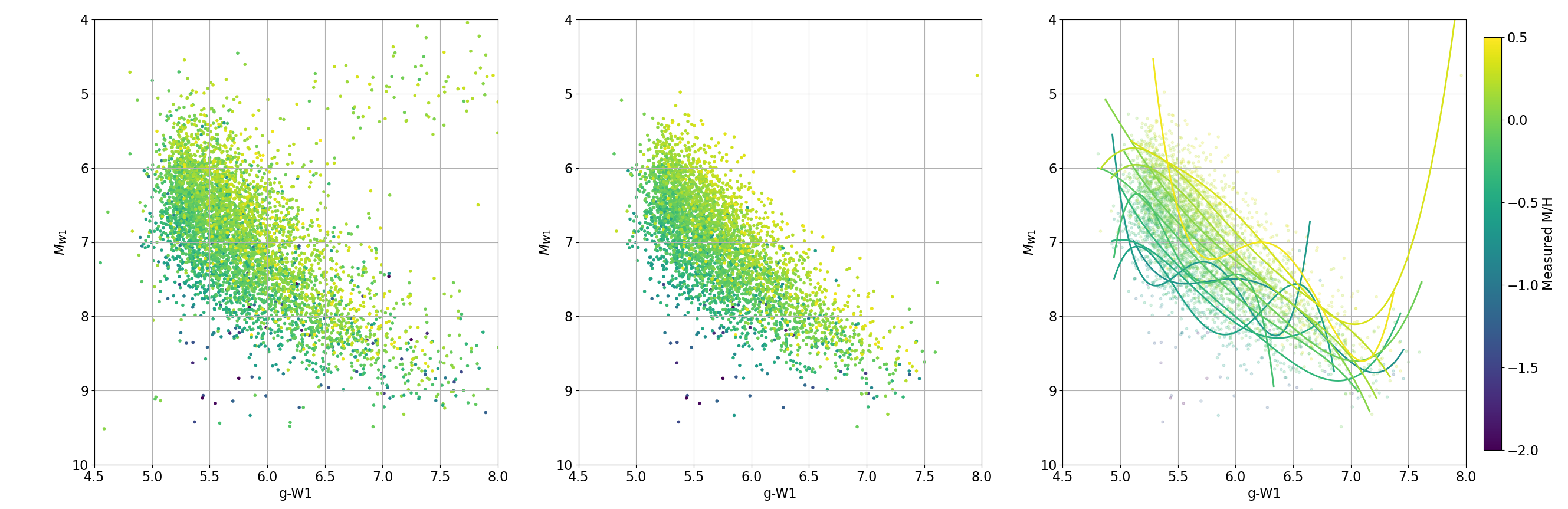}
	\caption{HR diagrams of $M_{W1}$ vs. $g-W1$ for our sample of M dwarf secondaries in common proper motion systems with a K or early M dwarf primary. The left panel is the original subset, the middle panel is the subset after unresolved binaries have been removed, using the method described in Section \ref{sec:KM_metals}, and the right panel shows the polynomial fits from Table \ref{tab:M_poly} overlayed on the subset after the removal of unresolved binaries, where the color of the line corresponds tp the bin of the polynomial fit.}
	\label{fig:M_single_star}
\end{figure*}

With this sample of 5,223 cooler M dwarfs now with ``observed" metallicity estimates, we train our Gaussian Process Regressor using the same procedure described in Section \ref{sec:KM_metals} for the K and early M dwarfs. After finding the color and absolute magnitude combination that is the best predictor of metallicity for the M dwarfs, and after removing possible unresolved binaries, our cleaned sample comprises 3,689 stars from the sample of wide binaries and 329 stars from \cite{hejazi2020}. The new polynomial fits to the single star main sequence for the M dwarfs used to clean the sample are shown in Table \ref{tab:M_poly}, and Figure \ref{fig:M_single_star} shows HR diagrams of the original (left panel) and cleaned (middle panel) samples. Additionally, the right panel of Figure \ref{fig:M_single_star} overlays the polynomials from Table \ref{tab:M_poly} on the HR diagram, where it is clear that many of the fits, especially in the metal-poor bins, did not converge on the single-star main sequence. In Section \ref{sec:M_dwarf_discuss}, we will discuss the overall impact of this on the final result.

\begin{deluxetable*}{cccccccc}
	\tablecaption{Polynomial fits to the single star main sequence for thirteen bins of metallicity for M dwarfs. Coefficients of the polynomial fits describe the relationship $y=a_0+a_1x+a_2x^2+a_3x^3+a_4x^4$ where $y=M_{W1}$ and $x=g-W1$. The last two columns of the table give the mean offset and standard deviation, in absolute magnitude, from the relationship; these are used to identify stars deemed over- or under-luminous.\label{tab:M_poly}}
	\tablehead{\colhead{M/H Range} & \colhead{$a_0$} & \colhead{$a_1$} & \colhead{$a_2$} & \colhead{$a_3$} & \colhead{$a_4$} & \colhead{$\overline{d_M}$} & \colhead{$\sigma_{d_M}$}}
	\startdata
	$-0.8<M/H\leq -0.7$ &-2509 & 1651 & -404.0 & 43.71 & -1.762 & -0.1254 & 0.5188 \\
	$-0.7<M/H\leq -0.6$ &-4509 & 3141 & -816.7 & 94.04 & -4.046 & 0.05895 & 0.5075\\
	$-0.6<M/H\leq -0.5$ &-4688 & 3153 & -790.2 & 87.59 & -3.622 & 0.08001 & 0.5806\\
	$-0.5<M/H\leq -0.4$ &-2098 & 1427 & -362.2 & 40.76 & -1.715 & -0.03318 & 0.3710\\
	$-0.4<M/H\leq -0.3$ &2894 & -2051 & 5434.0 & -63.86 & 2.802 & -0.03248 & 0.1901\\
	$-0.3<M/H\leq -0.2$ &4704 & -3303 & 867.9 & -101.0 & 4.397 & -0.01973 & 0.3114\\
	$-0.2<M/H\leq -0.1$ &-345.4 & 230.1 & -57.08 & 6.349 & -0.2657 & -0.02689 & 0.2953\\
	$-0.1<M/H\leq 0.0$ &395.7 & -254.7 & 61.24 & -6.422 & 0.2487 & -0.03397 & 0.2870\\
	$0.0<M/H\leq 0.1$ &233.0 & -170.5 & 46.54 & -5.521 & 0.2426 & -0.04784& 0.3043\\
	$0.1<M/H\leq 0.2$ &-349.8 & 247.6 & -65.07 & 7.620 & -0.3334 & -0.06066 & 0.2797\\
	$0.2<M/H\leq 0.3$ &770.2 & -492.8 & 117.8 & -12.38 & 0.4845 & -0.04733 & 0.3713\\
	$0.3<M/H\leq 0.4$ &665.1 & -403.2 & 91.19 & -9.037 & 0.3320 & -0.07541 & 0.5089\\
	$0.4<M/H\leq 0.5$ &-3466 & 2201 & -521.5 & 54.73 & -2.145 & 0.01033 & 0.5504\\
	\enddata
\end{deluxetable*}

\begin{figure*}
	\epsscale{0.95}
	\plotone{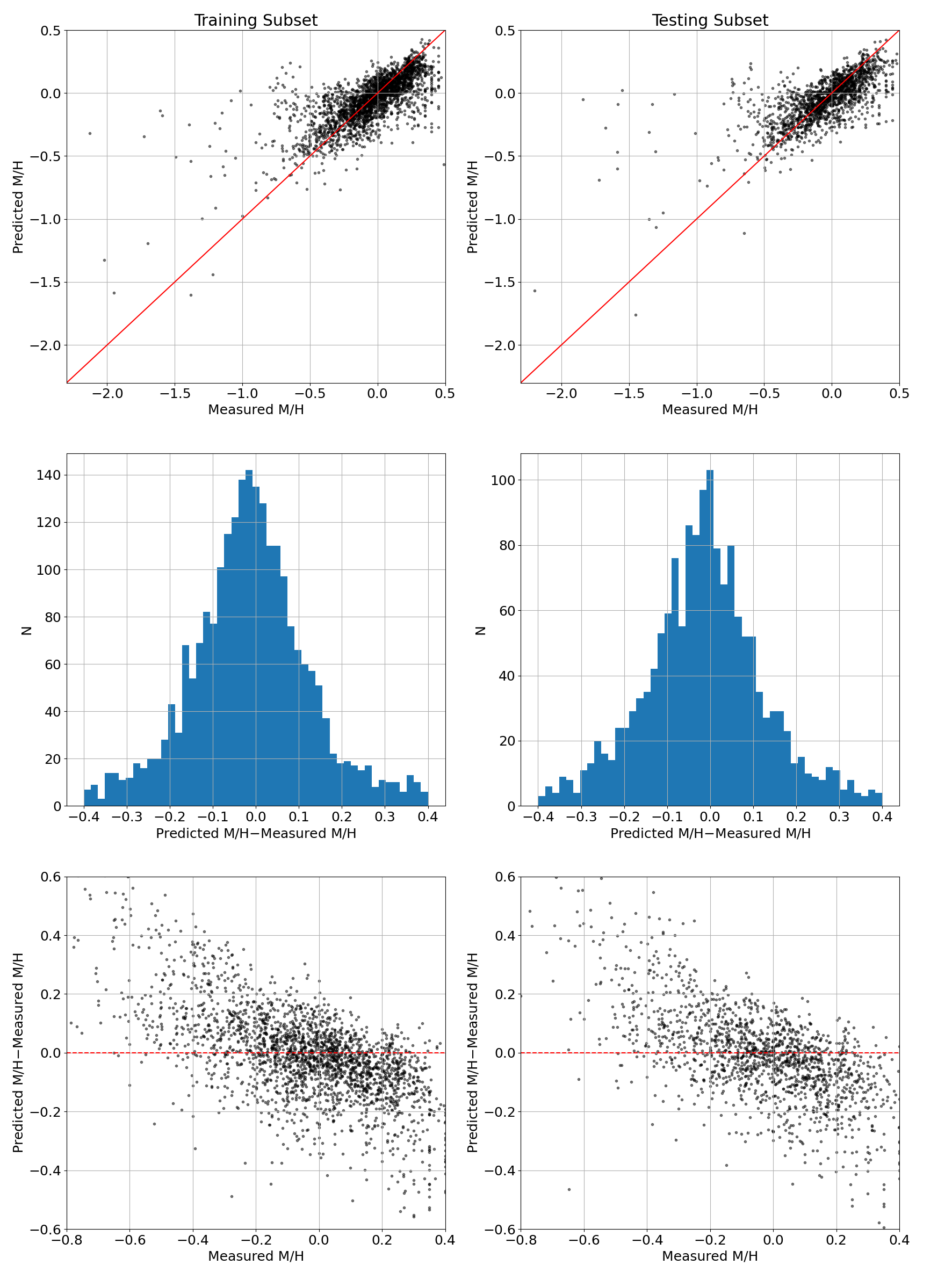}
	\caption{Results of the Gaussian Process Regressor on the single star data set for M dwarfs using the following combination of colors and absolute magnitudes: $M_r$, $r-W1$ and $i-K$. The first column shows the result for the training subset (the subset used to train the regressor) and the second column the testing subset (the subset used to evaluate the regressor). The top row compares the predicted metallicity from the regressor to the ``observed" metallicity, i.e., values inferred either from their brighter primaries using our K/early-M regressor or from the spectroscopic analysis of \cite{hejazi2020}, the distribution of the differences between the predicted and APOGEE metallicity, and the bottom row shows the residual error in the predicted versus the ``observed" metallicity.}
	\label{fig:M_regress_results}
\end{figure*}

Using this cleaned sample, a final regression is performed to identify the combination of colors and absolute magnitudes that yields the minimal mean squared error for the testing subset. For this subset, the combination of colors and absolute magnitudes that optimizes the final regressor is $M_r$, $r-W1$ and $i-K$. The results of the regressor for the training subset (the subset used to train the regressor) and the testing subset (the subset used to evaluate the regressor) are shown in Figure \ref{fig:M_regress_results}. A comparison of the predicted and ``observed" metallicity values (Figure \ref{fig:M_regress_results}, top row), suggest that the regressor is generally capable of predicting metallicities from the input colors and absolute with some accuracy. However the regressor struggles with more metal poor stars: the scatter in the predicted vs. ``observed" metallicities (Figure \ref{fig:M_regress_results}, middle row) shows a $1\sigma$ spread of 0.21 dex for the testing subset, though this does not account for the compounded error from basing this regressor on the results from the K and early M dwarf regressor. Finally, the residual error in predicted vs. ``observed" metallicity  (Figure \ref{fig:KM_regress_results}, bottom row) demonstrates that this scatter is not uniform over the full metallicity range, and that there are systematic offsets, with metal-poor stars having over-estimated predicted values, and metal-rich stars having under-estimated predicted values.

\section{Discussion}

\subsection{Cross-Match Efficiency}

The main motivation of this study was to improve the cross-match to external catalogs for \textit{Gaia} stars with the largest proper motions. Our overall matching rates to the high proper motion subset are listed in Table \ref{tab:best_match_results_compare}. We compare our percentage of high probability matches to that of the cross-match by \citet{gaia_cross_match}. By extracting our high-proper motion targets with external catalog matches, we found that the Best-Neighbor catalog from \citet{gaia_cross_match} for the same subset yields the match rates listed in Table \ref{tab:best_match_results_compare}. The comparison shows that our method, which combines astrometry and photometry, performs better for most catalogs, with the only exception being RAVE DR5. 

\begin{deluxetable}{lcc}
	\tablecaption{Number of high proper motion sources matched with a total probability $>95\%$\label{tab:best_match_results_compare}}
	\tablehead{\colhead{External Catalog} & \colhead{\% Matched} & \colhead{\% Matched} \\ \colhead{}  & \colhead{[\citet{gaia_cross_match}]} & \colhead{[This Study]}}
	\startdata
	2MASS  & 74.8\%& 75.40\% \\
	AllWISE  & 71.9\% & 75.17\%\\
	GALEX DR5 & \nodata & 8.36\% \\
	RAVE DR5 & 0.92\% & 0.63\% \\
	SDSS DR12  & 27.9\%$^\dagger$ & 30.69\%\\
	Pan-STARRS  & 14.3\% & 68.63\%\\
	\enddata
	\tablenotetext{\dagger}{The match done by \citet{gaia_cross_match} was to SDSS DR9 rather than DR12.}
\end{deluxetable}

\begin{figure}[t!]
	\plotone{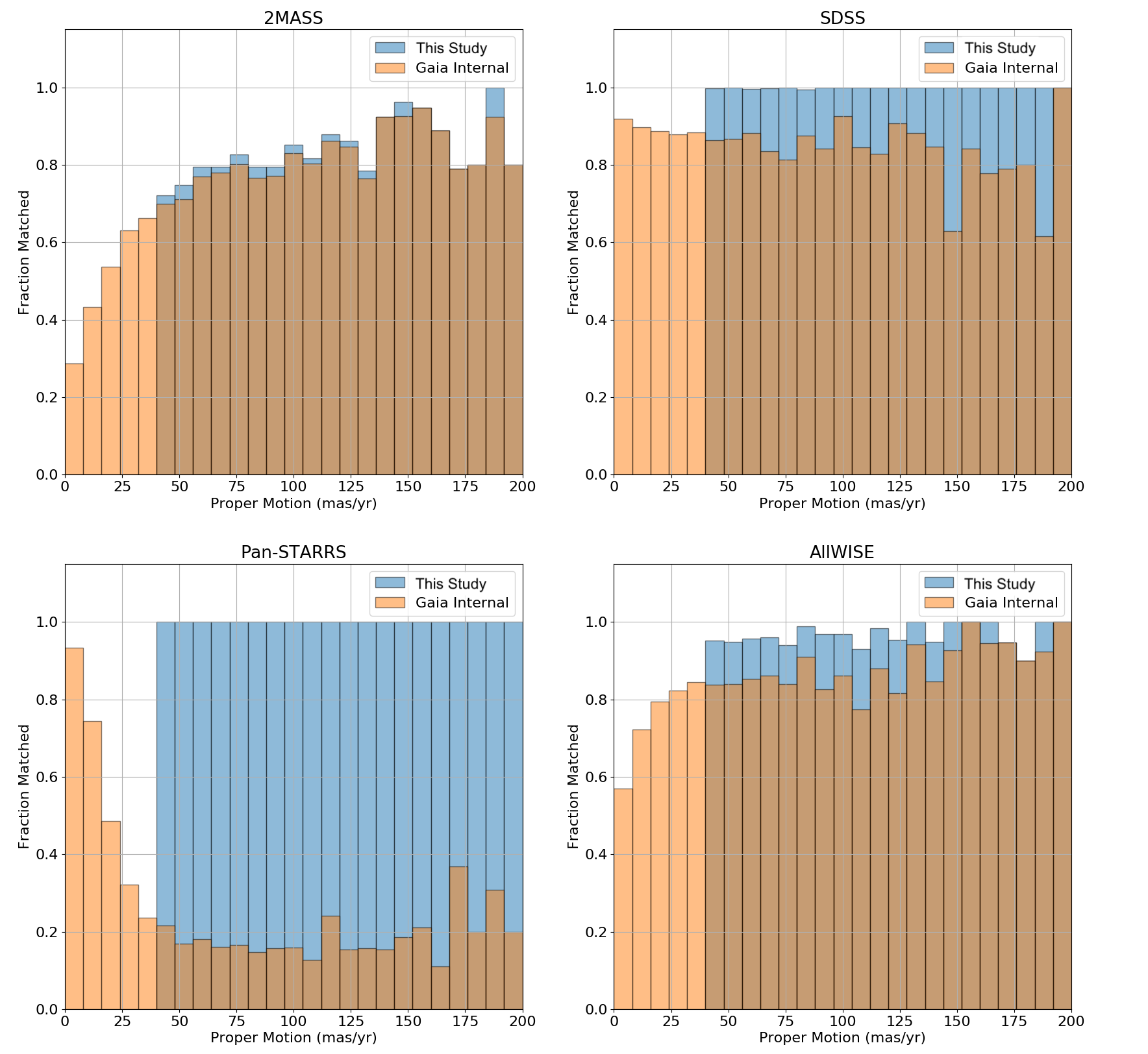}
	\caption{Recovered fraction of \textit{Gaia} stars within 3$^\circ$ of $\alpha,\delta=200^\circ,30^\circ$ that were matched to 2MASS, SDSS, Pan-STARRS and AllWISE by this study (blue bins) compared with the recovered fraction in the match by.\citet{gaia_cross_match} (orange bins). For all external catalogs, this region of sky is 100\% covered by the survey, and thus any deviations from a nearly 100\% match rate are either due to differences in magnitude limit or some systemic errors in the survey positions.}
	\label{fig:match_rate_vs_prop_motion}
\end{figure}

Another way to examine the improvement in the match is to look at the fraction of \textit{Gaia} sources matched as a function of proper motion for a small, uncrowded region of the sky where all external catalogs should have 100\% sky coverage when magnitude limits are not considered. Figure \ref{fig:match_rate_vs_prop_motion} shows this for the four external catalogs with the largest number of matched sources. For external catalogs that are not as deep as \textit{Gaia} (2MASS and AllWISE), we expect that the fraction matched should decrease with decreasing proper motions, because fainter sources beyond the magnitude limit of the external catalogs are more likely to have small proper motions. This pattern is indeed apparent in the 2MASS and AllWISE recovery rates both  from \citet{gaia_cross_match} and from the present study (see Figure \ref{fig:match_rate_vs_prop_motion}). For external catalogs deeper than \textit{Gaia} (SDSS and Pan-STARRS), we expect that nearly 100\% of the \textit{Gaia} sources should be matched in uncrowded fields. We find this to be the case for our cross-match results, but the \citet{gaia_cross_match} recovery rates tell a different story, as it is the high proper motion stars that have a lower match rate. In particular, the match to Pan-STARRS from \citet{gaia_cross_match} is proper motion dependent in a way that suggests significant issues in the Pan-STARRS pipeline in regards to the mean positioning or epoch determination, which then results in poor match rates for high proper motion stars unless these systematic errors are accounted for, as we have done in the present study. We note that these issues also seem present in \citet{gaia_cross_match} for stars with $\mu < 40 \ mas \ yr^{-1}$, demonstrating that additional improvements could be made using our method. Performing these matches are outside of the scope of the current study, but volume limited cross-matches are planned for future works.

This large improvement for matching high proper motion stars with Pan-STARRS, when taking into account the footprint of the survey (i.e. $\delta>-30^\circ$), results in a match rate of 99.8\% for our method, as compared to 20.8\% in the same region for the match by \citet{gaia_cross_match}. To demonstrate that our additional matches are genuine, Figure \ref{fig:ps1_hr_compare_match} shows the HR diagram using Pan-STARRS photometry combined with matched \textit{Gaia} parallaxes for three groups: ``Same Matches", consists of \textit{Gaia} stars that are matched to the same Pan-STARRS source by us and by \citet{gaia_cross_match}; ``Different Matches", consists of \textit{Gaia} stars that are matched to different Pan-STARRS source by us and by \citet{gaia_cross_match}; and ``Mutually Exclusive Matches", consists of \textit{Gaia} stars for which one method finds a match while the other does not. This figure shows that the large increase in the match rate is in fact real, as we are getting clean main-sequence, giant, and white dwarf loci for the stars in the mutually exclusive group. 

\begin{figure*}
	\plotone{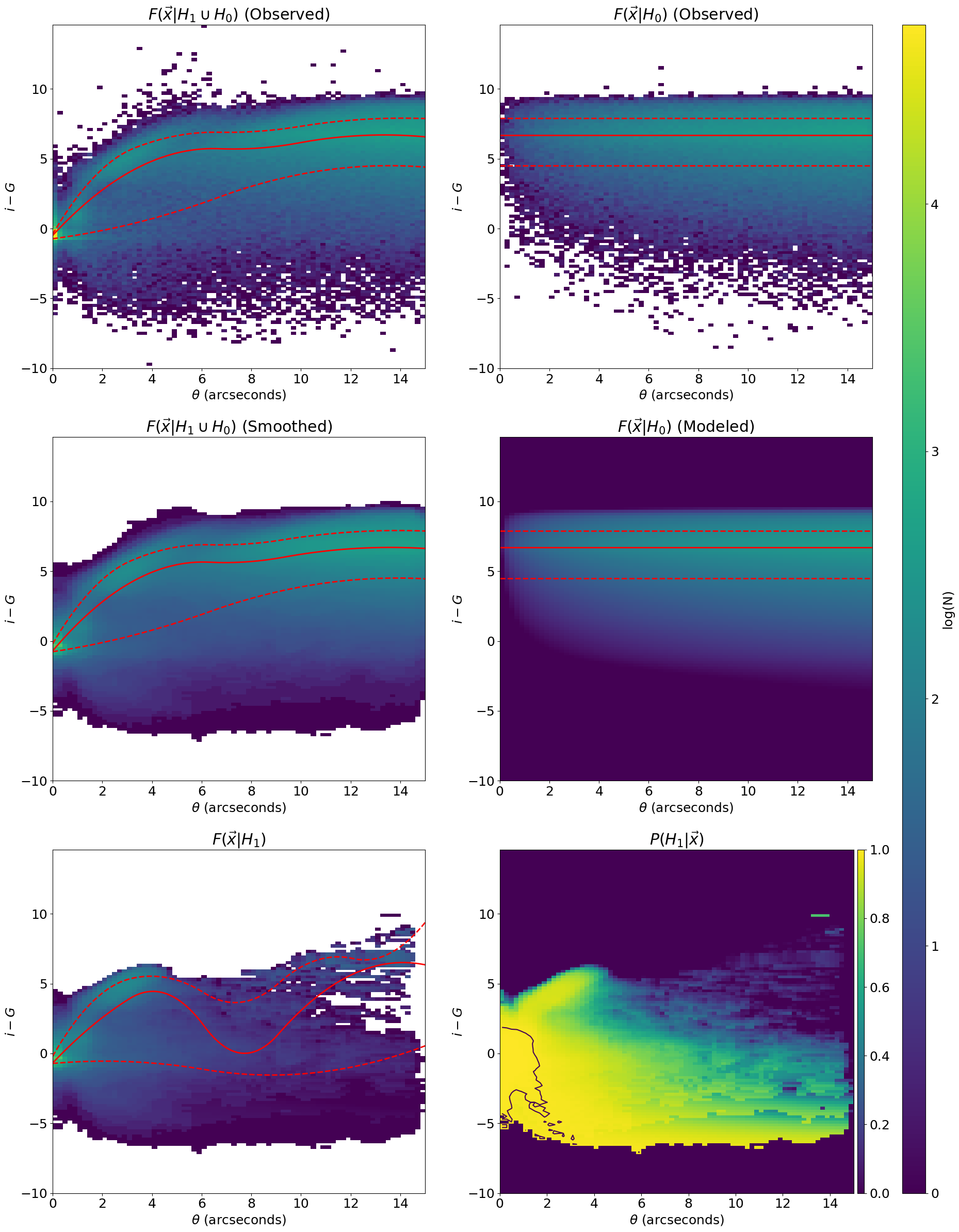}
	\caption{Probability distributions as discussed in Section \ref{sec:bayes} for Pan-STARRS where $12.5<G\leq 15$ and $19.5\leq |b|<41.8$. Top left panel: frequency distribution for the ``true" sample. Top right: frequency distribution for the ``displaced sample". Center left: smoothed frequency distribution for the ``true" sample. Center right: modeled frequency distribution for the ``displaced" sample. Bottom left: difference between the smoothed ``true" frequency distribution and the modeled ``displaced" frequency distribution using the correct scaling factor. Bottom right: resulting Bayesian probability distribution. All plots, except for the distribution for $P(H_1|\vec{x})$, show the number of sources in a bin on a logarithmic scale and they all share the same colormap range (indicated by the colorbar on the far right). The median and 68th percentile of $i-G$ as a function of $\theta$ are shown on these plots as the red solid and red dashed lines, respectively, where quantiles are found using COBS \citep{cobs}. The distribution for $P(H_1|\vec{x})$ is shown for a range of 0 to 1 and the overlaid, black solid line shows the 99\% line.}
	\label{fig:bright_result_ex}
\end{figure*}

\begin{figure*}
    \plotone{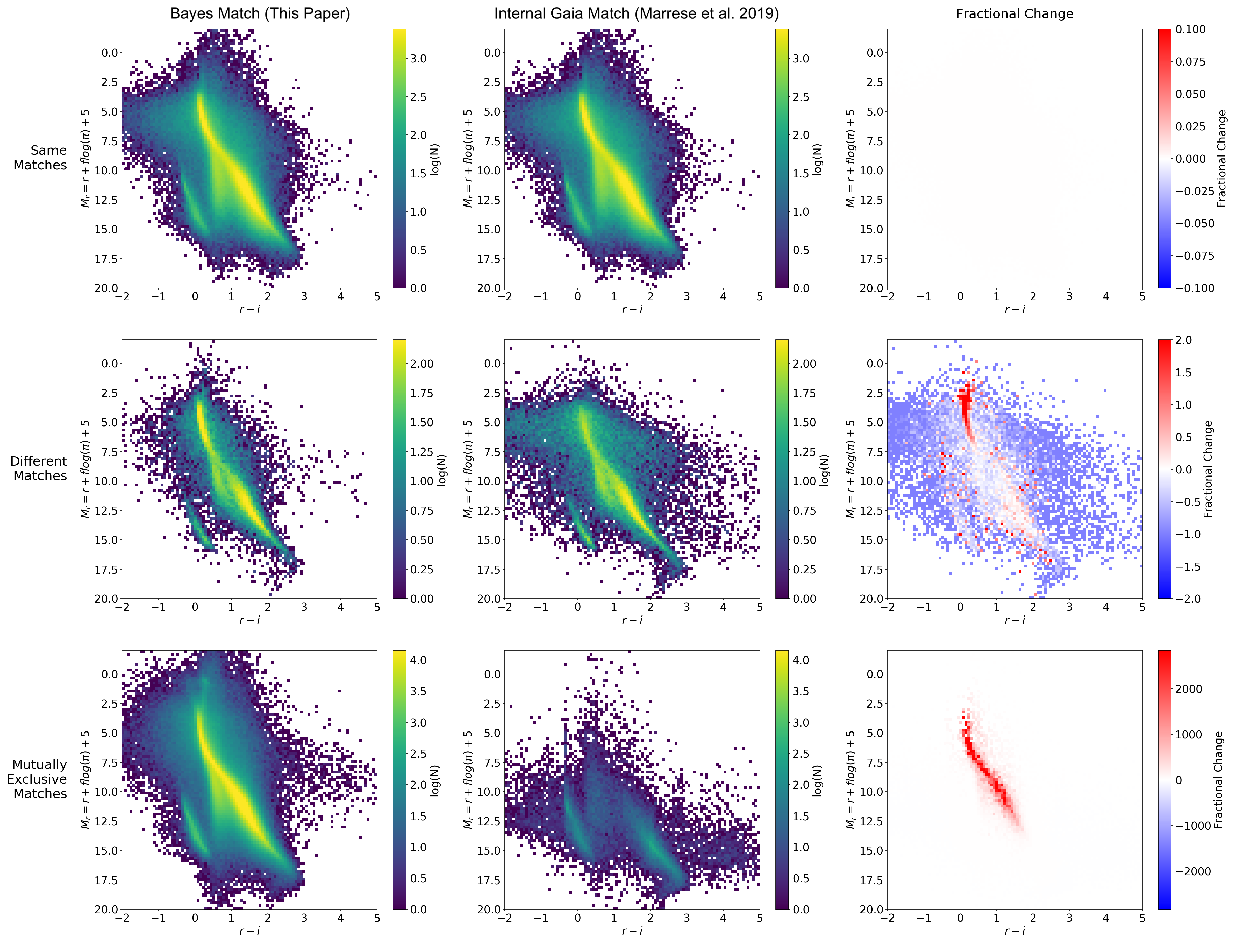}
    \caption{HR diagrams combining Pan-STARRS photometry with \textit{Gaia} parallaxes based on the cross-matches from this study (first column) and the cross-matches from \citet{gaia_cross_match} (second column). The third column shows the fractional difference between the HR diagrams in the first two columns. The top row shows the ``Same Matches", defined as when both studies find the same match to a Pan-STARRS source (both panels identical by definition); the center row displays the ``Different Matches", defined as when both studies find a match to a \textit{Gaia} source but determine the best match to be a different Pan-STARRS source; and finally the bottom row shows the ``Mutually Exclusive Matches", defined as when one study finds a match to a \textit{Gaia} source when the other study finds no suitable match.}
    \label{fig:ps1_hr_compare_match}
\end{figure*}

In addition, a look at the ``Different Matches" group shows two HR diagrams that look relatively similar in both studies (notably at the faint end), despite matching with different Pan-STARRS sources. We attribute this to the existence of a significant number of duplicates in the Pan-STARRS catalog. This appears to be especially common for high proper motion objects, likely because the Pan-STARRS pipeline is not perfectly able to match detections of  fast moving objects at different epochs. The HR diagram from our study, however, does show a ``cleaner'" main sequence (with less scatter) at the bright end, which suggests that our study generally finds the Pan-STARRS duplicate entry with the more reliable photometry. Part of this success comes from our more efficient accounting of epoch differences, but also from our Bayesian probability method having a preference for matches with smaller magnitude differences to the Gaia sources, as demonstrated when comparing the 99\% lines in the Bayesian probability distributions in Figures \ref{fig:result_ex} and \ref{fig:bright_result_ex}.

\begin{figure*}
    \plotone{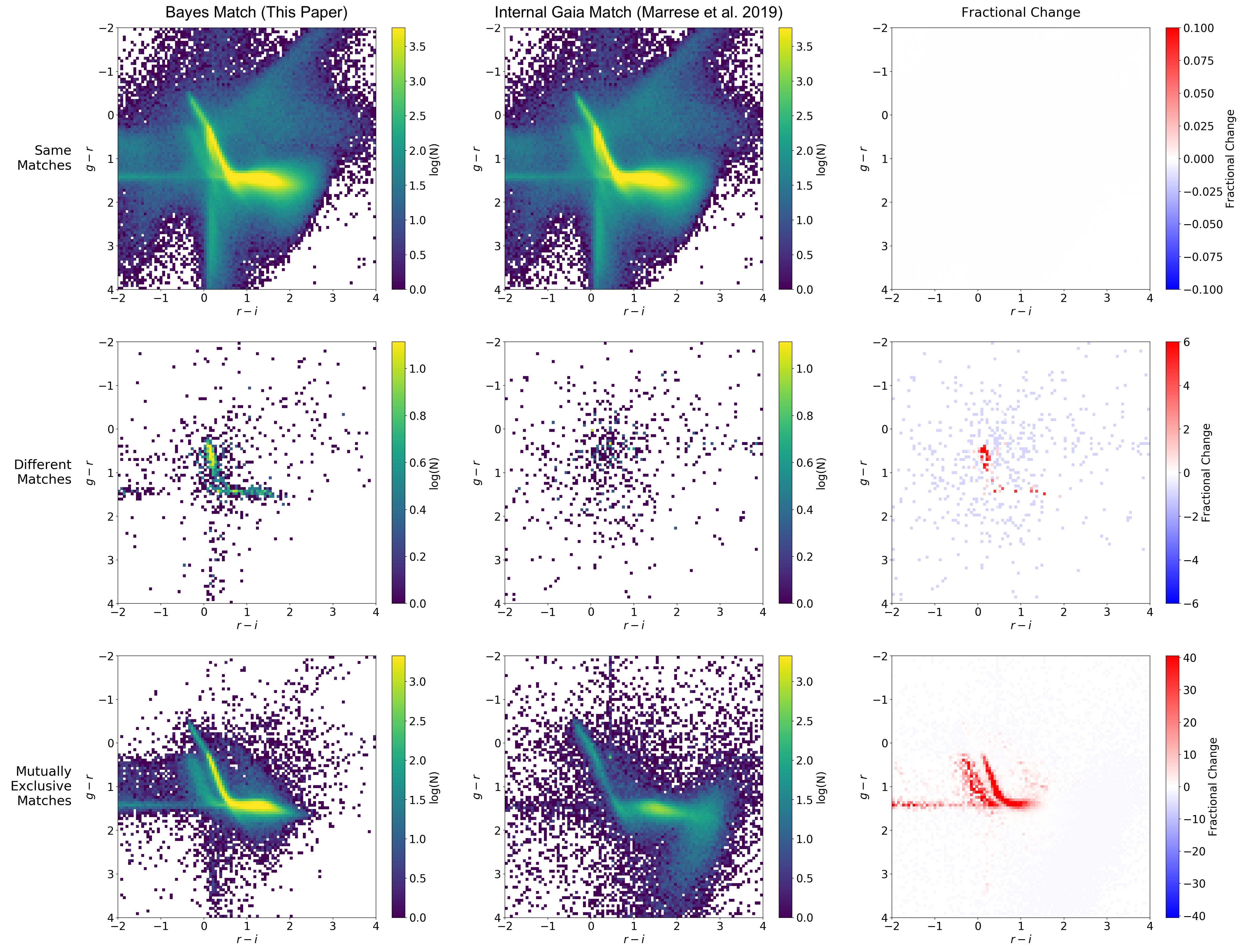}
    \caption{Standard $[g-r,r-i]$  color-color diagram of the SDSS DR12 sources matched in this study (first column) and of the SDSS DR9 match done by \citet{gaia_cross_match} (second column). The third column shows the fractional difference between the color-color diagrams in the first two columns, where all color maps are centered on zero, which is represented as white in the color map. Three subsets are represented: ``Same Matches", defined as when both matches find the same match to a Pan-STARRS source (top row; identical for both studies by definition); ``Different Matches", defined as when both matches find a match to a \textit{Gaia} source but we determine the best match to be a different Pan-STARRS source (center row); and ``Mutually Exclusive Matches", defined as when one match found a match to a \textit{Gaia} source when the other match did not (bottom row).}
    \label{fig:sdss_ccd_compare_match}
\end{figure*}

While we see the largest improvement in the Pan-STARSS cross match, we also see modest, but significant, improvements in the cross-match to other catalogs, like SDSS. Figure \ref{fig:sdss_ccd_compare_match} compares standard $[g-r,r-i]$ color-color diagrams for our match and the match by \citet{gaia_cross_match}, where the layout is the same as in Figure \ref{fig:ps1_hr_compare_match}. Two notable trends are apparent. First, we do find that for some matches that differ between the studies (``Different Matches" group) our cross-match again produces a cleaner subset with better photometry, i.e., more consistent with the expected main-sequence locus, which suggest that our match is more accurate. Second, we can see in the ``Mutually Exclusive Matches" group that our matches shows a denser main-sequence locus, and also show a clear ``white dwarf-M dwarf binary bridge", which is a well-known arc that connects the white dwarf and M dwarf loci on the color-color diagrams \citep[e.g.][]{augusteijn2008,liu2012}. In contrast, the ``Mutually Exclusive Matches" from \citet{gaia_cross_match}, show more stars with poor or inconsistent photometry, notably stars in a diagonal locus redder in $r-i$ than the expected M dwarf locus, and with inconsistent colors for a nearby main-sequence star.

Overall, this demonstrates that our cross-match method not only produces a more complete catalog of counterparts, but also provides cleaner photometry. Our catalog of cross-matches is also ``probabilistic", and users can lower the probability threshold in order to get a more complete catalog at the expense of having a higher rate of false matches.

\subsection{Photometric Metallicity Relationships}

\subsubsection{K and Early M Dwarf Relationship}

Using machine learning with a Gaussian Process Regressor, we show that it is possible to predict the metallicity of stars in a testing subset of K and early M dwarfs to within $\pm$0.12 dex, using a combination of absolute magnitudes and color terms. For comparison, past efforts to determine photometric metallicities in the same temperature range by \citet{Schmidt2016} and \citet{davenport2019} could predict metallicities with a claimed precision of 0.18 dex and 0.11 dex, respectively. However both studies used relatively metal-rich subsets, with the relationship from \citet{Schmidt2016} tested over $-1<M/H<0.2$ and \citet{davenport2019} over $-1<M/H<0.5$. The relationship from \citet{Schmidt2016} also covered a smaller temperature range of $3550<T_{eff}<4200$, while the temperature range for \citet{davenport2019} is unclear. In contrast, our regressor is calibrated over a broader range of metallicities ($-2.3<M/H<0.5$) and temperatures ($3500<T_{eff}<5280$ K).

\begin{figure*}
	\epsscale{0.95}
	\plotone{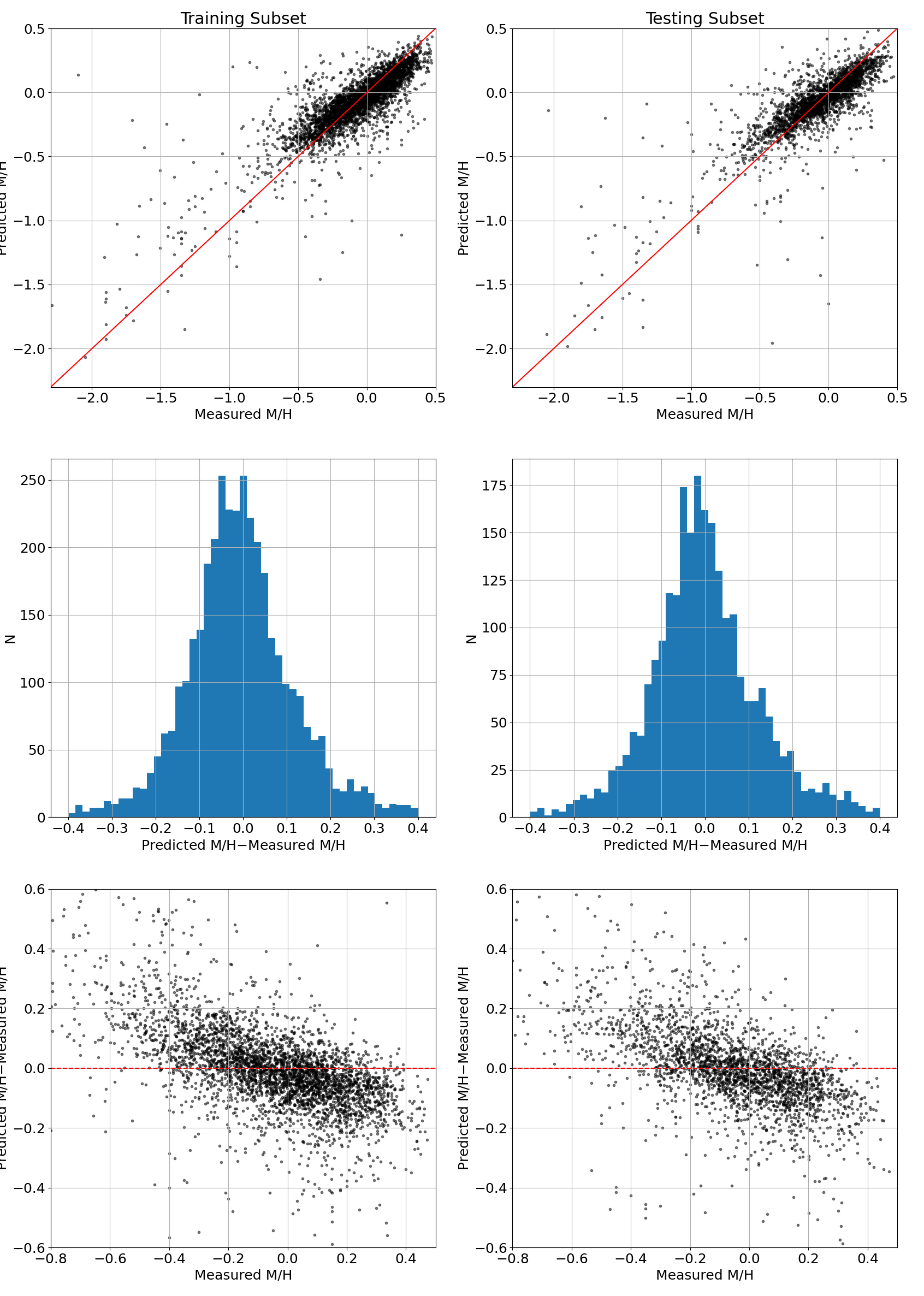}
	\caption{Results of the Gaussian Process Regressor on the training set of K and early M dwarfs but with no preliminary removal of suspected binaries (compare with Figure \ref{fig:KM_regress_results}), and again using the following combination of colors and absolute magnitudes: $M_g$, $g-y$, $y-W2$, $J-W2$, $W1-W2$. The first column shows the result for the training subset (the subset used to train the regressor) and the second column the testing subset (the subset used to evaluate the regressor). The top row compares the predicted metallicity from the regressor to the actual value from APOGEE, the middle row shows the distribution of the differences between the predicted and APOGEE metallicity, and the bottom row shows the residual error in metallicity versus the APOGEE metallicity. Compared with the training subset cleaned of suspected binaries (Figure \ref{fig:KM_regress_results}), this ``unclean" subset yields a regressor which introduces significantly larger systematic and random errors, thus demonstrating the need to use a ``cleaned" training subset.}
	\label{fig:KM_regress_results_w_overlum}
\end{figure*}

Compared with the relationship of \citet{Schmidt2016}, we find that our calibration, besides improving on precision, also significantly improves on known systematic errors. Figure 5 from \citet{Schmidt2016}, shows that predicted metallicity values for metal-poor stars are systematically over-estimated, while predicted values for metal-rich stars are systematically under-estimated. Such a trend is not observed with our regressor (Figure \ref{fig:KM_regress_results}, bottom row). These systematic errors are likely due to contamination from unresolved binaries in their sample, which is exactly what we have attempted to correct for in the present study. To demonstrate this effect, Figure \ref{fig:KM_regress_results_w_overlum} shows the results from our regressor when it is trained and evaluated on the K and early M dwarf data set before the unresolved binaries are removed. Not only is the scatter greater when the binaries are left in (Figure \ref{fig:KM_regress_results_w_overlum}, first and second row), but the residuals between the photometric metallicity estimates and the measured metallicity values show the same systematic errors as in previous studies (Figure \ref{fig:KM_regress_results_w_overlum}, third row). This demonstrates the importance of removing such contaminates from the training sample before extracting a relationship between photometry and metallicity. Another important distinction between our study and \citet{Schmidt2016} is that our regressor uses Pan-STARRS for the optical magnitudes, while \citet{Schmidt2016} uses those from SDSS. Due to the much larger sky coverage of Pan-STARRS (including large swaths of the Galactic plane overlooked by SDSS), our regressor can be used for a much larger number of stars than that of \citet{Schmidt2016}.

\begin{figure*}
	\plotone{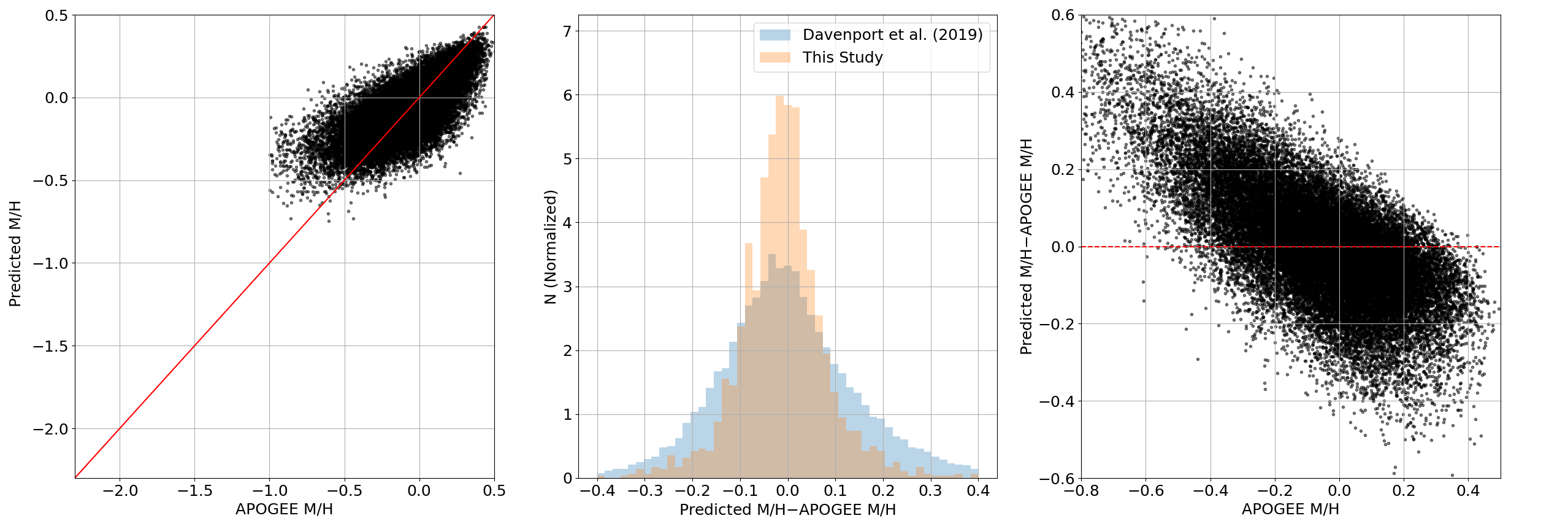}
	\caption{Results of the k-nearest neighbors regressor from \citet{davenport2019}. The left panel compares the predicted metallicity from the regressor to the actual value from APOGEE, the middle panel shows the distribution of the differences between the predicted and APOGEE metallicity for \citet{davenport2019} (blue bins) and this study (orange bins), and the right panel shows the residual error in metallicity versus the APOGEE metallicity.}
	\label{fig:KM_regress_results_davenport}
\end{figure*}

\citet{davenport2019} also used a machine learning technique, with a k-nearest neighbors regressor, in order to estimate photometric metallicities. They have provided their code and data set used for training\footnote{https://github.com/jradavenport/ingot}, and we have used these to predict metallicites for their training sample. We have also included $M_G$ as part of the inputs for their regressor, which they note improves the final results. The results from the \citet{davenport2019} regressor are shown in Figure \ref{fig:KM_regress_results_davenport} for comparison. The scatter in the distribution is found to be comparable to that from our own regressor, although the tails of our scatter distribution are much lower, the systematic trend of the metallicities of metal-poor stars being over-estimated, and the metallicities of metal-rich stars being underestimated is however stronger in the \citet{davenport2019} regressor results, suggesting that the sample used for training includes contaminates in the form of unresolved binaries which are skewing the results.

\subsubsection{M Dwarf Relationship}\label{sec:M_dwarf_discuss}

While our removal of unresolved binaries in the K and early M subset appears to be successful in reducing systematic errors, we cannot currently achieve the same results for our M dwarf calibration. Our current M dwarf regressor does show significant systematic errors consistent with binary star contamination (Figure \ref{fig:KM_regress_results_w_overlum}), which suggests that our attempted removal of unresolved binaries is less effective than in the K and early M subset. This result is affirmed by the polynomial fits in the right panel of Figure \ref{fig:M_single_star}, where it is clear the fits did not converge and effectively remove all of the unresolved binaries. This result is to be expected if we consider that the width of the main sequence is significantly wider for M dwarfs, thus making the removal of unresolved binaries more challenging with our polynomial method. Binary removal is especially challenging for metal-poor stars, which have minimal representation in the training subset in any case. 

\begin{figure*}
	\plotone{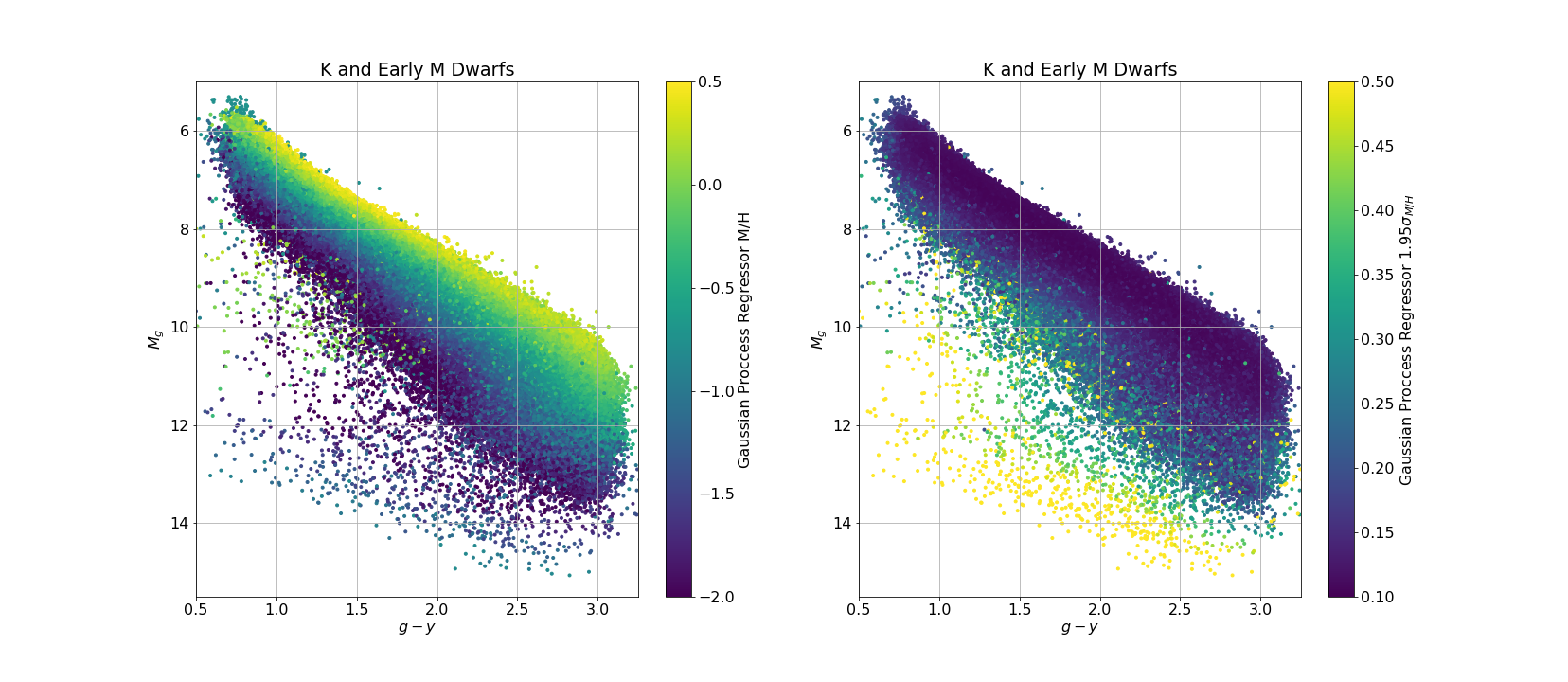}
	\plotone{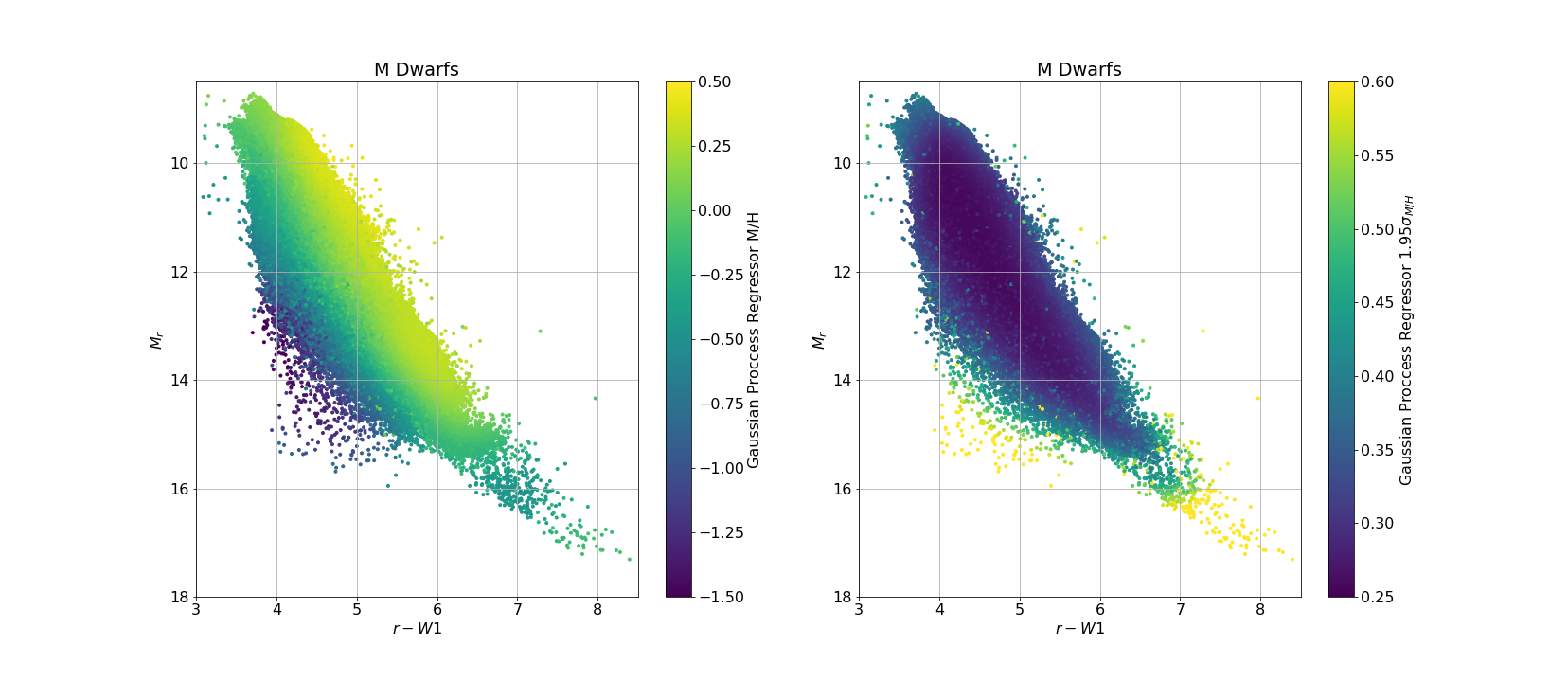}
	\caption{HR diagrams for the 716,651 K and early M dwarfs (top row), and for the 1,030,761 M dwarfs (bottom) for which we estimated metallicities using our Gaussian Process regressor, with the metallicity values shown as a color scale in the panels on the left, and the accuracy of the estimate shown as a color scale in the panels on the right. These represent all  stars with the required photometry in the cross matched high proper motion sample, that have a parallax error $<20\%$, have passed the single star polynomial fit pass, and have predicted metallicities within the range of the training subsets (i.e. $-2.3<M/H<0.5$) shown on respective HR diagrams. The panels on the right show the 95\% confidence interval for the metallicity measurements shown in the left column.}
	\label{fig:KM_all_metals_hr}
\end{figure*}

\begin{figure*}
	\plotone{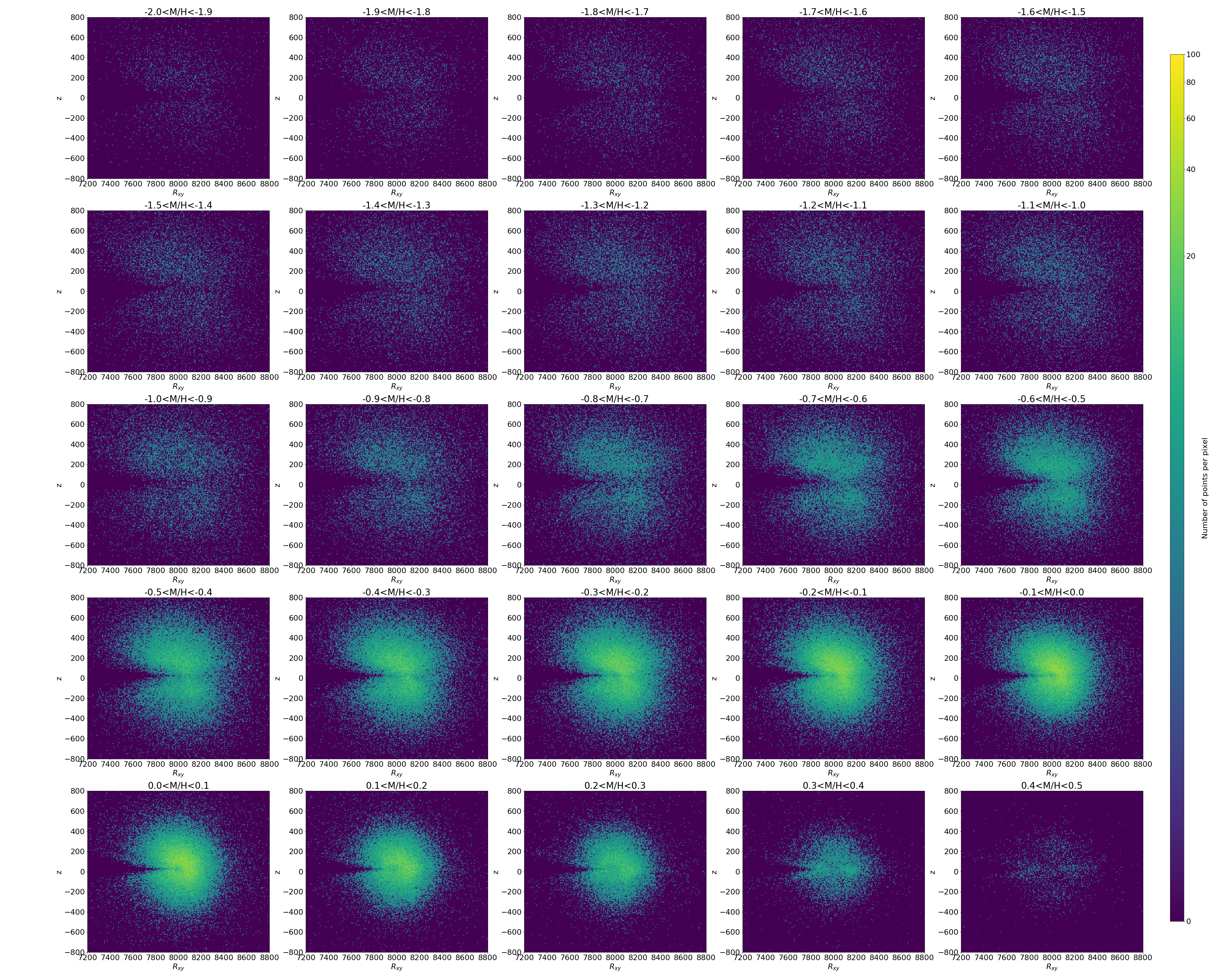}
	\caption{Height above the Galactic plane (z) versus Galactocentric radius for the 716,651 K and early M dwarfs with metallicity values estimated from our Gaussian Process Regressor (see Figure \ref{fig:KM_all_metals_hr}). Each panel shows a subset in a defined metallicity range, with more metal poor stars near the top and more metal rich stars near the bottom. Metal-poor stars show a more dispersed distribution consistent with old disk and halo membership, which metal-rich stars show a more pronounced concentration near the Galactic midplane.}
	\label{fig:K_all_metals_spatial}
\end{figure*}

\begin{figure*}
	\plotone{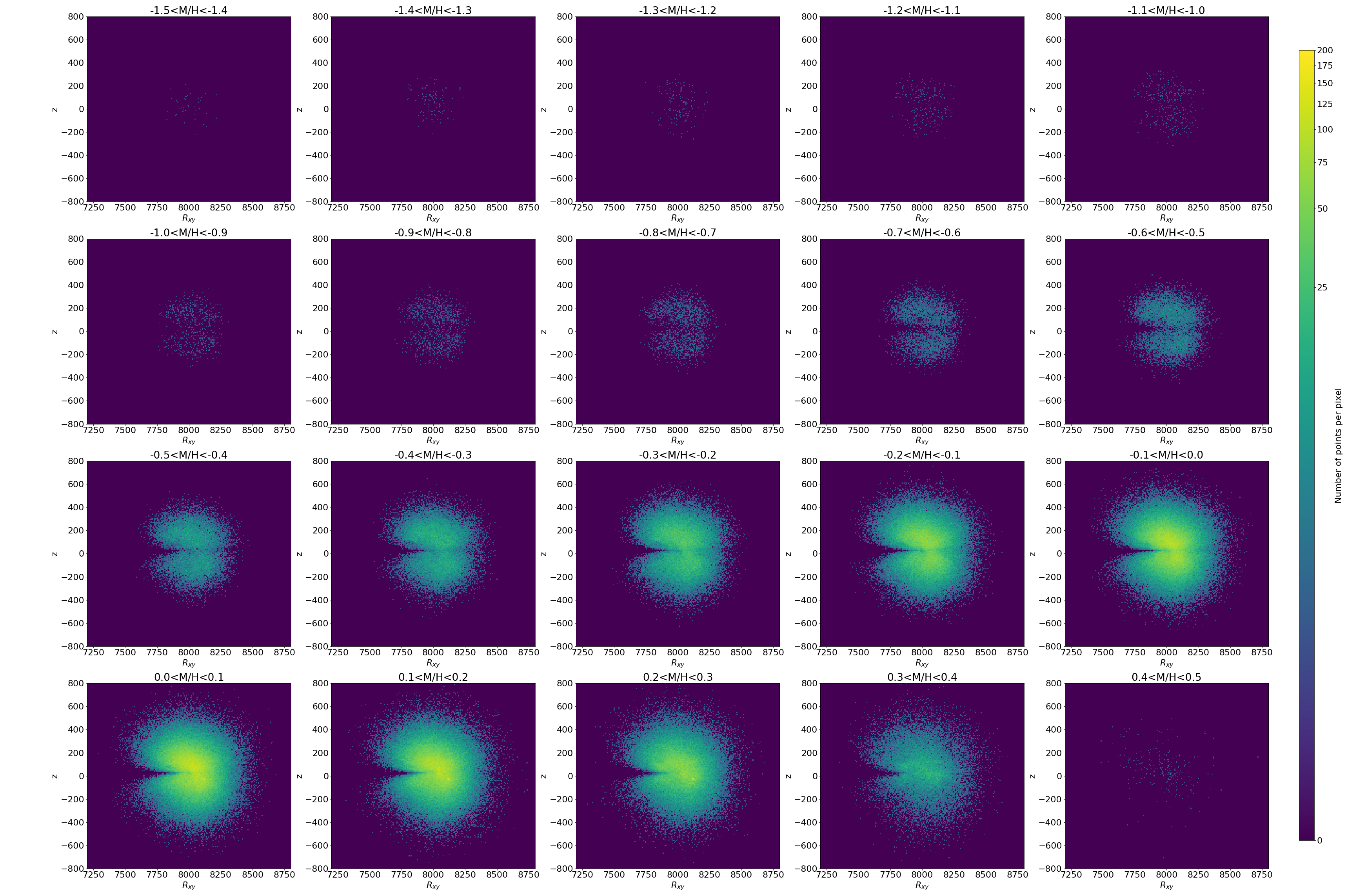}
	\caption{Height above the Galactic plane (z) versus Galactocentric radius for the 1,030,761 M with metallicity values estimated from our Gaussian Process Regressor (see Figure \ref{fig:KM_all_metals_hr}). Each panel shows a subset in a defined metallicity range, with more metal poor stars near the top and more metal rich stars near the bottom. Compared with the K and early M dwarf distribution shown in Figure \ref{fig:K_all_metals_spatial}, the M dwarf distribution appears significantly more distance limited, which is due to the Gaia magnitude limit. This prevents the inclusion of larger numbers of metal-poor objects.}
	\label{fig:M_all_metals_spatial}
\end{figure*}

To further illustrate the challenge of training a regressor for metal-poor M dwarfs, we applied both of the trained regressors to the entirety of the cross matched high proper motion sample in this study. This resulted in 716,651 K and early M dwarfs, and 1,030,761 M dwarfs that have the required photometry, have a parallax error $<20\%$, have passed the single star polynomial fit pass, and have predicted metallicities within the range of the training subsets (i.e. $-2.3<M/H<0.5$). The estimated metallicity values are listed in Tables \ref{tab:K_metals} and \ref{tab:M_metals}. Elimination of suspected binaries using our polynomial cut method removed 23\% of the K dwarfs and 10\% of the M dwarfs. HR diagrams of these samples are presented in Figure \ref{fig:KM_all_metals_hr}, and show a clear deficit of stars at the metal-poor (blue) edge of the main-sequence locus. Figure \ref{fig:K_all_metals_spatial} shows that the distribution of metal-poor K and early-M stars is very spread out compared with the distribution of metal-rich objects, which is more local. The difference comes from the metal-poor stars coming from the local thick disk and halo population, which has a much lower density near the Sun. Figure \ref{fig:M_all_metals_spatial} shows that this is a problem for locating metal-poor M dwarfs, because the Gaia catalog identifies them to a significantly shorter distance range, biasing the sample in favor of more metal-rich M dwarfs

\begin{deluxetable*}{lllll}
	\tablecaption{List of predicted metallicities for 716,651 K and early M dwarfs using the Gaussian Process Regressor outlined in Section \ref{sec:KM_metals}. The regressor uses the following data inputs, $M_g$, $g-y$, $y-W2$, $J-W2$, $W1-W2$, and stars are included only if they pass the single star polynomial cuts outline in Table \ref{tab:KM_poly}.\label{tab:K_metals}}
	\tablehead{\colhead{\textit{Gaia} ID} & \colhead{$\alpha_{Gaia}$} & \colhead{$\delta_{Gaia}$} & \colhead{M/H} & \colhead{$\sigma_{M/H}$} \\ \colhead{} & \colhead{[deg]} & \colhead{[deg]}& \colhead{[dex]}& \colhead{[dex]}}
	\startdata
       2738617902668633344 &  0.00059265319 &  1.41569388809  &0.182  &0.055\\
2420684907087327104  & 0.00064439500 &-14.44069806218  &0.175  &0.057\\
384458916557103744  & 0.00098644795&  42.92964411582 &-0.100 & 0.063\\
2422936569461883648  & 0.00149210844 & -9.59710946245 &-0.320 & 0.057\\
2422910181182822784 &  0.00181091004&  -9.91118104462  &0.138 & 0.054\\
2448241275523065856 &  0.00190792593 & -2.76638530879& -1.609 & 0.156\\
2335002057582474752  & 0.00194199280 &-25.60630079213  &0.302&  0.056\\
2334659529646655744  & 0.00206867958& -26.54384648213 &-0.268 & 0.065\\
2341416475275983872 &  0.00212583048& -21.25877753606  &0.424 & 0.057\\
2414490258576137984 &  0.00254114360& -17.08225464560& -0.129&  0.063\\
	\enddata
	\tablecomments{This table is published in its entirety in the machine-readable format. A portion is shown here for guidance regarding its form and content.}
\end{deluxetable*}

\begin{deluxetable*}{lllll}
	\tablecaption{List of predicted metallicities for 1,030,761 M dwarfs using the Gaussian Process Regressor outlined in Section \ref{sec:M_metals}. The regressor uses the following data inputs, $M_r$, $r-W1$ and $i-K$, and stars are included only if they pass the single star polynomial cuts outline in Table \ref{tab:M_poly}.\label{tab:M_metals}}
	\tablehead{\colhead{\textit{Gaia} ID} & \colhead{$\alpha_{Gaia}$} & \colhead{$\delta_{Gaia}$} & \colhead{M/H} & \colhead{$\sigma_{M/H}$} \\ \colhead{} & \colhead{[deg]} & \colhead{[deg]}& \colhead{[dex]}& \colhead{[dex]}}
	\startdata
        429379051804727296  & 0.00021864741  &60.89771179398  &0.051  &0.133\\
384318827606840448 & 0.00047931984 & 42.19465632646 & 0.308  &0.140\\
2772104109112208256 &  0.00051294158 & 15.56517501763  &0.039  &0.133\\
2846731811580119040 &  0.00115069265& 21.30261306611&  0.329 & 0.148\\
2441058204714858240 &  0.00121147731 & -8.90793754062 &-0.230&  0.138\\
386559121204819328 &  0.00193074605&  45.04225289803 & 0.173& 0.156\\
2313055461894623232 &  0.00204668844& -33.55516167237& -0.008 & 0.181\\
2421214317639640064&   0.00231957574& -12.53175806608& -0.155&  0.133\\
2334666126716440064 &  0.00256127213& -26.36534090966 &-0.071 & 0.133\\
2443194001755562624 &  0.00271533083 & -5.07308824110 & 0.281& 0.138\\
	\enddata
	\tablecomments{This table is published in its entirety in the machine-readable format. A portion is shown here for guidance regarding its form and content.}
\end{deluxetable*}

In the Gaussian Processor Regression method, standard deviations of predicted values can be computed, and these are represented as the color scale in the right panels of Figure \ref{fig:KM_all_metals_hr}. We find that for the highly sampled metallicity regime (i.e. $-0.6<M/H<0.5$ for both samples) the 95\% confidence of our sample is comparable to the $1\sigma$ spread we found in the testing subset for each regressor. However, the errors become much larger for lower metallicity regimes and regions of the HR diagram that are not as well sampled in the training subset. This is a good indicator that while there are limitations with both of these regressors, the predicted errors at least allow us to determine the range over which the method yields the best results.

\begin{figure*}
	\plotone{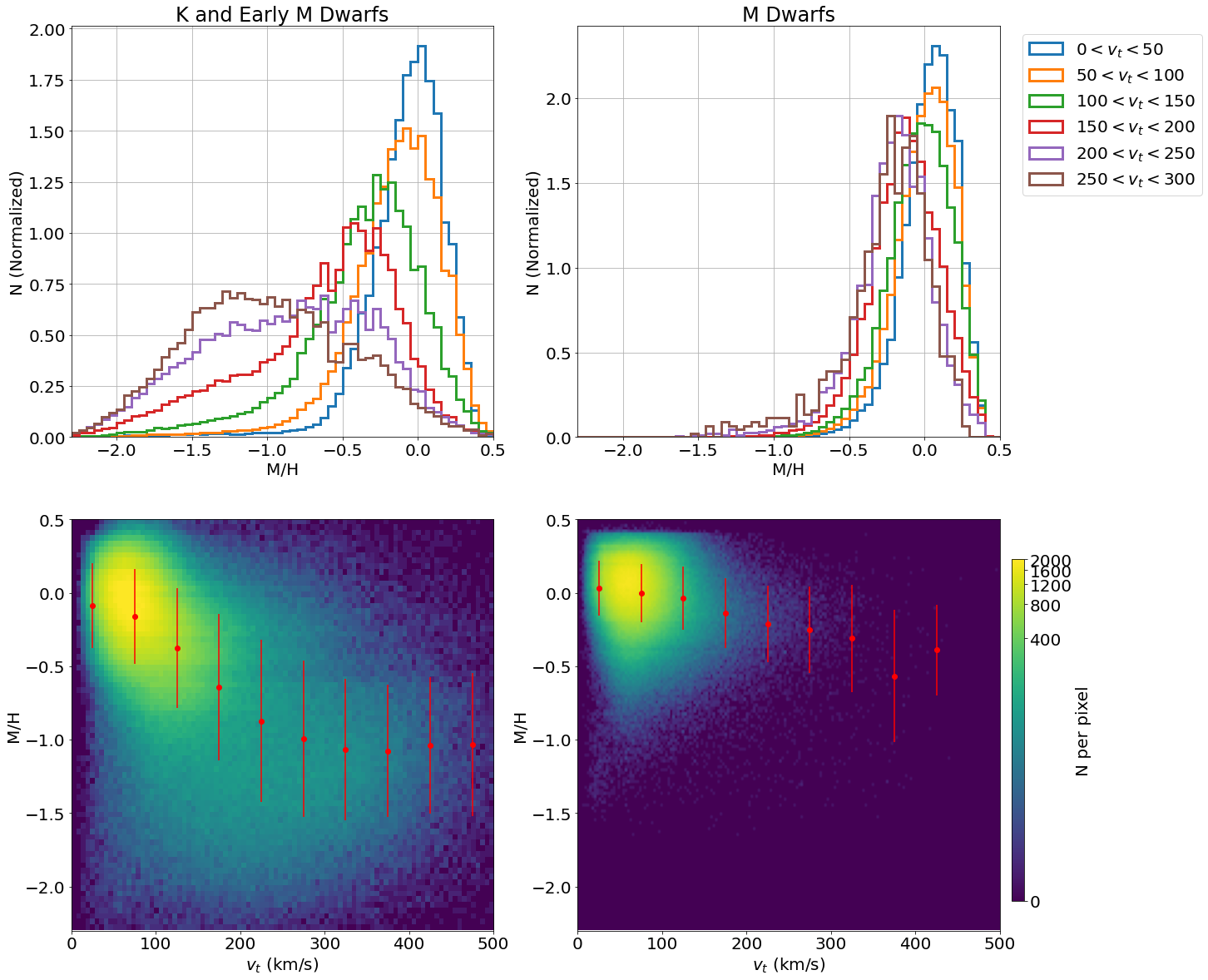}%fix this still!!!
	\caption{Distribution of estimated photometric metallicity values for different cuts of tangential velocity (top row) and photometric metallicity as a function of tangential velocity (bottom row) for the 716,651 K and early M dwarfs (bottom left column) and 1,030,761 M dwarfs (bottom right panel). The red data points and error bars represent the average and standard deviation, respectively, of the predicted metallicity in bins of 50 km/s of tangential velocity. The apparent lack of metal-poor M dwarfs is due to selection effects from the magnitude limit of of the \textit{Gaia} catalog.}
	\label{fig:KM_all_metals_z}
\end{figure*}

Knowing that stellar kinematics is locally correlated with chemical abundances, we validate our predicted metallicity values by examining the range of metallicity estimates for various subsets of K and M dwarfs selected by tangential velocity (Figure \ref{fig:KM_all_metals_z}). For the K and early M dwarfs (Figure \ref{fig:KM_all_metals_z}, left panels), we observe the trend noted in previous studies \citep[e.g.][]{nordstrom2004,ivezic2008,bensby2014,grieves2018} showing a steep negative metallicity gradient as a function of tangential velocity. The trend is however not observed for our subset of M dwarfs (Figure \ref{fig:KM_all_metals_z}, right panels) which shows few metal-poor stars, even at large tangential velocities. We suggest that this again is due to the inherent bias in the initial magnitude-limited, high proper motion \textit{Gaia} subset, which is efficient at selecting more distant, high velocity K dwarfs of the thick disk and halo, but does a poor job at identifying their M dwarf equivalents  because of magnitude limit. In order to improve our metallicity calibration for the M dwarfs, we would need to include a larger sample of nearby, metal-poor M dwarfs in our training subset, which likely will depend on the use of deeper surveys like SDSS, Pan-STARRS, and the upcoming LSST \citep{lsst}.

\subsubsection{Photometric Metallicities of Nearby Clusters: Hyades and Praesepe}

\begin{figure*}
	\epsscale{0.95}
	\plotone{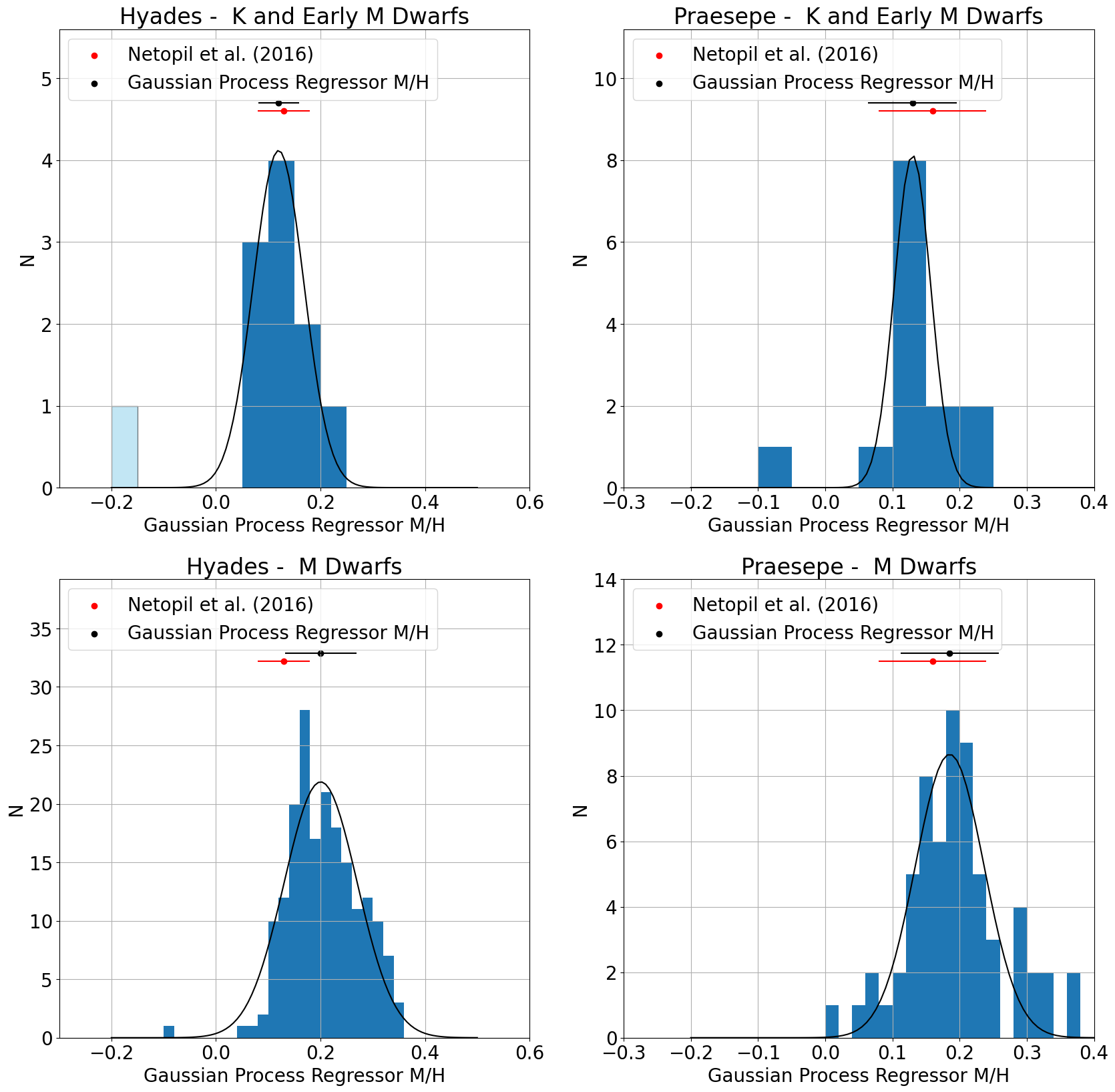}
	\caption{Distributions of estimated photometric metallicities for the subset of K and early M dwarfs (top row), and for the subset of later-type M dwarfs (bottom row) for members of the Hyades (left column) and Praesepe (right column), where membership of these open clusters is from \citet{gaiaHR}. The solid black line is a fitted Gaussian distribution to the histogram. Gaussian fits are shown with the black line, and the measured mean and dispersion values are noted with the black dots and their errobars. For the K and early M dwarf members of the Hyades, the one member at $[M/H]\approx -0.2$ dex (light blue histogram) has not been included in the mean for the cluster as it presented as a clear outlier on an HR diagram. The cluster averages are compared to the spectroscopically derived cluster metallicity estimates from \citet{netopil2016} (red data points and errorbars); our average and RMS errors for the Hyades are $\overline{M/H}_{RMS}=0.120\pm0.039$ for the K and early M Dwarfs, and $\overline{M/H}_{RMS}=0.200\pm0.068$ for the M dwarfs, as compared to $\overline{M/H}_{RMS}=0.13\pm0.05$ from \citet{netopil2016}. The average and RMS errors for Praesepe are $\overline{M/H}_{RMS}=0.130\pm0.064$ for the K and early M Dwarfs, and $\overline{M/H}_{RMS}=0.185\pm0.073$ for the M dwarfs, as compared to $\overline{M/H}_{RMS}=0.16\pm0.08$ from \citet{netopil2016}.}
	\label{fig:KM_metals_clusts}
\end{figure*}

As a final check on our calibrations, we verify the average metallicity of stars in our high proper motion sample (from Section \ref{sec:M_dwarf_discuss}) that are members of nearby open clusters. Specifically, we consider stars that are members of the Hyades and Praesepe open clusters. We use the open cluster members from \citet{gaiaHR}, where members were chosen using an iterative process that determines the cluster velocity and location centroid, as detailed in \citet{gaia_HR_analysis}. We use the \textit{Gaia} IDs from \citet{gaiaHR} to identify members in our high proper motion sample. The distributions of photometric metallicities from both calibrations are shown in Figure \ref{fig:KM_metals_clusts}. The number of cluster members displayed in Figure \ref{fig:KM_metals_clusts} is lower than the count from \citet{gaiaHR}, especially for the K and early M dwarfs, because we only estimate metallicities for stars with $g_{PS1}<13.5$ mag to avoid saturation errors.

A Gaussian fit to the K/early-M metallicity distribution of Hyades members shows an average value of $\overline{M/H}=0.120$ and a dispersion of $\sigma_{M/H} = 0.048$. For the M dwarfs, we find $\overline{M/H}=0.200$ and a dispersion of $\sigma_{M/H} = 0.070$. A Gaussian fit to the K/early-M dwarf in Praesepe yields a average value $\overline{M/H}=0.130$ with a dispersion $\sigma_{M/H}=0.027$, while the M dwarf distribution yields $\overline{M/H}=0.185$ with a dispersion $\sigma_{M/H}=0.051$. We compare these values to the metallicity estimates from \citet{netopil2016}, which provides metallicities derived from open cluster members with high quality spectra. To be consistent with their results, we present our averages, from the fitted Gaussian distributions, along with the RMS error, which are shown as the data points in Figure \ref{fig:KM_metals_clusts}. Additionally, for the K and early M dwarf members of the Hyades, the one member at $[M/H]\approx -0.2$ dex (light blue histogram in Figure \ref{fig:KM_metals_clusts}) has not been included in the mean for the cluster as it presented as a clear outlier on an HR diagram of the cluster members presented here. The average and RMS errors for the Hyades are $\overline{M/H}_{RMS}=0.121\pm0.406$ for the K and early M Dwarfs, and $\overline{M/H}_{RMS}=0.200\pm0.068$ for the M dwarfs, as compared to $\overline{M/H}_{RMS}=0.13\pm0.05$ from \citet{netopil2016}. The average and RMS errors for Praesepe are $\overline{M/H}_{RMS}=0.130\pm0.064$ for the K and early M Dwarfs, and $\overline{M/H}_{RMS}=0.185\pm0.073$ for the M dwarfs, as compared to $\overline{M/H}_{RMS}=0.16\pm0.08$ from \citet{netopil2016}. It seems that, within the uncertainties, our photometric calibrations agree with the spectroscopic estimates from \citet{netopil2016}.

From this test alone, we cannot however rule out systematic errors in our calibrations. This is because the metallicity probed by these two clusters is in a region in metallicity where the results from our calibration indicate either no systematic errors (K and early M dwarfs, Figure \ref{fig:KM_regress_results}) or considerably lessened systematic errors (M Dwarfs, Figure \ref{fig:M_regress_results}), as compared to more metal-poor and metal-rich regimes. A test of the metal-poor and metal-rich regimes may prove difficult due to the limited number of nearby stars clusters, and may have to rely on more direct spectroscopic metallicity measurements from individual stars. With this in mind, we remain cautious about the application of our regressors for metal-poor systems, but with the results in this section do feel confident about the metallicity estimates provided for M dwarfs in the local disk population of the Milky Way.

%\begin{figure}
%	\plotone{M_dwarf_hr_pleides.png}
%	\caption{HR diagram showing the M dwarf members of the Pleiades defined as ``normal" ($M/H<0.178 \ dex$, red data points) and ``metal-rich" ($M/H>0.178 \ dex$, black data points). For comparison, the sample 1,082,527 M dwarfs (right panel) that had the required photometry in the cross matched high proper motion sample, had a parallax error $<20\%$, passed the single star polynomial fit pass and had predicted metallicities within the range of the training subsets (i.e. $-0.8<M/H<0.4$) are also plotted and colored according to predicted metallicity.}
%	\label{fig:M_hr_pleiades}
%\end{figure}

\section{Conclusions}

In this study we sought to develop a Bayesian framework for cross-matching high proper motion stars in \textit{Gaia} to photometric external catalogs of various size and wavelength coverage. This was accomplished by comparing two dimensional distributions that take into account photometry and astrometry of a ``true" initial catalog query and a ``displaced" catalog query, used to infer the distribution of random field stars. Using these distributions, we are able to find match probabilities for all stars within 15" of our \textit{Gaia} sources at the mean epoch of the external survey. These sources and their Bayesian match probabilities are included in an accompanying electronic table.

When comparing the counterparts with the highest match probabilities ($>$95\%) to the counterparts found in the internal \textit{Gaia} cross-match presented in \citet{gaia_cross_match}, we find that our method produces a better match rate, with the largest improvement achieved for Pan-STARRS. Additionally, we show that the photometry produced by our method creates a ``cleaner" catalog overall, with fewer stars having clearly incorrect colors due to mismatches.

With this more complete cross matched catalog of high proper motion sources, we identify a training subset of 4,378 K and early M dwarfs with metallicity values measured from SDSS-APOGEE spectra, and recover their optical and infrared photometry from Pan-STARRS, 2MASS, and AllWISE. We use this training subset to derive a Gaussian Process Regressor that yields metallicity estimates based on the optical and infrared photometry. Our regressor applies to K and early M dwarfs in the color range $0.98<BP-RP<2.39$ mag. In addition, we apply the results of the first regressor on a set of 3,689 common proper motion binaries and infer metallicity for the later-type M dwarf companions in these systems. We use this as a training subset to create a second Gaussian Process Regressor, which we use to estimate metallicities in later type M dwarfs, in the color range $2.39<BP-RP<5$ mag.

We found that our K and early M dwarf regressor performs better than previously calibrated photometric metallicity relationships, with noticeably smaller random and systematic errors, and an effective range extending to significantly more metal-poor populations than in previous work. For the M dwarf regressor, we find larger random errors for the testing subset along with systematic offsets, which we attribute to the contaminating presence of unresolved binaries in our training subset. 

When applying these regressors to the full catalog of high proper motion stars, we find that the metallicity estimates are consistent with that we would expect from the stellar populations probed by our sample based on their location (galactic height) and kinematics (old/young population). Additionally, when we examine the metallicity distributions of stars in our sample that are members of the Hyades and Praesepe open clusters, we find average metallicity values that are consistent with values measured with high resolution spectroscopy of the more massive cluster members. While we do not notice any systematic errors for these clusters, we are not yet able to test the relationship for the more critical metal-poor metallicity regime, due to limitations with the high proper motion/low-mass sample which restricts the census  to very local populations. The validation of the relationship in the metal-poor regime might require direct metallicity measurements through spectroscopic observations. Additionally, for the M dwarf regressor specifically, we note that systematic errors are still present to some degree throughout the metallicity range probed, which is indicative of unresolved binaries in our training sample. Future work will need to be done in order to better remove unresolved binaries from the training subset in order to improve the regression and mitigate these issues.

\acknowledgments

Mr.~Medan gratefully acknowledges support from a Georgia State University Second Century Initiative (2CI) Fellowship.

This work has made extensive use of an astronomical web-querying package in Python, \textit{astroquery} \citep{astroquery}, in the querying of external catalogs.

This work has made use of data from the European Space Agency (ESA) mission \textit{Gaia} (\url{https://www.cosmos.esa.int/gaia}), processed by the Gaia Data Processing and Analysis Consortium (DPAC, \url{https://www.cosmos.esa.int/web/gaia/dpac/consortium}). Funding for the DPAC has been provided by national institutions, in particular the institutions participating in the Gaia Multilateral Agreement.

This work has made use of data from Sloan Digital Sky Survey IV. Funding for the Sloan Digital Sky Survey IV has been provided by the Alfred P. Sloan Foundation, the U.S. Department of Energy Office of Science, and the Participating Institutions. SDSS acknowledges support and resources from the Center for High-Performance Computing at the University of Utah. The SDSS web site is \url{www.sdss.org}. SDSS is managed by the Astrophysical Research Consortium for the Participating Institutions of the SDSS Collaboration, who are listed at \url{www.sdss.org/collaboration/affiliations/}.

This work has made use of data from Pan-STARRS. The Pan-STARRS Surveys (PS1) and the PS1 public science archive have been made possible through contributions by the Institute for Astronomy, the University of Hawaii, the Pan-STARRS Project Office, the Max-Planck Society and its participating institutes, the Max Planck Institute for Astronomy, Heidelberg and the Max Planck Institute for Extraterrestrial Physics, Garching, The Johns Hopkins University, Durham University, the University of Edinburgh, the Queen's University Belfast, the Harvard-Smithsonian Center for Astrophysics, the Las Cumbres Observatory Global Telescope Network Incorporated, the National Central University of Taiwan, the Space Telescope Science Institute, the National Aeronautics and Space Administration under Grant No. NNX08AR22G issued through the Planetary Science Division of the NASA Science Mission Directorate, the National Science Foundation Grant No. AST-1238877, the University of Maryland, Eotvos Lorand University (ELTE), the Los Alamos National Laboratory, and the Gordon and Betty Moore Foundation.

This work makes use of data products from the Two Micron All Sky Survey, which is a joint project of the University of Massachusetts and the Infrared Processing and Analysis Center/California Institute of Technology, funded by the National Aeronautics and Space Administration and the National Science Foundation.

This work makes use of data products from the Wide-field Infrared Survey Explorer, which is a joint project of the University of California, Los Angeles, and the Jet Propulsion Laboratory/California Institute of Technology, funded by the National Aeronautics and Space Administration.

This work makes use of data from GALEX. GALEX is a NASA SMEX program, whose French contribution is funded by the Centre National d’Etudes Spatiales and the Centre National de la Recherche Scientifique.

The work makes use of data from RAVE. Funding for RAVE has been provided by: the Australian Astronomical Observatory; the Leibniz-Institut fuer Astrophysik Potsdam (AIP); the Australian National University; the Australian Research Council; the French National Research Agency; the German Research Foundation (SPP 1177 and SFB 881); the European Research Council (ERC-StG 240271 Galactica); the Istituto Nazionale di Astrofisica at Padova; The Johns Hopkins University; the National Science Foundation of the USA (AST-0908326); the W. M. Keck foundation; the Macquarie University; the Netherlands Research School for Astronomy; the Natural Sciences and Engineering Research Council of Canada; the Slovenian Research Agency; the Swiss National Science Foundation; the Science \& Technology Facilities Council of the UK; Opticon; Strasbourg Observatory; and the Universities of Groningen, Heidelberg and Sydney. The RAVE web site is at https://www.rave-survey.org.

\bibliography{cross_match_bib.bib}

\end{document}